\documentclass[runningheads]{llncs}

\usepackage{graphicx}
\usepackage{comment}
\usepackage{amsmath,amssymb} 
\usepackage{color}
\usepackage{biblatex}
\usepackage{csvsimple}
\usepackage{multirow}
\usepackage[table]{xcolor}
\usepackage{hhline,booktabs}
\usepackage{subfig}
\usepackage{microtype}
\usepackage[font=small]{caption}

\newcommand{\Lagr}{\mathcal{L}}

\newcommand{\etal}{\textit{et al}. }

\newcommand{\eg}{\textit{e}.\textit{g}., }
\newcommand{\vs}{\textit{vs}.\ }
\newcolumntype{L}[1]{>{\raggedright\let\newline\\\arraybackslash\hspace{0pt}}m{#1}}

\usepackage{pdfpages}

\usepackage{times}
\usepackage{microtype}

\usepackage[width=122mm,left=12mm,paperwidth=146mm,height=193mm,top=12mm,paperheight=217mm]{geometry}

\usepackage[breaklinks=true,bookmarks=false]{hyperref}

\begin{document}

\pagestyle{headings}
\mainmatter
\def\ECCVSubNumber{570}  
    
\title{Quantization Guided JPEG Artifact Correction}

\author{Max Ehrlich \\ \texttt{maxehr@umiacs.umd.edu} \and
    Ser-Nam Lim \\ \texttt{sernamlim@fb.com} \and
    Larry Davis \\ \texttt{lsd@umiacs.umd.edu} \and
    Abhinav Shrivastava \\ \texttt{abhinav@cs.umd.edu} 
}

%
\author{Max Ehrlich\inst{1,2} \and
Larry Davis\inst{1} \and
Ser-Nam Lim\inst{2} \and
Abhinav Shrivastava\inst{1}}
\authorrunning{M. Ehrlich et al.}
%
\institute{University of Maryland, College Park, MD 20742, USA \\
\email{\{maxehr,lsd\}@umiacs.umd.edu}, \email{abhinav@cs.umd.edu} \and
Facebook AI, New York, NY 10003, USA \\
\email{sernamlim@fb.com}}

\maketitle

\begin{abstract}   
The JPEG image compression algorithm is the most popular method of image compression because of it’s ability for large compression ratios. However, to achieve such high compression, information is lost. For aggressive quantization settings, this leads to a noticeable reduction in image quality. Artifact correction has been studied in the context of deep neural networks for some time, but the current methods delivering state-of-the-art results require a different model to be trained for each quality setting, greatly limiting their practical application. We solve this problem by creating a novel architecture which is parameterized by the JPEG file’s quantization matrix. This allows our single model to achieve state-of-the-art performance over models trained for specific quality settings.
\dots
\end{abstract}

\section{Introduction}

The JPEG image compression algorithm~\cite{wallace1992jpeg} is ubiquitous in modern computing. Thanks to 
its high compression ratios, it is extremely popular in bandwidth constrained applications. The JPEG algorithm is a lossy compression algorithm, so by 
using it, some information is lost for a corresponding gain in saved space. This is most noticable for low quality settings 

For highly space-constrained scenarios, it may be desirable to use aggressive compression. Therefore, 
algorithmic restoration of the lost information, referred to as artifact correction, has been well studied both in classical literature and in the context of deep neural networks.

While these methods have enjoyed academic success, their practical application is limited by a single 
architectural defect:  
they train a single model per JPEG quality level. The JPEG quality level is an 
integer between 0 and 100, where 100 indicates very little loss of information and 0 
indicates the maximum loss of information. Not only is this expensive to train and deploy, but the quality setting is not known at inference time (it 
is not stored with the JPEG image~\cite{wallace1992jpeg}) making it impossible to use these models in 
practical applications. Only recently have methods begun considering the ``blind'' restoration scenario \cite{kim2019pseudo, kim2020agarnet} with a single network, with mixed results compared to non-blind methods.

\begin{figure}[t]
    \centering
    \includegraphics[width=0.9\linewidth]{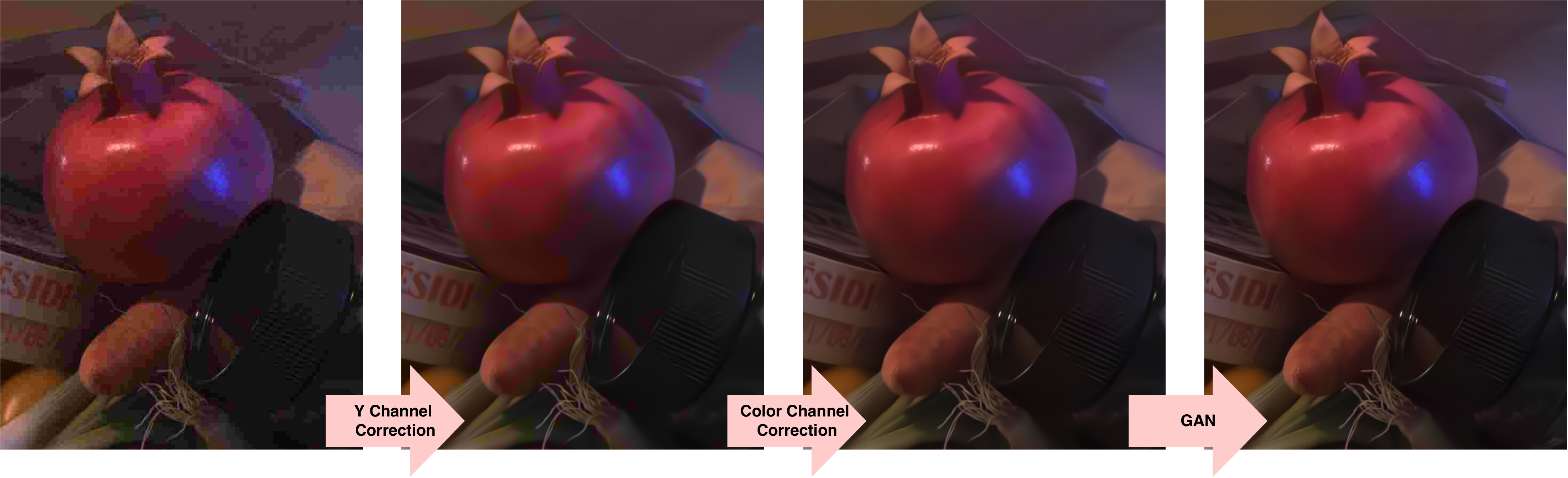}
    \caption{\textbf{Correction process.} Excerpt from ICB RGB 8bit dataset ``hdr.ppm''. Input was compressed at quality 10.}
    \label{fig:header}
\end{figure}

We solve this problem by creating a single model that uses quantization data, which is stored in the JPEG file. Our CNN model processes the image entirely in the DCT~\cite{ahmed1974discrete} domain. While 
previous works have recognized that the DCT domain is less likely to spread quantization errors 
\cite{wang2016d3, zhang2018dmcnn}, DCT domain-based models alone have historically not been successful unless 
combined with pixel domain models (so-called ``dual domain'' models). Inspired by recent methods 
\cite{ehrlich2019deep, deguerre2019fast,dong2015compression,gueguen2018faster}, we formulate fully DCT domain
regression. This allows our model to be parameterized by the quantization matrix, 
an $8 \times 8$ matrix that directly determines the quantization applied to each DCT coefficient. We develop
a novel method for parameterizing our network called Convolution Filter Manifolds, an 
extension of the Filter Manifold technique~\cite{kang2016crowd}. By adapting our network weights to the input 
quantization matrix, our single network is able to handle a wide range of quality settings. Finally, since JPEG images are stored in
the YCbCr color space, with the Y channel containing more information than the subsampled color channels, we
use the reconstructed Y channel to guide the color channel reconstructions. As in~\cite{zini2019deep}, we observe that using the Y channel in this way achieves good color correction
results. Finally, since regression results for artifact correction are often blurry, as a result of lost
texture information, we fine-tune our model using a GAN loss specifically designed to restore texture. This allows 
us to generate highly realistic reconstructions. See Figure \ref{fig:header} for an overview of the correction flow.
 
To summarize, our contributions are:
\begin{enumerate}
    \item A single model for artifact correction of JPEG images at any quality, parameterized by the 
    quantization matrix, which is state-of-the-art in color JPEG restoration.
    \item A formulation for fully DCT domain image-to-image regression.
    \item Convolutional Filter Manifolds for parameterizing CNNs with spatial side-channel information.
\end{enumerate}

\section{Prior Work}

Pointwise 
Shape-Adaptive DCT~\cite{foi2006pointwise} is a standard classical technique which uses thresholded
DCT coefficients reconstruct local estimates of the input signal. Yang~\etal~\cite{yang2000blocking} use a 
lapped transform to approximate the inverse DCT on the quantized coefficients.

More recent techniques use convolutional neural networks~\cite{lecun1990handwritten, sutskever2012imagenet}.
ARCNN~\cite{dong2015compression} is a regression model inspired 
by superresolution techniques; L4/L8~\cite{svoboda2016compression} continues this work. CAS-CNN~\cite{cavigelli2017cas} adds hierarchical skip connections and a 
multi-scale loss function. Liu~\etal~\cite{liu2018multi} use a wavelet-based network 
for general denoising and artifact correction, which is extended by Chen~\etal~\cite{chen2018dpw}. Galteri~\etal~\cite{galteri2017deep} use a GAN formulation to
achieve more visually appealing results. S-Net~\cite{zheng2018s} introduces a scalable architecture that can produce
different quality outputs based on the desired computation complexity. Zhang~\etal~\cite{zhang2020residual} use a dense residual formulation for image enhancement. Tai~\etal 
\cite{tai2017memnet} use persistent memory in their restoration network.

Liu~\etal~\cite{liu2015data} introduce the dual domain idea in the sparse coding setting. Guo and Chao~\cite{guo2016building}
use convolutional autoencoders for both domains. DMCNN~\cite{zhang2018dmcnn} extends this with DCT rectifier to
constrain errors. Zheng~\etal~\cite{zheng2019implicit} target color images and use an implicit DCT layer
to compute DCT domain loss using pixel information. D3~\cite{wang2016d3} extends Liu~\etal~\cite{liu2015data}
by using a feed-forward formulation for parameters which were assumed in~\cite{liu2015data}. Jin \etal \cite{jin2020dual} extend the dual domain concept to separate streams processing low and high frequencies, allowing them to achieve competitive results with a fraction of the parameters. 

The latest works examine the "blind" scenario that we consider here. Zhang~\etal~\cite{zhang2017beyond} formulate general image denoising and apply it to JPEG artifact correction with a single network. DCSC uses convolution features in their sparse coding scheme~\cite{fu2019jpeg} with a single network. Galteri~\etal~\cite{galteri2019deep} extend their GAN work with an ensemble of GANs where each GAN in the ensemble is trained to correct artifacts of a specific quality level. They train an auxiliary network to classify the image into the quality level that it was compressed with. The resulting quality level is used to pick a GAN from the ensemble to use for the final artifact correction. Kim \etal \cite{kim2019pseudo} also use an ensemble method based on quality factor estimation. AGARNET \cite{kim2020agarnet} uses a single network by learning a per-pixel quality factor extending the concept \cite{galteri2019deep} from a single quality factor to a per-pixl map. This allows them to avoid the ensemble method and using a single network with two inputs.

\section{Our Approach}

Our goal is to design a single model capable of JPEG artifact correction at any quality. Towards this, we formulate an architecture that is parameterized by the quantization matrix.

Recall that a JPEG quantization matrix captures the amount of rounding applied to DCT coefficients and is indicative of information lost during compression. A key contribution of our approach is utilizing this quantization matrix directly to guide the restoration process using a fully DCT domain image-to-image regression network. JPEG stores color data in the YCbCr colorspace. The compressed Y channel is much higher quality compared to CbCr channels since human perception is less sensitive to fine color details than to brightness details. Therefore, we follow a staged approach: first restoring artifacts in the Y channel and then using the restored Y channel as guidance to restore the CbCr channels. 

\begin{figure}[t]
  \centering
  \includegraphics[width=0.95\linewidth]{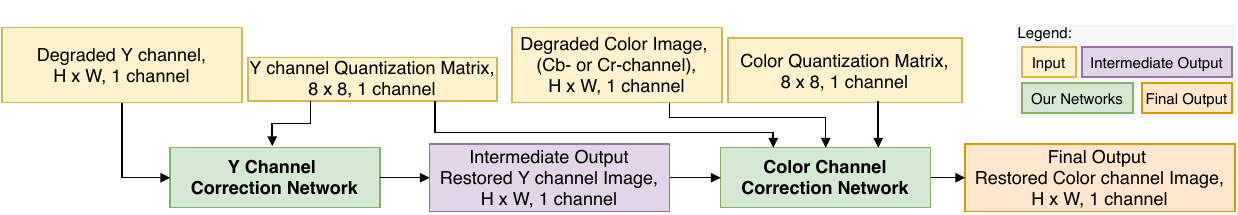}
  \caption{\textbf{Overview.} We first restore the Y channel of the input image, then use the restored Y channel to correct
  the color channels which have much worse input quality.}
  \label{fig:overview}
\end{figure}

An illustrative overview of our approach is presented in Figure~\ref{fig:overview}. 
Next, we present building blocks utilized in our architecture in $\S$\ref{sec:app:bb}, that allow us to parameterize our model using the quantization matrix and operate entirely in the DCT domain.  Our Y channel and color artifact correction networks are described in $\S$\ref{sec:app:y} and $\S$\ref{sec:app:c} respectively, and finally the training details in $\S$\ref{sec:app:train}.

\subsection{Building Blocks}
\label{sec:app:bb}
By creating a single model capable of JPEG artifact correction at any quality, our model solves a significantly harder problem than previous works. To solve it, we parameterize our network using the $8 \times 8$ quantization matrix available with every JPEG file. We first describe Convolutional Filter Manifolds (CFM), our solution for adaptable convolutional kernels parameterized by the quantization matrix. Since the quantization matrix encodes the amount of rounding per each DCT coefficient, this parameterization is most effective in the DCT domain, a domain where CNNs have previously struggled. Therefore, we also formulate artifact correction as fully DCT domain image-to-image regression and describe critical frequency-relationships-preserving operations.

\noindent\textbf{Convolutional Filter Manifold (CFM).}
Filter Manifolds~\cite{kang2016crowd} were introduced as a way to parameterize a deep CNN using side-channel scalar data. The method learns a manifold of convolutional kernels, which is a function of a scalar input. The manifold is modeled as a three-layer multilayer perceptron. The input to this network is the scalar side-channel data, and the output vector is reshaped to the shape of the desired convolutional kernel and then convolved with the input feature map for that layer.

Recall that in the JPEG compression algorithm, a quantization matrix is derived from a scalar quality setting to determine the amount of rounding to apply, and therefore the amount of information removed from the original image. This quantization matrix is then stored in the JPEG file to allow for correct scaling of the DCT coefficients at decompression time. This quantization matrix is then a strong signal for the amount of information lost. However, the quantization matrix is an $8 \times 8$ matrix with spatial structure, applying the Filter Manifold technique to it has the same drawbacks as processing images with multilayer perceptrons,
\eg a large number of parameters and a lack of spatial relationships.

\begin{figure}[t]
    \begin{minipage}{0.45\linewidth}
        \centering
        \includegraphics[width=\linewidth]{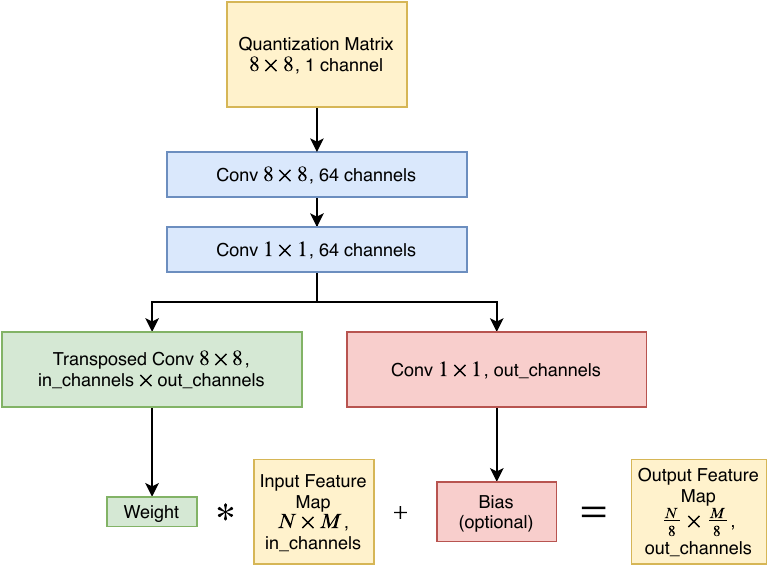}
        \caption{\textbf{Convolutional Filter Manifold}, as used in our network. Note that the convolution with the input
        feature map is done with stride-8.}
        \label{fig:cmf}
    \end{minipage}
    \hspace{0.03\linewidth}
    \begin{minipage}{0.47\linewidth}
        \centering
        \includegraphics[width=\linewidth]{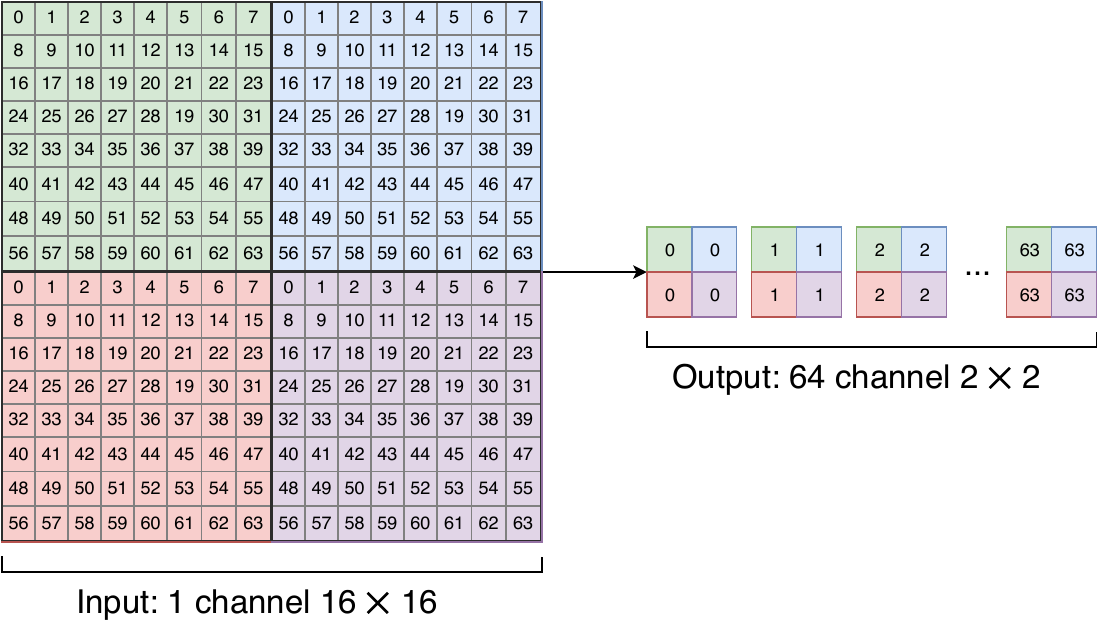}
        \caption{\textbf{Coefficient Rearrangement}. Frequencies are arranged channelwise giving an image with 64 times the number of channels
        at $\frac{1}{8}$th the size. This can then be convolved with 64 groups per convolution to learn per-frequency filters.}
        \label{fig:coef}
    \end{minipage}
\end{figure}

To solve this, we propose an extension to create Convolutional Filter Manifolds (CFM), replacing the multilayer perceptron by a lightweight three-layer CNN. The input to the CNN is our quantization matrix, and the output is reshaped to the desired convolutional kernel shape and convolved with the input feature map as in the Filter Manifold method. For our problem, we follow the network structure in Figure~\ref{fig:cmf} for each CFM layer. However, this is a general technique and can be used with a different architecture when spatially arranged side-channel data is available. 

\noindent\textbf{Coherent Frequency Operations.}
In prior works, DCT information has been used in dual-domain models~\cite{wang2016d3,zhang2018dmcnn}. These models used standard $3 \times 3$ convolutional kernels with U-Net~\cite{ronneberger2015u} structures to process the coefficients. Although the DCT is a linear map on image pixels~\cite{smith1994fast, ehrlich2019deep}, ablation studies in prior work show that the DCT network alone is not able to surpass even classical artifact correction techniques.

Although the DCT coefficients are arranged in a grid structure of the same shape as the input image, that spatial structure does not have the same meaning as pixels. Image pixels are samples of a continuous signal in two dimensions. DCT coefficients, however, are samples from different, orthogonal functions and the two-dimensional arrangement indexes them. This means that a $3 \times 3$ convolutional kernel is trying to learn a relationship not between spatially related samples of the same function as it was designed to do, but rather between samples from completely unrelated functions. Moreover, it must maintain
this structure throughout the network to produce a valid DCT as output. This is the root cause of CNN's poor performance on DCT coefficients for image-to-image regression, semantic segmentation, and object detection (Note that this should not affect whole image classification performance as in~\cite{gueguen2018faster,ghosh2016deep}).

A class of recent techniques~\cite{deguerre2019fast,lo2019exploring}, which we call Coherent Frequency Operations for their preservation of frequency relationships, are used as the building block for our regression network. The first layer is an $8 \times 8$ stride-8 layer~\cite{deguerre2019fast}, which computes a representation for each block (recall that JPEG blocks are non-overlapping $8\times8$ DCT coefficients). This block representation, which is one eighth the size of the input, can then be processed with a standard CNN. 

The next layer is designed to process each frequency in isolation. Since each of the 64 coefficients in an $8\times8$ JPEG block corresponds to a different frequency, the input DCT coefficients are first rearranged so that the coefficients corresponding to different frequencies are stored channelwise (see Figure~\ref{fig:coef}). This gives an input, which is again one eighth the size of the original image, but this time with 64 channels (one for each frequency). This was referred to as Frequency Component Rearrangement in~\cite{lo2019exploring}. We then use convolutions with 64 groups to learn per-frequency convolutional weights. 

Combining these two operations (block representation using $8\times8$ 8-stride and frequency component rearrangement) allows us to match state-of-the-art pixel and dual-domain results using only DCT coefficients as input and output.

\subsection{Y Channel Correction Network}
\label{sec:app:y}
Our primary goal is artifact correction of full color images, and we again leverage the JPEG algorithm to do this. JPEG stores color data in the YCbCr colorspace. The color channels, which contribute less to the human visual response, are both subsampled and more heavily quantized. Therefore, we employ a larger network to correct only the Y channel, and a smaller network which uses the restored Y channel to more effectively correct the Cb and Cr color channels.

\begin{figure}[t]
    \begin{minipage}[t]{0.32\linewidth}
        \centering
        \includegraphics[height=0.9\linewidth]{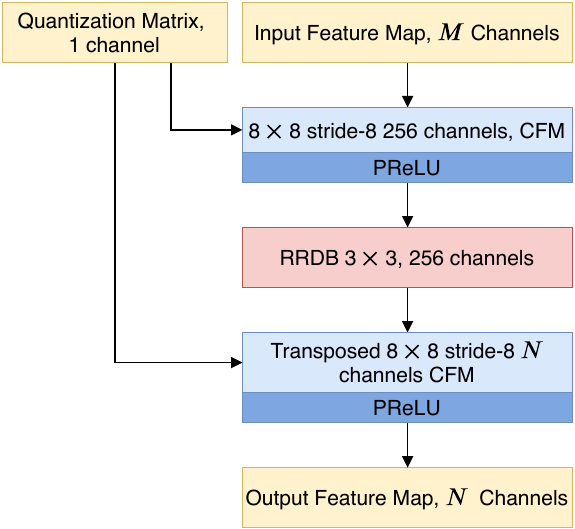}
        \caption{\textbf{BlockNet.} Both the block generator and decoder
        are parameterized by the quanitzation matrix.}
        \label{fig:blocknet}
    \end{minipage}
    \hfill
    \begin{minipage}[t]{0.3\linewidth}
        \centering
        \includegraphics[height=\linewidth]{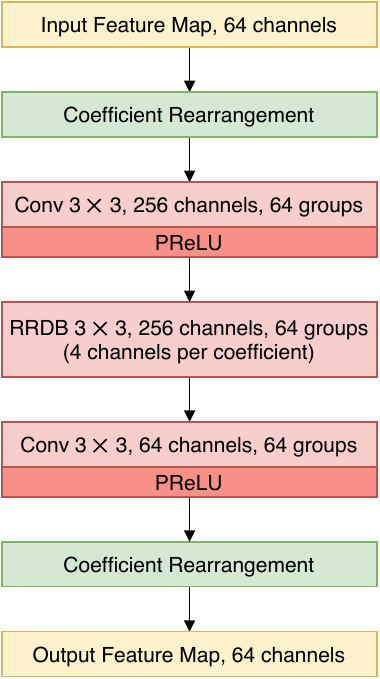}
        \caption{\textbf{FrequencyNet.} Note that the 256 channels in the
        RRDB layer actually compute 4 channels per frequency.}
        \label{fig:frequencynet}
    \end{minipage}
        \hfill
    \begin{minipage}[t]{0.32\linewidth}
        \centering
        \includegraphics[height=\linewidth]{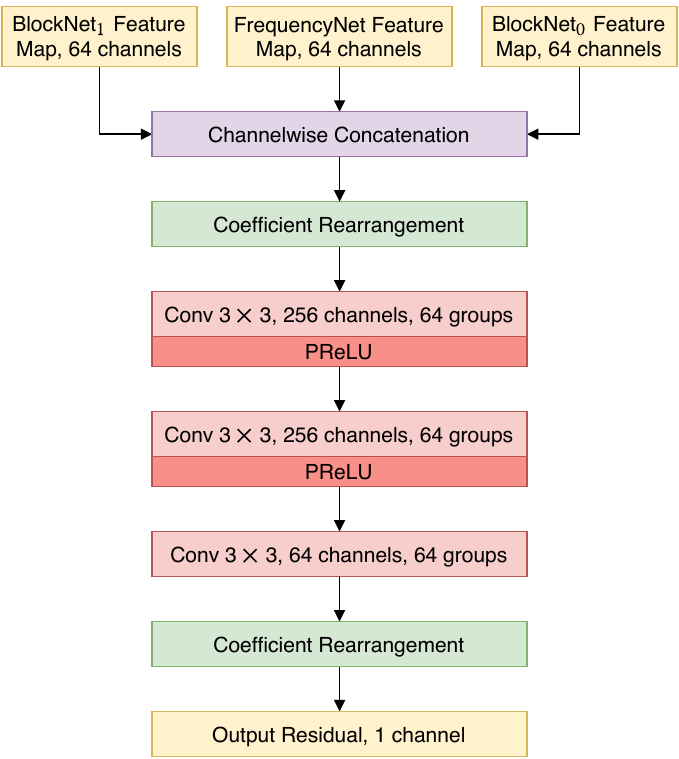}
        \caption{\textbf{Fusion subnetwork.} Outputs from all three subnetworks are fused to produce the final residual.}
        \label{fig:fusion}
    \end{minipage}   
\end{figure}

\noindent\textbf{Subnetworks.}
Utilizing the building blocks developed earlier, our network design proceeds in two phases: block enhancement, which learns a quantization invariant representations for each JPEG block, and frequency enhancement, which tries to match each frequency reconstruction to the regression target. These phases are fused to produce the final residual for restoring the Y channel. We employ two purpose-built subnetworks: the block network (BlockNet) and the frequency network (FrequencyNet). Both of these networks can be thought of as separate image-to-image regression models with a structure inspired by ESRGAN~\cite{wang2018esrgan}, which allows sufficient low-level information to be preserved as well as allowing sufficient gradient flow to train these very deep networks. Following recent techniques \cite{wang2018esrgan}, we remove batch normalization layers. While recent works have largely replaced PReLU~\cite{he2015delving} with LeakyReLU~\cite{maas2013rectifier, wang2018esrgan,galteri2017deep,galteri2019deep}, we find that PReLU activations give much higher accuracy.

\noindent\textbf{BlockNet.} This network processes JPEG blocks to restore the Y channel (refer to Figure~\ref{fig:blocknet}). We use
the $8 \times 8$ stride-8 coherent frequency operations to create a block representation. Since this layer is computing a block representation from all the
input DCT coefficients, we use a Convolutional Filter Manifold (CFM) for this layer
so that it has access to quantization information. This allows the layer to learn the quantization table entry to DCT coefficient correspondence with the goal to output a quantization-invariant block representation. Since there is a one to one correspondence between the quantization table entry and rounding applied to a DCT coefficient, this motivates our choice to operate entirely in the DCT domain. We then process these quantization-invariant block representations with Residual-in-Residual Dense Blocks (RRDB) from~\cite{wang2018esrgan}. RRDB layers are an extension of the commonly used residual block~\cite{he2016deep} and define several recursive and highly residual layers. Each RRDB has 15 convolution layers, and we use a single RRDB for the block network with 256 channels. The network terminates with another 
$8 \times 8$ stride-8 CFM, this time transposed, to reverse the block representation back to its original form so that it can be used for later tasks.

\noindent\textbf{FrequencyNet.} This network, shown in Figure \ref{fig:frequencynet}, processes the individual frequency coefficients using the Frequency Component Rearrangement technique (Figure~\ref{fig:coef}). The architecture of this network is similar to BlockNet. We use a single $3 \times 3$ convolution to change the number of channels from the 64 input channels to the 256 channels used by the RRDB layer. The single RRDB layers processes feature maps with 256 channels and 64 groups yielding 4 channels per frequency. An output $3 \times 3$ convolution transforms the 4 channel output to the 64 output channels, and the coefficients are rearranged back into blocks for later tasks.

\begin{figure}[t]
    \begin{minipage}[t]{0.43\linewidth}
        \centering
        \includegraphics[width=0.9\columnwidth]{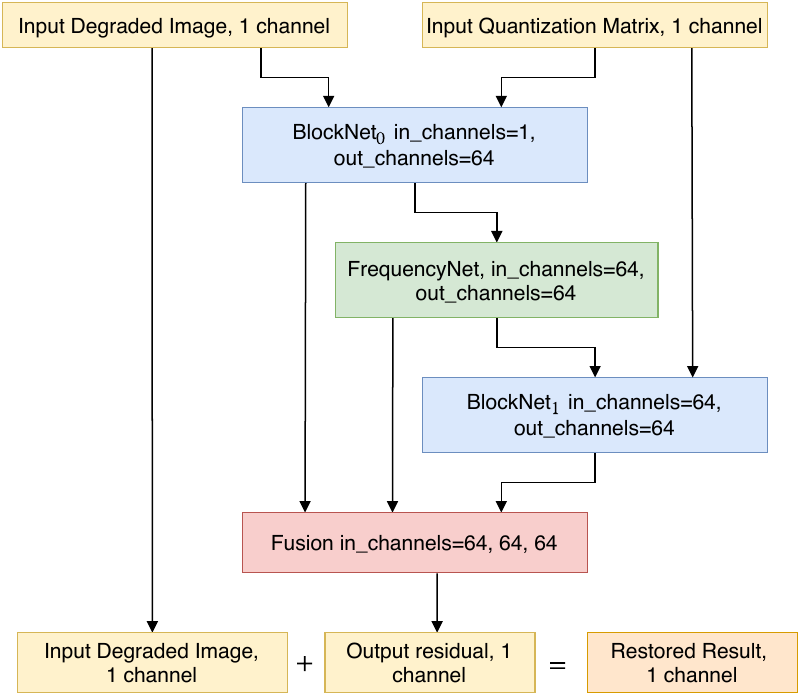}
        \caption{\textbf{Y Channel Network.} We include two copies of the BlockNet, one to perform early restoration of
        high frequency coefficients, and one to work on the restored frequencies. All three
        subnetworks contribute to the final result using the fusion subnetwork.}
        \label{fig:ychannel}
    \end{minipage}
\hfill
    \begin{minipage}[t]{0.54\linewidth}
        \centering
        \includegraphics[width=0.9\columnwidth]{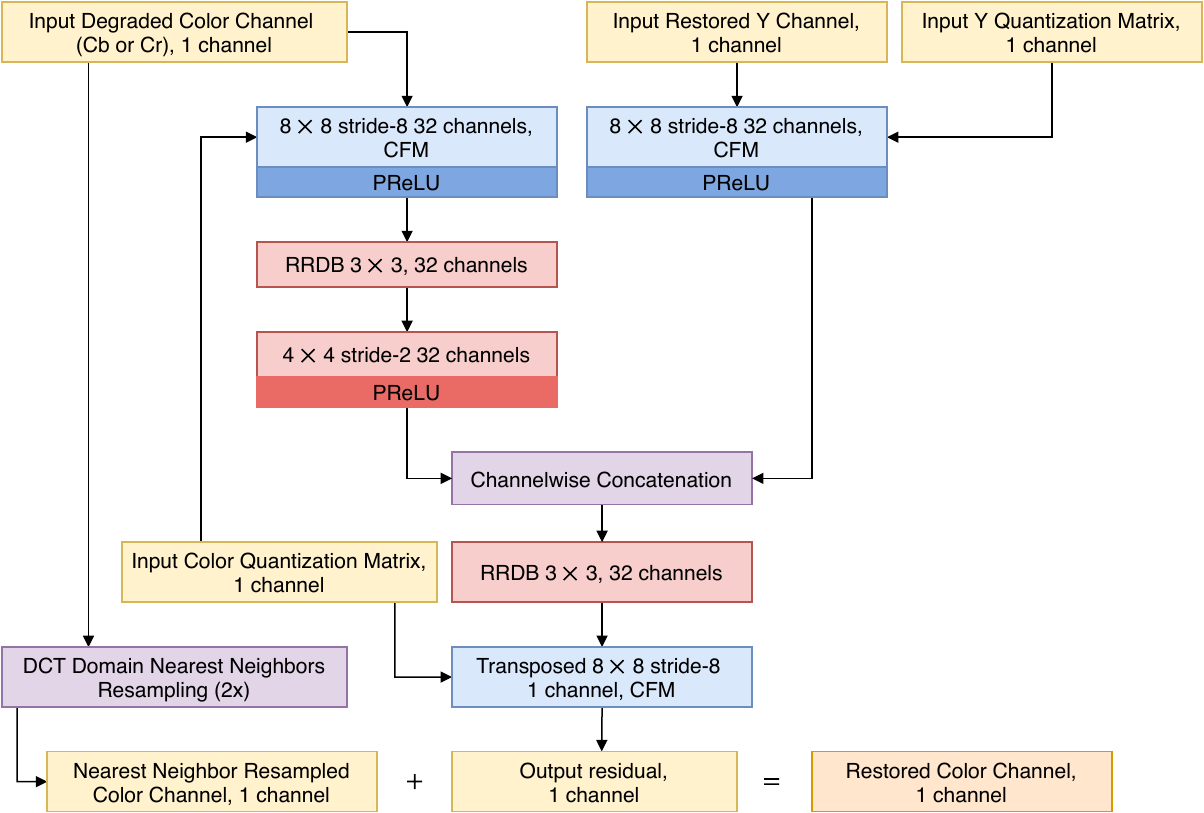}
        \caption{\textbf{Color Channel Network.} Color channels are downsampled, so the block representation is upsampled using a learned upsampling. 
        The Y and color channel block representations are then concatenated to guide the color channel restoration. Cb and Cr channels are processed independently with the
        same network.}
        \label{fig:color}
    \end{minipage}
\end{figure}

\noindent\textbf{Final Network.}
The final Y channel artifact correction network is shown in Figure~\ref{fig:ychannel}. We observe that since the FrequencyNet processes frequency coefficients in isolation, if those coefficients were zeroed out by the compression 
process, then it can make no attempt at restoring them (since they are zero valued they would be set to the layer bias). This is common with high frequencies by design, since they have larger quanitzation table entries and they contribute less to the human visual response. We, therefore, lead with the BlockNet to restore high frequencies. We then pass the result to the FrequencyNet, and its result is then processed by a second block network to restore more information. Finally, a three-layer fusion network (see Figure~\ref{fig:fusion} and~\ref{fig:ychannel}) fuses the output of all three subnetworks into a final result. Having all three subnetworks contribute to the final result in this way allows for better gradient flow. The effect of fusion, as well as the three subnetworks, is tested in our ablation study. The fusion output is treated as a residual and added to the input to produce the final corrected coefficients for the Y channel.

\subsection{Color Correction Network}
\label{sec:app:c}
The color channel network (Figure~\ref{fig:color}) processes the Cb and Cr DCT coefficients. Since the color channels are subsampled with respect to the Y channel by half, they incur a much higher loss of information and lose the structural information which is preserved in the Y channel. We first compute the block representation of the downsampled color channel coefficients using a CFM layer, then process them with a single RRDB layer. The block representation is then upsampled using a $4 \times 4$ stride-2  convolutional layer. We compute the block representation of the restored Y channel, again using a CFM layer. The block representations are concatenated channel-wise and processed using a single RRDB layer before being transformed back into coefficient space using a transposed $8 \times 8$ stride-8 CFM. By concatenating the Y channel restoration, we give the network structural information that may be completely missing in the color channels. The result of this network is the color channel residual. This process is repeated individually for each color channel with a single network learned on Cb and Cr. The output residual is added to nearest-neighbor upsampled input coefficients to give the final restoration.

\subsection{Training}
\label{sec:app:train}
\noindent\textbf{Objective.} We use two separate objective functions to train, an error loss and a GAN loss. Our error loss is based on prior works which minimize the $l_1$ error of the result and the target image. We additionally
maximize the Structural Similarity (SSIM)~\cite{wang2004image} of the result since SSIM is generally regarded as a closer metric to human perception than PSNR. This gives our final objective function as 
\begin{equation}
    \label{eq:regressionloss}
    \Lagr_{\text{JPEG}}(x, y) =  \|y - x\|_1 - \lambda\text{SSIM}(x, y)
\end{equation}
where $x$ is the network output, $y$ is the target image, and $\lambda$ is a balancing hyperparameter.

A common phenomenon in JPEG artifact correction and superresolution is the production of a blurry or textureless result. 
To correct for this, we fine tune our fully trained regression network with a GAN loss. For this objective, we use the relativistic 
average GAN loss $\Lagr^{Ra}_G$~\cite{jolicoeur2018relativistic}, we use $l_1$ error to prevent the image from moving too far away from the regression result, and we use preactivation network-based loss~\cite{wang2018esrgan}. Instead of a perceptual loss that tries to keep the outputs close in ImageNet-trained VGG feature space used in prior works, we use a network trained on the MINC dataset~\cite{bell2015material},
for material classification. This texture loss provided only marginal benefit in ESRGAN \cite{wang2018esrgan} for super-resolution. We find it to be critical in our task for restoring texture to blurred regions, since JPEG compression destroys these fine details. The texture loss is defined as 
\begin{equation}
    \Lagr_{\text{texture}}(x, y) = \|\text{MINC}_{5,3}(y) - \text{MINC}_{5,3}(x)\|_1
\end{equation}
where MINC$_{5,3}$ indicates that the output is from layer 5 convolution 3.
The final GAN loss is 
\begin{equation}
    \label{eq:ganloss} 
    \Lagr_\text{GAN}(x, y) = \Lagr_{\text{texture}}(x, y) + \gamma\Lagr^{Ra}_{G}(x, y) + \nu\|x - y\|_1
\end{equation} 
with $\gamma$ and $\nu$ balancing hyperparameters. We note that the texture restored using the GAN model is, in general,
not reflective of the regression target at inference time and actually produces worse numerical results than the regression model despite the images looking more realistic.

\noindent\textbf{Staged Training.} Analogous to our staged restoration, Y channel followed by color channels, we follow a staged training approach. We first train the Y channel correction network using $\Lagr_{\text{JPEG}}$. We then train the color correction network using $\Lagr_{\text{JPEG}}$ keeping the Y channel network weights frozen. Finally, we train the entire network (Y and color correction) with $\Lagr_\text{GAN}$.

\section{Experiments}

We validate the theoretical discussion in the previous sections with experimental results. We first describe the datasets
we used along with the training procedure we followed. We then show artifact correction results and compare them with previous state-of-the-art methods. Finally, we perform an ablation study.  Please see our supplementary material for further
results and details.

\begin{table}[t]
    \centering
    \caption{\textbf{Color Artifact Correction Results.} PSNR / PSNR-B / SSIM format. Best result in bold, second best underlined. JPEG column gives input error. For ICB, we used the RGB 8bit dataset.}
    \footnotesize    
    \renewcommand{\arraystretch}{1.2}
	\renewcommand{\tabcolsep}{1.2mm}
    \resizebox{\textwidth}{!}{
        \begin{tabular}{@{}l *{7}{c}@{}}
            \toprule
            Dataset & Quality & JPEG & ARCNN\cite{dong2015compression} &  MWCNN \cite{liu2018multi} & IDCN \cite{zheng2019implicit} & DMCNN \cite{zhang2018dmcnn} & Ours \\
            \midrule
            \multirow{3}{*}{Live-1} & \begin{filecontents*}{data/comparisons/live1_color.csv}
Quality,JPEG,ARCNN,MWCNN,IDCN,DMCNN,OURS
10,25.60 / 23.53 / 0.755,26.66 / 26.54 / 0.792,27.21 / 27.02 / 0.805,\underline{27.62} / \underline{27.32} / \underline{0.816},27.18 / 27.03 / 0.810,\textbf{27.65 / 27.40 / 0.819}
20,27.96 / 25.77 / 0.837,28.97 / 28.65 / 0.860,29.54 / 29.23 / 0.873,\textbf{30.01} / \underline{29.49} / \underline{0.881},29.45 / 29.08 / 0.874,\underline{29.92} / \textbf{29.51 / 0.882}
30,29.25 / 27.10 / 0.872,30.29 / 29.97 / 0.891,\underline{30.82} / \underline{30.45} / \underline{0.901},-,-,\textbf{31.21 / 30.71 / 0.908}
\end{filecontents*}
\csvreader[late after line=\\]{data/comparisons/live1_color.csv}{}{\csviffirstrow{}{&}\csvlinetotablerow}
            \midrule
            \multirow{3}{*}{BSDS500} & \begin{filecontents*}{data/comparisons/BSDS500_color.csv}
Quality,JPEG,ARCNN,MWCNN,IDCN,DMCNN,OURS
10,25.72 / 23.44 / 0.748,26.83 / 26.65 / 0.783,27.18 / 26.93 / 0.794,\underline{27.61} / \underline{27.22} / \underline{0.805},27.16 / 26.95 / 0.799,\textbf{27.69 / 27.36 / 0.810}
20,28.01 / 25.57 / 0.833,29.00 / 28.53 / 0.853,29.45 / 28.96 / 0.866,\textbf{29.90} / \underline{29.20} / \underline{0.873},29.35 / 28.84 / 0.866,\underline{29.89} / \textbf{29.29 / 0.876}
30,29.31 / 26.85 / 0.869,30.31 / 29.85 / 0.887,\underline{30.71} / \underline{30.09} / \underline{0.895},-,-,\textbf{31.15 / 30.37 / 0.903}
\end{filecontents*}
\csvreader[late after line=\\]{data/comparisons/BSDS500_color.csv}{}{\csviffirstrow{}{&}\csvlinetotablerow}
            \midrule
            \multirow{3}{*}{ICB} & \begin{filecontents*}{data/comparisons/ICB-RGB8.csv}
Quality,JPEG,ARCNN,MWCNN,IDCN,DMCNN,OURS
10,29.31 / 28.07 / 0.749,30.06 / 30.38 / 0.744,30.76 / 31.21 / 0.779,\underline{31.71} / \underline{32.02} / \underline{0.809},30.85 / 31.31 / 0.796,\textbf{32.11 / 32.47 / 0.815}
20,31.84 / 30.63 / 0.804,32.24 / 32.53 / 0.778,32.79 / 33.32 / 0.812,\underline{33.99} / \underline{34.37} / \underline{0.838},32.77 / 33.26 / 0.830,\textbf{34.23 / 34.67 / 0.845}
30,33.02 / 31.87 / 0.830,33.31 / 33.72 / 0.807,\underline{34.11} / \underline{34.69} / \underline{0.845},-,-,\textbf{35.20 / 35.67 / 0.860}
\end{filecontents*}
\csvreader[late after line=\\]{data/comparisons/ICB-RGB8.csv}{}{\csviffirstrow{}{&}\csvlinetotablerow} 
            \bottomrule
        \end{tabular}
    }
    \label{tab:colorresults}
\end{table}

\subsection{Experimental Setup}

\noindent\textbf{Datasets and Metrics.} For training, we use the DIV2k and Flickr2k~\cite{agustsson2017ntire} datasets. DIV2k consists of 900 images, and the Flickr2k dataset contains 2650 images. We preextract $256 \times 256$ patches from these images taking 30 random patches from each image and compress them using quality in $[10, 100]$ in steps of 10. This gives a total training set of 1,065,000 patches.
For evaluation, we use the Live1~\cite{sheikh2006statistical, sheikh2006live}, Classic-5~\cite{foi2006pointwise}, 
BSDS500~\cite{arbelaez2010contour}, and ICB datasets~\cite{icb}. ICB is a new dataset which provides 15 high-quality lossless images designed specifically to measure compression quality. It is our hope that the community will gradually begin including ICB dataset results. Where previous works have provided code and models, we reevaluate their methods and provide results here for comparison. As with all prior works, we report PSNR, PSNR-B~\cite{tadala2012novel}, and SSIM~\cite{wang2004image}.

\noindent\textbf{Implementation Details.}  All training uses the Adam \cite{kingma2014adam} optimizer with a batch size of 32 patches. Our network is implemented using the PyTorch \cite{NEURIPS2019_9015} library. We normalize the DCT coefficients using per-frequency and per-channel mean and standard deviations. Since the DCT coefficients are measurements of different signals, by computing the statistics per-frequency we normalize the distributions so that they are all roughly the same magnitude. We find that this greatly speeds up the convergence of the network. Quantization table entries are
normalized to [0, 1], with 1 being the most quantization and 0 the least. We use libjpeg~\cite{libjpeg} for compression with the baseline quantization setting. 

\noindent\textbf{Training Procedure.} 
As described in Section~\ref{sec:app:train}, we follow a staged training approach by first training the Y channel or grayscale artifact correction network, then training the color (CbCr) channel network, and finally training both networks using the GAN loss.

For the first stage, the Y channel artifact correction network, the learning rate starts at $1 \times 10^{-3}$ and decays by a factor of 2 every 100,000 batches. We stop training after 400,000 batches. We set $\lambda$ in Equation~\ref{eq:regressionloss} to 0.05. 

For the next stage, all color channels are restored. The weights for the Y channel network are initialized from the previous stage and frozen during training. The color channel network weights are trained using a cosine annealing learning rate schedule \cite{loshchilov2016sgdr} decaying from $1 \times 10^{-3}$ to $1 \times 10^{-6}$ over 100,000 batches. 

Finally, we train both Y and color channel artifact correction networks (jointly referred to as the generator model) using a GAN loss to improve qualitative textures. The generator model weights are initialized to the pre-trained models from the previous stages. We use the DCGAN \cite{radford2015unsupervised} discriminator. The model is trained for 100,000 iterations using cosine annealing~\cite{loshchilov2016sgdr}
with the learning rate starting from $1 \times 10^{-4}$ and ending at $1 \times 10^{-6}$.  We set $\gamma$ and $\nu$ in Equation~ \ref{eq:ganloss} to $5 \times 10^{-3}$ and $1 \times 10^{-2}$ respectively.  

\subsection{Results: Artifact Correction}

\noindent\textbf{Color Artifact Correction.} 
We report the main results of our approach, color artifact correction, on Live1, BSDS500, and ICB in Table~\ref{tab:colorresults}. Our model consistently outperforms recent baselines on all datasets. Note that of all the approaches, only ours and IDCN~\cite{zheng2019implicit} include native processing of color channels. For the other models, we convert input images to YCbCr and process the channels independently. 

\begin{figure}[t]
    \begin{minipage}[t][][b]{0.48\linewidth}
        \centering
        \includegraphics[height=0.7\linewidth]{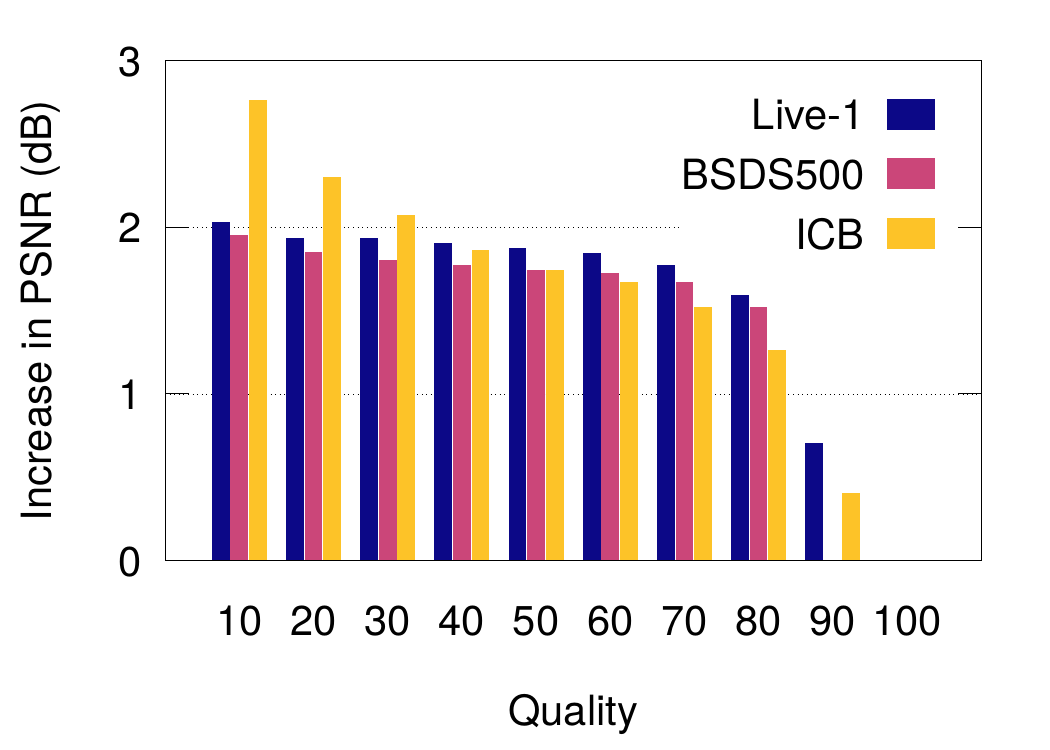}
        \caption{\textbf{Increase in PSNR on color datasets.} For all three datasets we show the average improvement in PSNR values on qualities 10-100. Improvement drops off steeply at quality 90.}
        \label{fig:ipsnr}
    \end{minipage}
    \hfill
    \begin{minipage}[t][][b]{0.48\linewidth}
        \centering
        \includegraphics[height=0.65\linewidth]{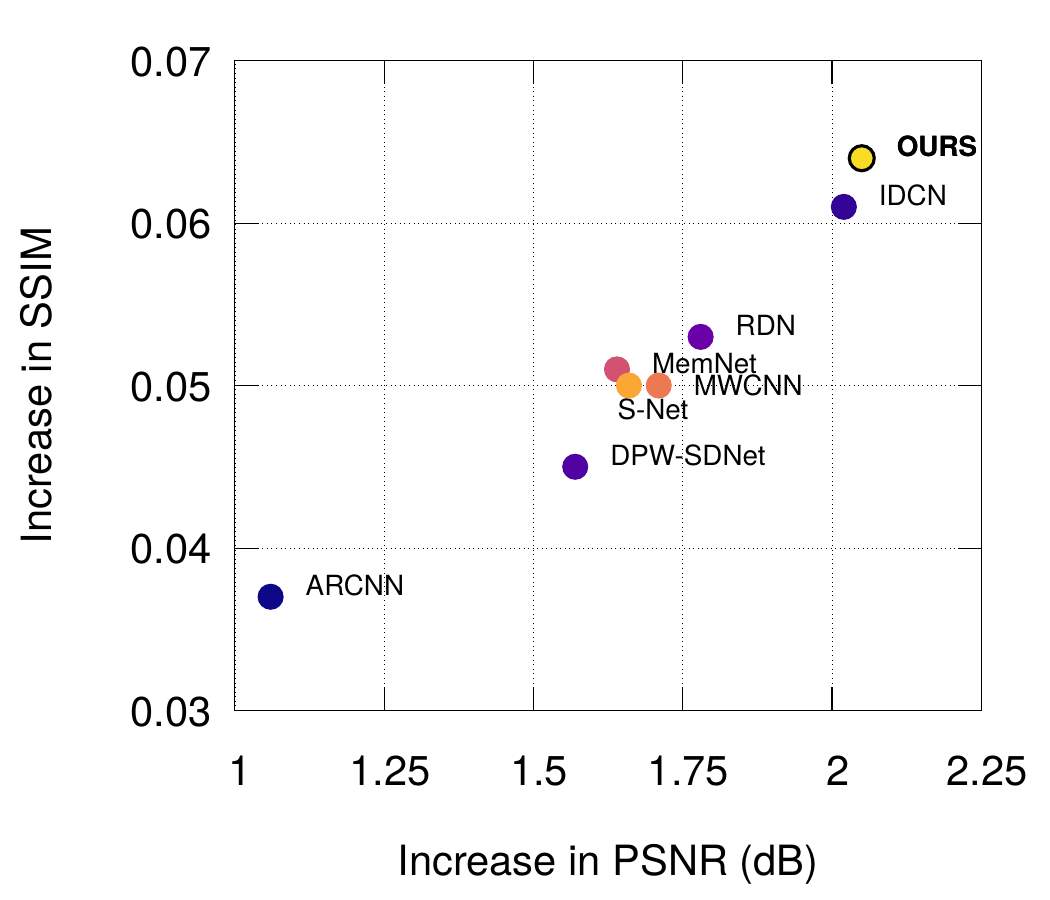}
        \caption{\textbf{Comparison for Live-1 quality 10.} Where
        code was available we reevaluated, otherwise we used published numbers.}
        \label{fig:compare}
    \end{minipage}
\end{figure}

\begin{figure}[t]
    \centering
    \resizebox{0.85\columnwidth}{!}{
    \begin{tabular}{ccccccc}
        Original & JPEG & IDCN & Ours & Ours-GAN \\
        \includegraphics[width=0.2\linewidth]{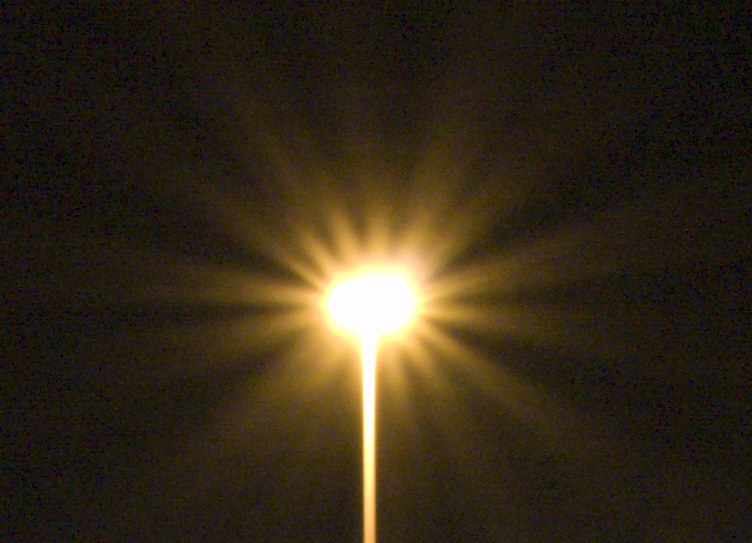} & 
        \includegraphics[width=0.2\linewidth]{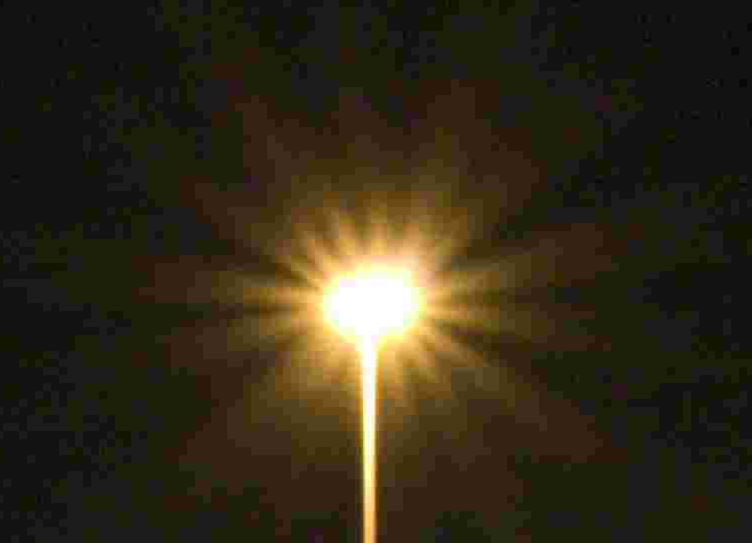} &
        \includegraphics[width=0.2\linewidth]{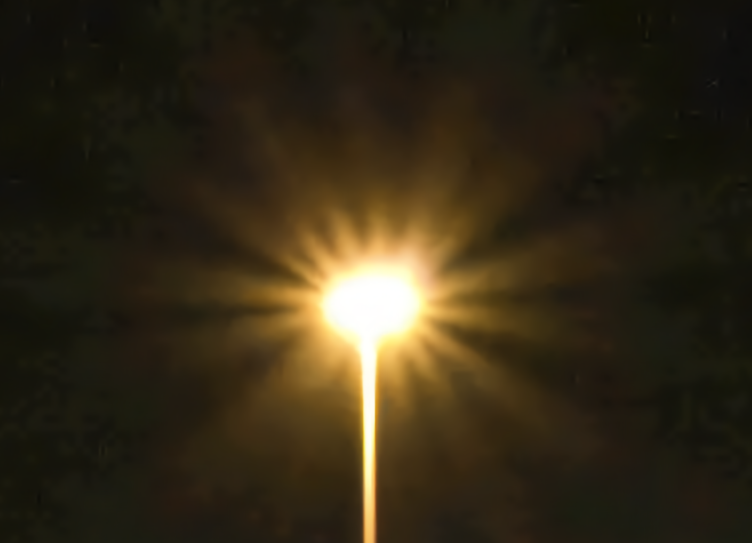} &
        \includegraphics[width=0.2\linewidth]{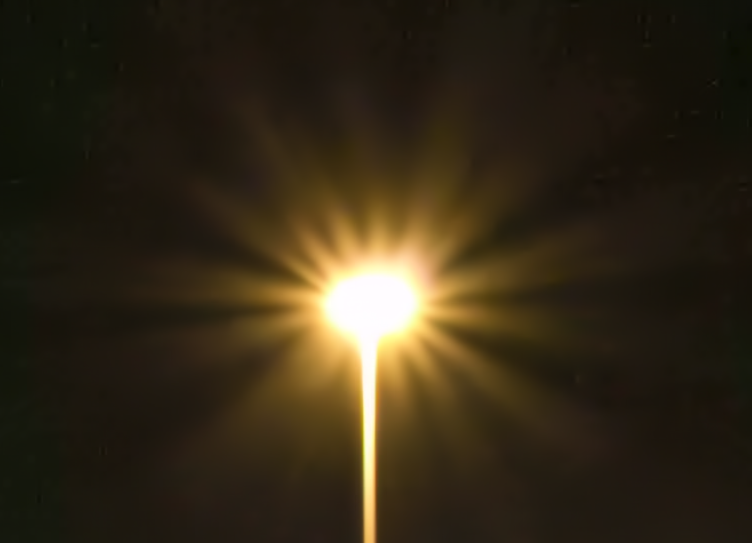} & 
        \includegraphics[width=0.2\linewidth]{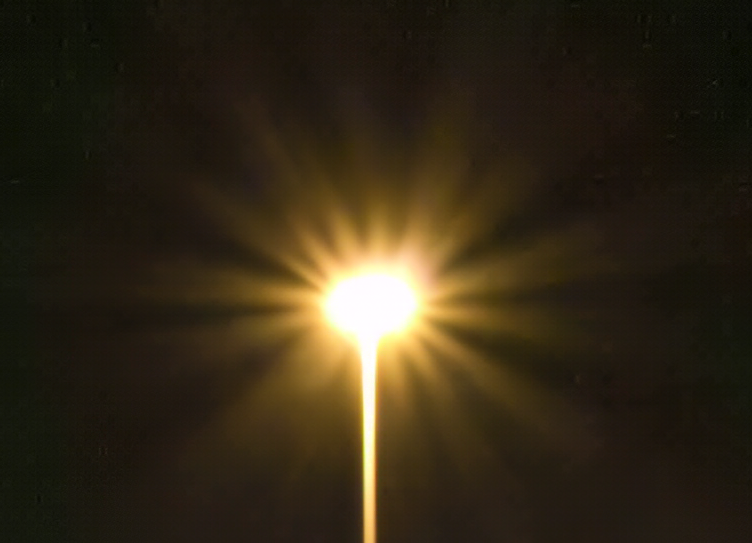} \\
        \includegraphics[width=0.2\linewidth]{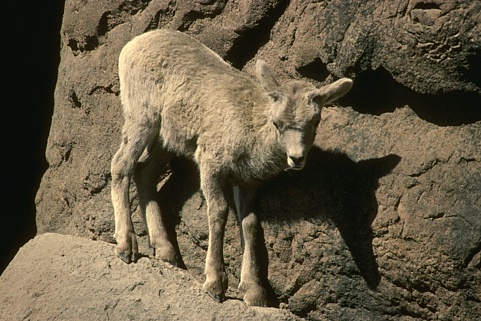} & 
        \includegraphics[width=0.2\linewidth]{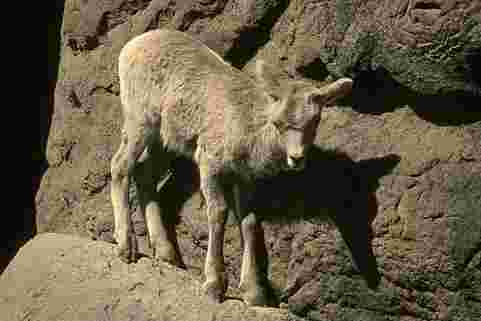} &
        \includegraphics[width=0.2\linewidth]{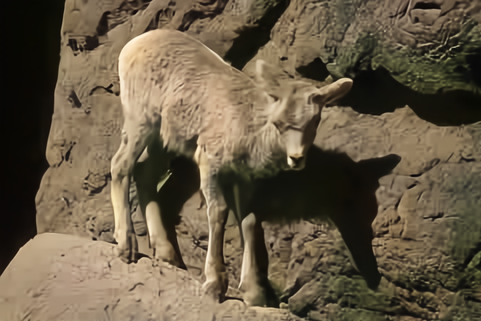} &
        \includegraphics[width=0.2\linewidth]{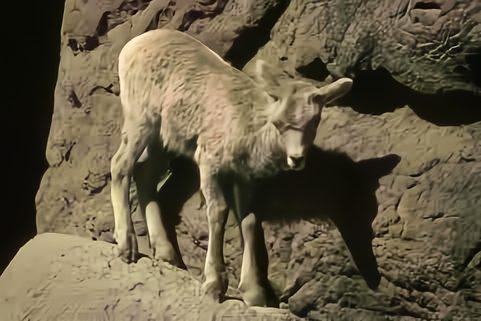} &
        \includegraphics[width=0.2\linewidth]{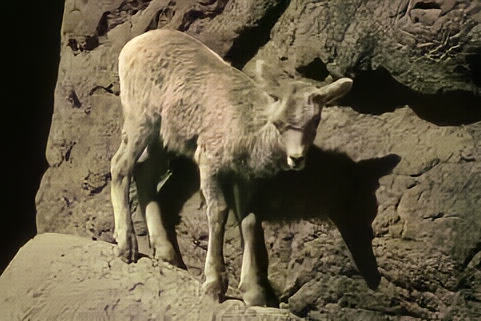} \\
        \includegraphics[width=0.2\linewidth]{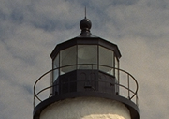} & 
        \includegraphics[width=0.2\linewidth]{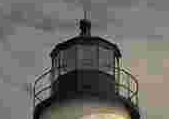} &
        \includegraphics[width=0.2\linewidth]{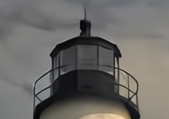} &
        \includegraphics[width=0.2\linewidth]{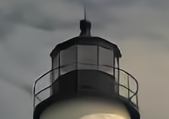} &
        \includegraphics[width=0.2\linewidth]{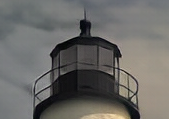} \\
        \includegraphics[width=0.2\linewidth]{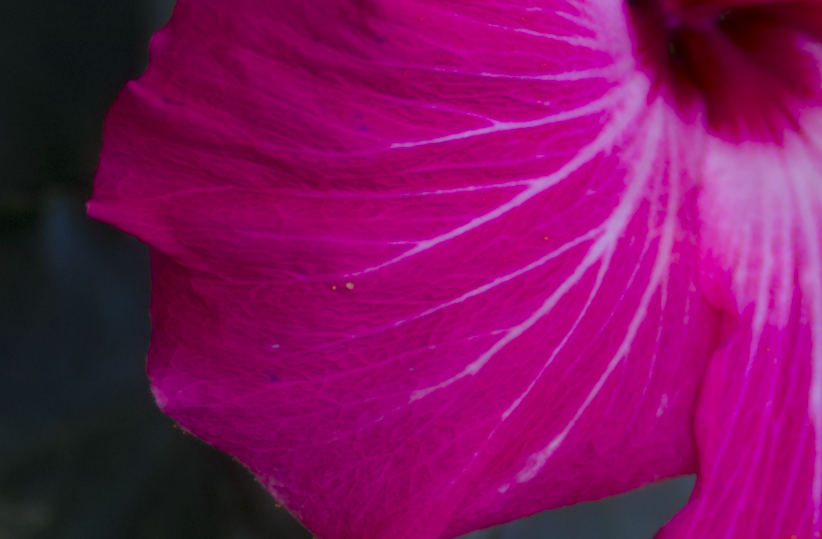} & 
        \includegraphics[width=0.2\linewidth]{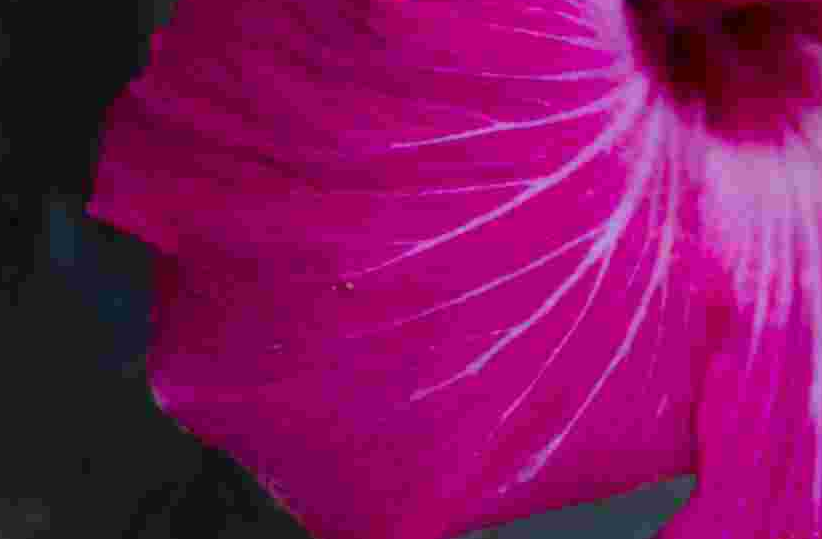} &
        \includegraphics[width=0.2\linewidth]{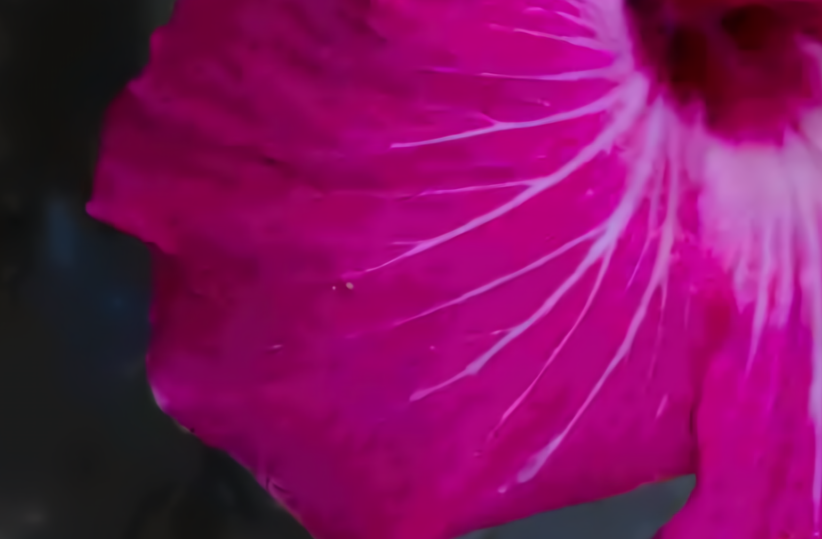} &
        \includegraphics[width=0.2\linewidth]{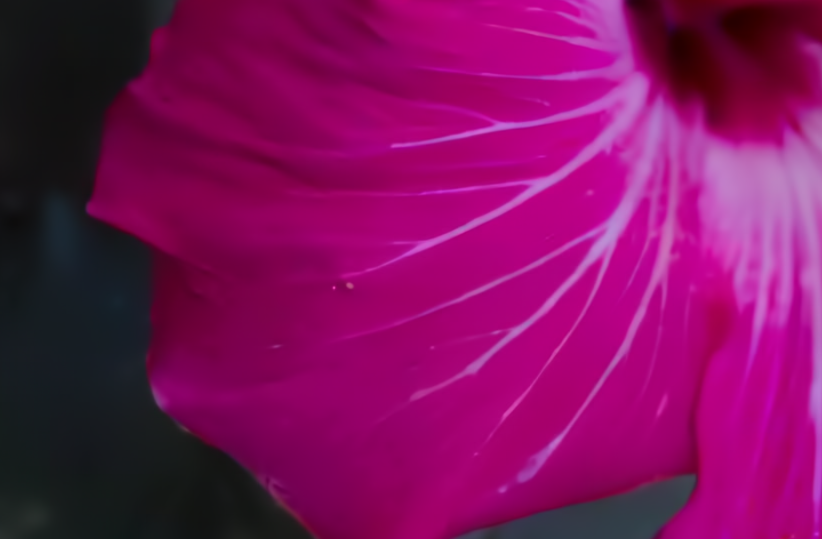} & 
        \includegraphics[width=0.2\linewidth]{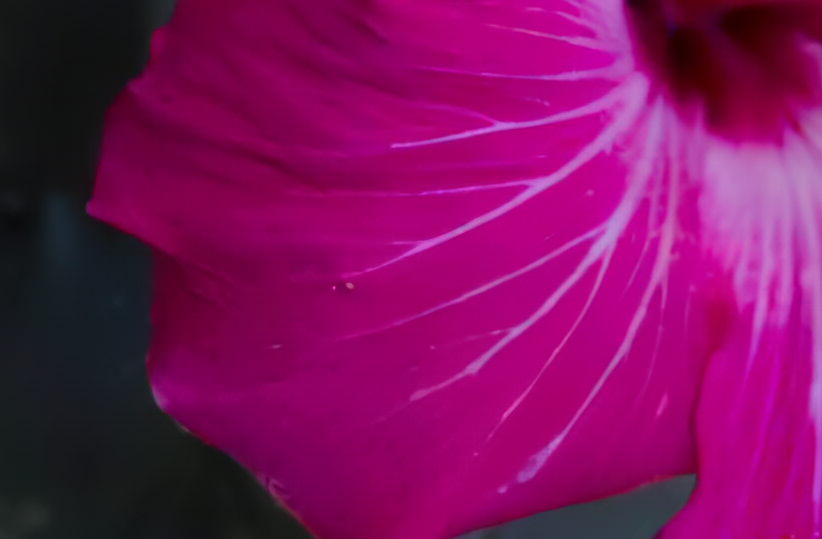} \\
        \includegraphics[width=0.2\linewidth]{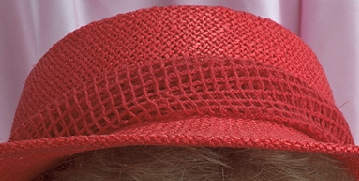} & 
        \includegraphics[width=0.2\linewidth]{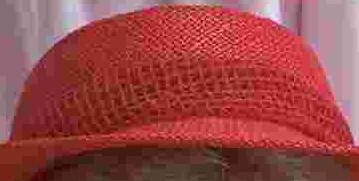} &
        \includegraphics[width=0.2\linewidth]{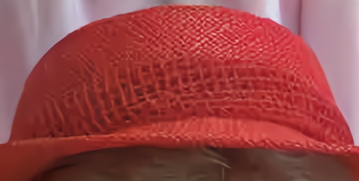} &
        \includegraphics[width=0.2\linewidth]{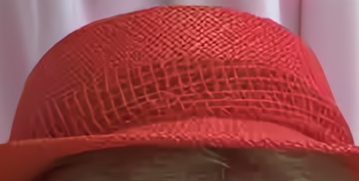} & 
        \includegraphics[width=0.2\linewidth]{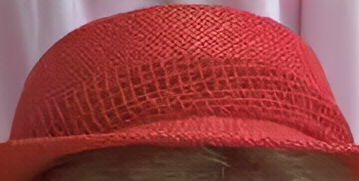} \\
    \end{tabular}}
    \caption{\textbf{Qualitative Results.} All images were compressed at Quality 10. Please zoom in to view details.}
    \label{fig:gan}
\end{figure}

For quantitative comparisons to more methods on Live-1 dataset, at compression quality 10, refer to Figure \ref{fig:compare}. We present qualitative results from a mix of all three datasets in Figure~\ref{fig:gan} (``Ours''). Since our model is not restricted by which quality settings it can be run on, we also show the increase in PSNR for qualities 10-100 in Figure \ref{fig:ipsnr}.

\begin{table}[t]
    \centering
    \caption{\textbf{Y Channel Correction Results.} PSNR / PSNR-B / SSIM format, the best result is highlighted in bold, second best is underlined. The JPEG column gives with input error of the images. For ICB, we used the Grayscale 8bit dataset. We add Classic-5, a grayscale only dataset.}
    \footnotesize    
    \renewcommand{\arraystretch}{1.2}
	\renewcommand{\tabcolsep}{1.2mm}
    \resizebox{\textwidth}{!}{
        \begin{tabular}{@{}l *{7}{c}@{}}
            \toprule
            Dataset & Quality & JPEG & ARCNN\cite{dong2015compression} &  MWCNN \cite{liu2018multi} & IDCN \cite{zheng2019implicit} & DMCNN \cite{zhang2018dmcnn} & Ours \\
            \midrule
            \multirow{3}{*}{Live-1} & \begin{filecontents*}{data/comparisons/live1.csv}
Quality,JPEG,ARCNN,MWCNN,IDCN,DMCNN,OURS
10,27.76  /  25.32  /  0.790,28.96  /  28.68  /  0.821,\underline{29.68}  /  29.30  /  0.839,\underline{29.68}  /  \underline{29.32}  /  0.838,\textbf{29.73}  /  \textbf{29.43}  /  0.839,29.53  /  29.15  /  \textbf{0.840}
20,30.05  /  27.55  /  0.868,31.26  /  30.73  /  0.887,32.00  /  \underline{31.47}  /  \textbf{0.901},\underline{32.05}  /  31.46  /  \underline{0.900},\textbf{32.07}  /  \textbf{31.49}  /  \textbf{0.901},31.86  /  31.27  /  \textbf{0.901}
30,31.37  /  28.90  /  0.900,32.64  /  32.11  /  0.916,\textbf{33.40}  /  \textbf{32.76}  /  \textbf{0.926},-,-,\underline{33.23}  /  \underline{32.50}  /  \underline{0.925}
\end{filecontents*}
\csvreader[late after line=\\]{data/comparisons/live1.csv}{}{\csviffirstrow{}{&}\csvlinetotablerow}
            \midrule
            \multirow{3}{*}{Classic-5} & \begin{filecontents*}{data/comparisons/classic5.csv}
Quality,JPEG,ARCNN,MWCNN,IDCN,DMCNN,OURS
10,27.82 / 25.21 / 0.780,29.03 / 28.76 / 0.811,\textbf{30.01} / \underline{29.59} / \textbf{0.837},29.83 / 29.48 / 0.833,\underline{29.98} / \textbf{29.65} / 0.836,29.84 / 29.43 / \textbf{0.837}
20,30.12 / 27.50 / 0.854,31.15 / 30.59 / 0.869,\textbf{32.16 / 31.52 / 0.886},31.99 / 31.46 / 0.884,\underline{32.11} / \underline{31.48} / \underline{0.885},31.98 / 31.37 / \underline{0.885}
30,31.48 / 28.94 / 0.884,32.51 / 31.98 / 0.896,\textbf{33.43 / 32.62 / 0.907},-,-,\underline{33.22} / \underline{32.42} / \textbf{0.907}
\end{filecontents*}
\csvreader[late after line=\\]{data/comparisons/classic5.csv}{}{\csviffirstrow{}{&}\csvlinetotablerow}
            \midrule
            \multirow{3}{*}{BSDS500} & \begin{filecontents*}{data/comparisons//BSDS500.csv}
Quality,JPEG,ARCNN,MWCNN,IDCN,DMCNN,OURS
10,27.86 / 25.18 / 0.785,29.14 / 28.76 / 0.816,\underline{29.63} / \underline{29.16} / \underline{0.831},29.60 / 29.13 / 0.829,\textbf{29.66} / \textbf{29.27} / \underline{0.831},29.54 / 29.04 / \textbf{0.833}
20,30.08 / 27.28 / 0.864,31.27 / 30.52 / 0.881,\underline{31.88} / \underline{31.12} / \textbf{0.894},\underline{31.88} / 31.05 / 0.893,\textbf{31.91 / 31.13 / 0.894},31.79 / 30.96 / \textbf{0.894}
30,31.37 / 28.56 / 0.896,32.64 / 31.90 / 0.912,\textbf{33.23 / 32.29 / 0.920},-,-,\underline{33.12} / \underline{32.07} / \textbf{0.920}
\end{filecontents*}
\csvreader[late after line=\\]{data/comparisons//BSDS500.csv}{}{\csviffirstrow{}{&}\csvlinetotablerow}
            \midrule
            \multirow{3}{*}{ICB} & \begin{filecontents*}{data/comparisons//ICB-GRAY8.csv}
Quality,JPEG,ARCNN,MWCNN,IDCN,DMCNN,OURS
10,32.08 / 29.92 / 0.856,31.13 / 30.97 / 0.794,34.12 / 34.06 / 0.884,32.50 / 32.42 / 0.826,\underline{34.18} / \underline{34.15} / \underline{0.874},\textbf{34.73 / 34.58 / 0.896}
20,35.04 / 32.72 / 0.905,32.62 / 32.31 / 0.821,\underline{36.56} / \underline{36.44} / 0.902,34.30 / 34.18 / 0.851,35.93 / 35.79 / \underline{0.918},\textbf{37.12 / 36.88 / 0.924}
30,36.66 / 34.22 / 0.927,33.79 / 33.52 / 0.841,\underline{38.20} / \underline{37.96} / \underline{0.927},-,-,\textbf{38.43 / 38.05 / 0.938}
\end{filecontents*}
\csvreader[late after line=\\]{data/comparisons//ICB-GRAY8.csv}{}{\csviffirstrow{}{&}\csvlinetotablerow}
            \bottomrule
        \end{tabular}
    }
    \label{tab:yresults}
\end{table}

\begin{figure}[t]
    \centering
    \resizebox{0.85\columnwidth}{!}{
    \begin{tabular}{ccccc}
        Original & JPEG & IDCN Q=10 & IDCN Q=20 & Ours \\
        \includegraphics[width=0.2\linewidth]{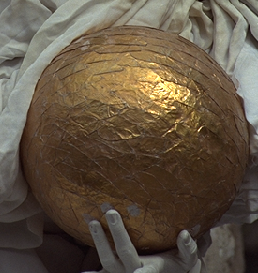} & 
        \includegraphics[width=0.2\linewidth]{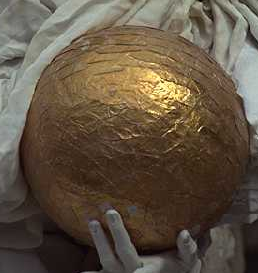} &
        \includegraphics[width=0.2\linewidth]{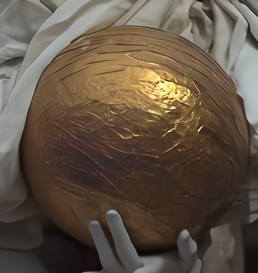} &
        \includegraphics[width=0.2\linewidth]{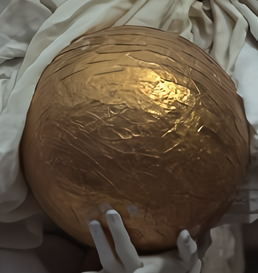} & 
        \includegraphics[width=0.2\linewidth]{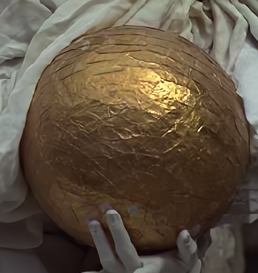} 
    \end{tabular}}
    \caption{\textbf{Generalization Example}. Input was compressed at quality 50. Please zoom in to view details.}
    \label{fig:gen}
\end{figure}

\begin{table}[t]
    \centering
    \caption{\textbf{Generalization Capabilities}. Live-1 dataset (PSNR / PSNR-B / SSIM).}
    \footnotesize    
    \renewcommand{\arraystretch}{1.2}
	\renewcommand{\tabcolsep}{1.2mm}
    \resizebox{0.85\columnwidth}{!}{
        \begin{tabular}{@{} *{5}{c}@{}}
            \toprule
            Model Quality & Image Quality & JPEG & IDCN \cite{zheng2019implicit} & Ours \\
            \midrule
            %
10 & \multirow{2}{*}{50} & 30.91 / 28.94 / 0.905 & 30.19 / 30.14 / 0.889 & \multirow{2}{*}{\textbf{32.78 / 32.19 / 0.932}} \\
20 & & 30.91 / 28.94 / 0.905 & 31.91 / 31.65 / 0.916 & \\
\midrule
10 & 20 & 27.96 / 25.77 / 0.837 & 29.25 / 29.08 / 0.863 & \textbf{29.92 / 29.51 / 0.882} \\
20 & 10 & 25.60 / 23.53 / 0.755 & 26.95 / 26.24 / 0.804 & \textbf{27.65 / 27.40 / 0.819} \\
            \bottomrule
        \end{tabular}
    }
    \label{tab:genresults}
\end{table}

\begin{table}[t]
    \footnotesize    
    \renewcommand{\arraystretch}{1.1}
    \renewcommand{\tabcolsep}{1mm}
  	\centering
    \begin{minipage}[t][][b]{0.47\linewidth}
        \centering
        \caption{\textbf{GAN FID Scores.}}
        \resizebox{0.6\columnwidth}{!}{
        \begin{tabular}{@{}l *{3}{c}@{}}
            \toprule
            Dataset & Quality & Ours & Ours-GAN \\
            \midrule 
            \multirow{3}{*}{Live-1} & \begin{filecontents*}{data/fid/fid_live1.csv}
Quality,Ours,Ours-GAN
10,69.57,\bfseries 35.86
20,36.32,\bfseries 16.99
30,24.72,\bfseries 12.20
\end{filecontents*}
\csvreader[late after line=\\]{data/fid/fid_live1.csv}{}{\csviffirstrow{}{&}\csvlinetotablerow}
            \midrule
            \multirow{3}{*}{BSDS500} & \begin{filecontents*}{data/fid/fid_bsds500.csv}
Quality,Ours,Ours-GAN
10,75.15,\bfseries 34.80
20,42.46,\bfseries 18.74
30,29.04,\bfseries 13.03
\end{filecontents*}
\csvreader[late after line=\\]{data/fid/fid_bsds500.csv}{}{\csviffirstrow{}{&}\csvlinetotablerow}
            \midrule
            \multirow{3}{*}{ICB} & \begin{filecontents*}{data/fid/fid_icb.csv}
Quality,Ours,Ours-GAN
10,33.37,\bfseries 26.08
20,17.23,\bfseries 13.53
30,11.66,\bfseries 10.13
\end{filecontents*}
\csvreader[late after line=\\]{data/fid/fid_icb.csv}{}{\csviffirstrow{}{&}\csvlinetotablerow}
            \bottomrule
        \end{tabular} }
        \label{tab:fid}
    \end{minipage}
    \hfill
    \begin{minipage}[t][][b]{0.51\linewidth}
        \centering
        \caption{\textbf{Ablation Results.} (refer to Section~\ref{sec:res:ablation} for details).}
        \resizebox{0.75\columnwidth}{!}{
        \begin{tabular}{@{}L{1.8cm} L{1.8cm} *{7}{c}@{}}
            \toprule
            Experiment & Model & PSNR & PSNR-B & SSIM \\
            \midrule
            \multirow{3}{*}{CFM} & \begin{filecontents*}{data/ablation/cmf.csv}
Model,PSNR,PSNR-B,SSIM
None,29.38,28.9,0.825
Concat,29.32,28.94,0.823
CFM,\textbf{29.46},\textbf{29.05},\textbf{0.827}
\end{filecontents*}
\csvreader[late after line=\\]{data/ablation/cmf.csv}{}{\csviffirstrow{}{&}\csvlinetotablerow}
            \midrule
            \multirow{2}{*}{Subnetworks} & \begin{filecontents*}{data/ablation/bvf.csv}
Model,PSNR,PSNR-B,SSIM
FrequencyNet,28.03,25.58,0.787
BlockNet,\textbf{29.45},\textbf{29.04},\textbf{0.827}
\end{filecontents*}
\csvreader[late after line=\\]{data/ablation/bvf.csv}{}{\csviffirstrow{}{&}\csvlinetotablerow}
            \midrule
            \multirow{2}{*}{Fusion} & \begin{filecontents*}{data/ablation/ft.csv}
Model,PSNR,PSNR-B,SSIM
No Fusion,27.82,25.21,0.78
Fusion,\textbf{29.22},\textbf{28.76},\textbf{0.822}
\end{filecontents*}
\csvreader[late after line=\\]{data/ablation/ft.csv}{}{\csviffirstrow{}{&}\csvlinetotablerow}
            \bottomrule
        \end{tabular} }  
        \label{tab:ablation}
\end{minipage}
\end{table}

\noindent\textbf{Intermediate Results on Y Channel Artifact Correction.}
Since the first stage of our approach trains for grayscale or Y channel artifact correction, we can also compare the intermediate results from this stage with other approaches. We report results in Table~\ref{tab:yresults} for Live1, Classic-5, BSDS500, and ICB. As the table shows, intermediate results from our model can match or outperform previous state-of-the-art models in many cases, consistently providing high SSIM results using a single model for all quality factors.

\noindent\textbf{GAN Correction}
\label{sec:res:gan}
Finally, we show results from our model trained using GAN correction. We use model interpolation~\cite{wang2018esrgan}
and show qualitative results for the interpolation parameter ($\alpha$) set to 0.7 in Figure \ref{fig:gan}. (``Ours-GAN'') Notice that the GAN loss is able to restore texture to blurred, flat regions and sharpen edges, yielding a more visually pleasing result. We provide additional qualitative results in the supplementary material. Note that we do not show error metrics using the GAN model as it produces higher quality images, at the expense of quantitative metrics, by adding texture details that are not present in the original images. We instead show
FID scores for the GAN model compared to our regression model in Table~\ref{tab:fid}, indicating that the GAN model generates significantly more 
realistic images.

\subsection{Results: Generalization Capabilities}
The major advantage of our method is that it uses a single model to correct JPEG images at any quality, while prior works train a model for each quality factor. Therefore, we explore if other methods are capable of generalizing or if they really require this ensemble of quality-specific models. To evaluate this, we use our closest competitor and prior state-of-the-art, IDCN~\cite{zheng2019implicit}. IDCN does not provide a model for quality higher than 20, we explore if their model generalizes by using their quality 10 and quality 20 models to correct
quality 50 Live-1 images. We also use the quality 20 model to correct quality 10 images and use the quality
10 model to correct quality 20 images. These results are shown in Table~\ref{tab:genresults} along with our result.

As the table shows, the choice of model is critical for IDCN, and there is a significant quality drop when choosing the wrong model. Neither their quality 10 nor their quality 20 model is able to effectively correct images that it was not trained on, scoring significantly lower than if the correct model were used. At quality 50, the quality 10 model produces a result worse than the input JPEG, and the quality 20 model makes only a slight improvement. In comparison, our single model provides consistently better results across image quality factors. We stress that the quality setting is not stored in the JPEG file, so a deployed system has no way to pick the correct model. We show an example of a quality 50 image and artifact correction results in Figure \ref{fig:gen}.

\subsection{Design and Ablation Analysis}
\label{sec:res:ablation}
Here we ablate many of our design decisions and observe their effect on network accuracy. The results are reported in Table~\ref{tab:ablation}, we report metrics on quality 10 classic-5. 

\noindent\textbf{Implementation details:}\smallskip For all ablation experiments, we keep the number of parameters approximately the same between tested models to alleviate the concern that a network performs better simply because it has a higher capacity. All models are trained for 100,000 batches on the grayscale training patch set using cosine
annealing~\cite{loshchilov2016sgdr} from a learning rate of $1 \times 10^{-3}$ to $1 \times 10^{-6}$.

\noindent\textbf{Importance of CFM layers.} We emphasized the importance of adaptable weights in the CFM layers, which can be adapted using the quantization matrix. However, there are other simpler methods of using side-channel information. We could simply concatenate the quantization matrix channelwise with the input, or we could ignore the quantization matrix altogether. As shown in the ``CFM'' experiment in Table~\ref{tab:ablation}, the CFM unit performs better than both of these alternatives by a considerable margin. We further visualize the filters learned by the CFM layers and the underlying embeddings in the supplementary material which validate that the learned filters follow a manifold structure.

\noindent\textbf{BlockNet \vs FrequencyNet.} We noted that the FrequencyNet should not be able to perform without a preceding BlockNet because high-frequency information will be zeroed out from the compression process. To test this claim, we train individual BlockNet and FrequencyNet in isolation and report the results in Table~\ref{tab:ablation} (``Subnetworks''). We can see that BlockNet alone attains significantly higher performance than FrequencyNet alone.

\noindent\textbf{Importance of the fusion layer.} Finally, we study the necessity of the fusion layer presented. We posited that the fusion layer was necessary for gradient flow to the early layers of our network. As demonstrated in Table~\ref{tab:ablation} (``Fusion''), the network without fusion fails to learn, matching the input PSNR of classic-5 after full training, whereas the network with fusion makes considerable progress.

\section{Conclusion}

We showed a design for a quantization guided JPEG artifact correction network. Our single network is able to
achieve state-of-the-art results, beating methods which train a different network for
each quality level. Our network relies only on information that is available at inference time, and solves
a major practical problem for the deployment of such methods in real-world scenarios. 

\subsubsection{Acknowledgement} This project was partially supported by Facebook AI and Defense Advanced Research Projects Agency (DARPA) MediFor program (FA87501620191). There is no collaboration between Facebook and DARPA.

{\small
    \printbibliography
}

\appendix
    
\title{Quantization Guided JPEG Artifact Correction: Appendices}
    
\titlerunning{Quantization Guided JPEG Artifact Correction}

\author{}
\institute{}

\authorrunning{M. Ehrlich et al.}
\maketitle

\section{Additional Evaluation Details}

In this section we elaborate on the evaluation procedure for prior works as well as discuss a number of hyperparameters critical to correct evaluation. In our results section, three of the four prior works did not have native handeling of color channels. To evaluate them on color images, we applied their Y channel network to both Y, Cb, and Cr channels separately as well as R, G, and B channels separately. In all cases, using the Y, Cb, and Cr channels performed the best, so these are the results we report (\eg we report the scheme that gives prior works the best numbers). Note that we do not modify the published network structure to take a three channel input as was done in IDCN. We do this to remain as faithful to the published methods as possible, and we note that by examining the numbers reported in IDCN, the ranking of the methods does not change. Altering the network structures to take a three channel input does, however, improve their results on color images even if it is a small improvement. 

Next, we note important evaluation hyperparameters. We defer to the ARCNN evaluation code for these settings, although they are not objectively correct. SSIM evaluation in particular uses an $8 \times 8$ window with uniform weighting in contrast to the default $11 \times 11$ gaussian window. Setting this correctly is critical to producing a fair comparison and we have found prior works are not uniform in correctly setting it. ARCNN uses a strict definition of the Y channel giving an output in the range $[16, 240]$, this was intended to match the YCbCr transform used in the JPEG standard, however it is incorrect and stems from the default MATLAB settings. JPEG uses the full-frame Y channel conversion giving outputs in $[0, 255]$. We would like to see this corrected in future works, however it seems unlikely as it changes the comparisons quite a bit. Finally, we note that PSNR-B is an assymetric measure, \eg the blocking effect factor (BEF) is only computed on the degraded image, so the order of the arguments is critical. We have seen at least one prior work that passes these arguments in reverse order resulting in nearly perfect PSNR-B (defined as PSNR-B very close to PSNR).

We have made our model and evaluation code as well as pretrained weights avaible at \href{https://gitlab.com/Queuecumber/quantization-guided-ac}{https://gitlab.com/Queuecumber/quantization-guided-ac}. The evaluation code is reimplemented in PyTorch using ARCNN MATLAB code as a reference and checked for accuracy. We invite future work to use this framework for correct evaluation.

\section{Further Analysis}

In this section we provide futher analysis of our model. We start by examining the Convolution Filter Manifold layers in more detail, providing visualizations of what they learn in order to better understand their contribution to our result. Next, we examine model interpolation in more detail by showing qualitative comparisons for varying interpolation strengths between the regression and GAN model. We then conduct a study that shows how much space can be saved by storing low quality JPEG images and using our method to restore them. We then examine the frequency domain qualitative results and show that our GAN model is capabile of generating images that have more high frequency content than the regression model alone. We conclude by examining the runtime throughput of our model compared to the other methods we tested against. 

\subsection{Understanding Convolutional Filter Manifolds}

\begin{figure}
    \begin{minipage}{0.49\linewidth}
        \centering
        \begin{tabular}{ccc}
            \includegraphics[width=0.33\linewidth]{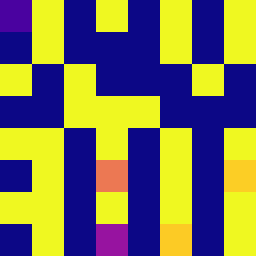} & \includegraphics[width=0.33\linewidth]{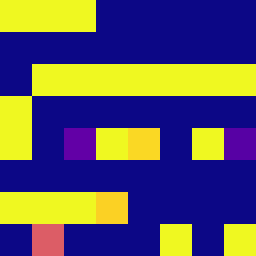} & \includegraphics[width=0.33\linewidth]{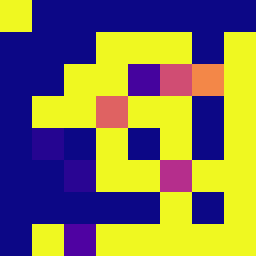} \\
            \includegraphics[width=0.33\linewidth]{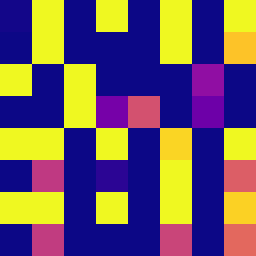} & \includegraphics[width=0.33\linewidth]{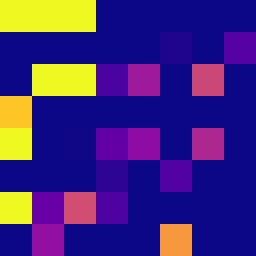} & \includegraphics[width=0.33\linewidth]{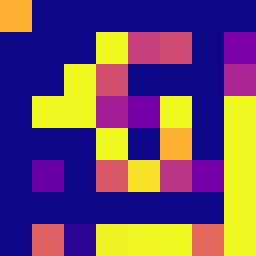} \\
            \includegraphics[width=0.33\linewidth]{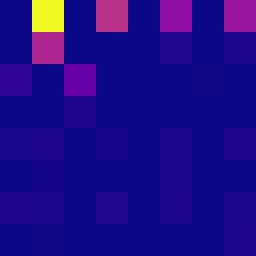} & \includegraphics[width=0.33\linewidth]{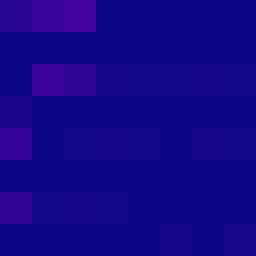} & \includegraphics[width=0.33\linewidth]{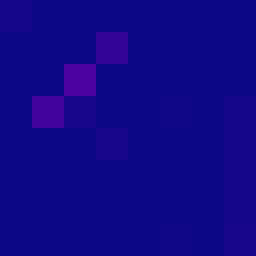}
        \end{tabular}
        \caption{\textbf{CFM Weight Visualization.} Horizontal axis shows different channels of the weight, vertical
        axis shows quality. Quality levels shown are Top: 10, Middle: 50, Bottom: 100.}
        \label{fig:weight_vis}
    \end{minipage}
    \hspace{0.01\linewidth}
    \begin{minipage}{0.49\linewidth}
        \centering
        \begin{tabular}{ccc}
            \includegraphics[width=0.33\linewidth]{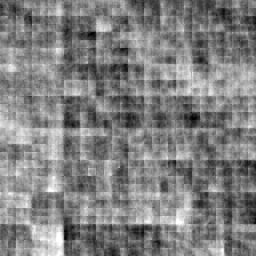} & \includegraphics[width=0.33\linewidth]{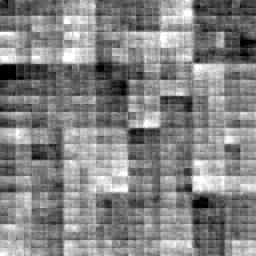} & \includegraphics[width=0.33\linewidth]{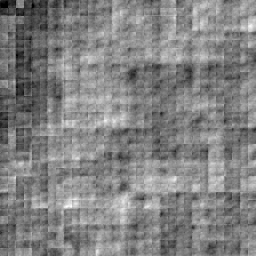} \\
            \includegraphics[width=0.33\linewidth]{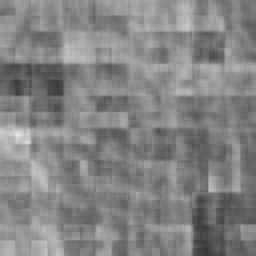} & \includegraphics[width=0.33\linewidth]{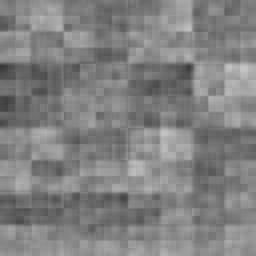} & \includegraphics[width=0.33\linewidth]{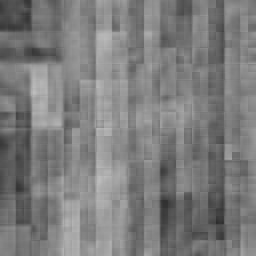} \\
            \includegraphics[width=0.33\linewidth]{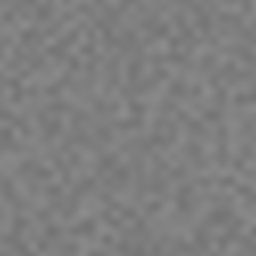} & \includegraphics[width=0.33\linewidth]{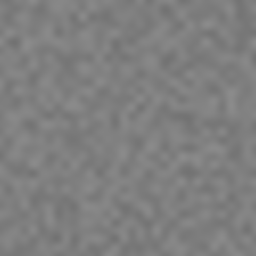} & \includegraphics[width=0.33\linewidth]{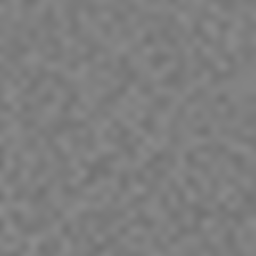}
        \end{tabular}
        \caption{\textbf{Images Which Maximally Activate CFM Weights.} Horizontal axis shows
        different channels from the weight, vertical axis shows quality. Quality levels shown are Top: 10, Middle: 50, Bottom: 100.}
        \label{fig:max_activation}
    \end{minipage}
\end{figure}

\begin{figure}[t]
    \centering
    \includegraphics[width=\linewidth]{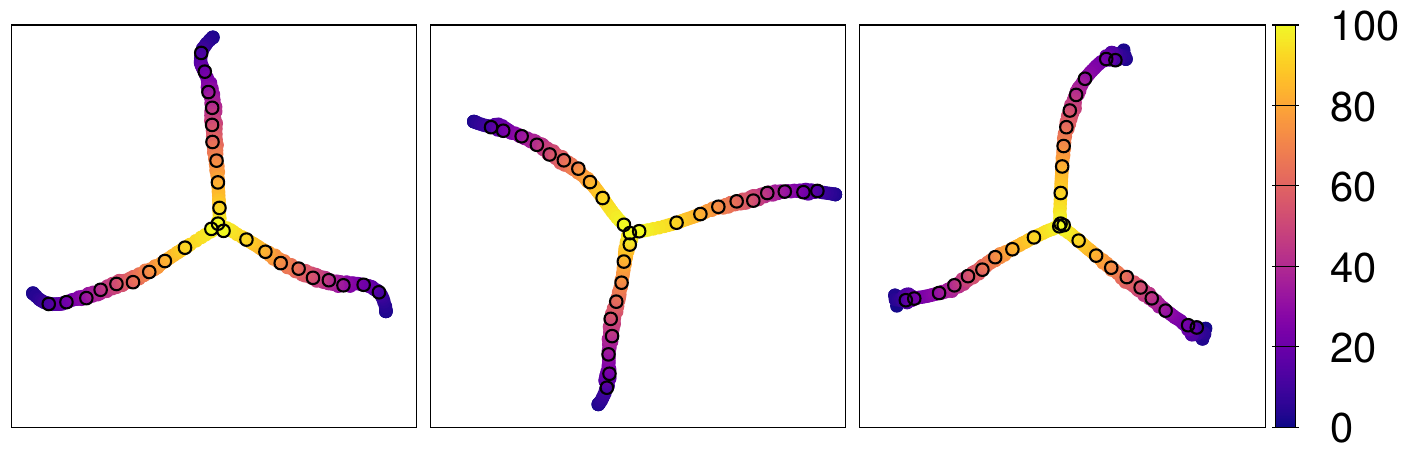}   
    \caption{\textbf{Embeddings for Different CFM Layers.} 3 channels are taken from each embedding, color shows JPEG quality
    setting that produced the input quantization matrix. Circled
    points indicate quantization matrices that were seen during training.}
    \label{fig:quality_tsne}
\end{figure}

CFM layers are both our largest departure
from a vanilla CNN and also quite important to learning quality invariant features, so it is a natural result to try to visualize 
their operation.  In Figure \ref{fig:weight_vis}, we 
compute the final $8 \times 8$ convolution weight for different quality levels. The quality levels, on the vertical axis, are 
10, 50, and 100. The horizontal axis shows three different channels from the weight. What we see makes intuitive sense: 
the filters in different channels have different patterns, but for the same channel, the pattern is roughly the same as the 
quality increases. Furthermore, the filter response becomes smaller as the quality increases since the filters have
to do less ``work'' to correct a high quality JPEG. 

Next we visualize compression artifacts learned by the weight. To do this we find the image that maximally activates a single channel of the
CFM weight. The result
of this is shown in Figure \ref{fig:max_activation}. Again the horizontal axis shows different channels of the weight and the vertical axis 
shows quality levels 10, 50, and 100. The result shows clear images of 
JPEG artifacts. At quality 10, the local blocking artifacts are extremely prominant. By 
quality 50, the blocking artifacts are suppressed, while structural artifacts remain. The qualtiy 100 images are almost untouched, leaving only the
input noise pattern. It makes sense that 
quality 100 filters are only minmally activated since there is not much correction to do on a quality 100 JPEG. Note that we only show Y
channel response for this figure and that Figures \ref{fig:weight_vis} and \ref{fig:max_activation} use the same channels from the same 
layer.

Finally we examine the manifold structure of the CFM. We claim in Section 3.1 (and the name implies) that the CFM learns a smooth manifold
of filters through quantization space. If this is true, then a quality 25 quantization matrix should generate a weight halfway inbetween
a qualty 20 and a quality 30 one. To show that this happens, we generate weights
for all 101 quanitzation matrices (0 to 100 inclusive) and then compute t-SNE 
embeddings to reduce the dimensionality to 2. We plot 3 channels from the weight embeddings with the quality level that was used to
generate the weight given 
as the color of the point. This plot is shown in Figure \ref{fig:quality_tsne}. What see is a smooth line through the space starting from 
dark (low quality) to bright (high quality) showing that the CFM has not only separated the different quality levels but has 
ordered them as well.
Futhermore we see that the low quality filters are separated in space, indicating that they are quite different (and
perform different functions), a property that is important for effective neural networks. As the quality increases and the problem becomes
easier, the filters tend to converge on a single point where they are all doing very little to correct the image.

\subsection{Model Interpolation}

Here we show more model interpolation results. Model interpolation creates a new model by linearly interpolating the GAN
and regresion model parameters as follows

\begin{equation}
    \Theta_I = (1 - \alpha)\Theta_R + \alpha\Theta_G
\end{equation}
where $\Theta_I$ are the interpolated parameters, $\Theta_R$ are the regression model parameters and $\Theta_G$ are the 
GAN model parameters with $\alpha \in [0, 1]$ being the interpolation parameter. The new model blends the result of the GAN and regression results. We observe that using the GAN model alone
can introduce artifacts (see Figure \ref{fig:model_interp}), blending the models in this way helps surpress those artifacts. Note that in this scheme, $\alpha=0$ gives the regression model and $\alpha = 1$ gives the GAN
model. Model interpolation has been shown to produce cleaner results than image interpolation, and has the
added benefit of not needing to run two models to produce a result. In Figure \ref{fig:model_interp} we show the model interpolation
results for $\alpha \in \{0.0, 0.7, 0.9, 1.0\}$ for several images from
the Live-1 dataset. This figure also serves as additional qualitative results for our method. These results were
generated from quality 10 JPEGs.

\begin{figure}
    \centering
    \resizebox{\columnwidth}{!}{
        \begin{tabular}{cc}
            Regression & $\alpha=0.7$ \\ 
            \includegraphics[width=0.4\linewidth]{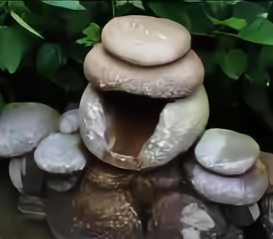} & 
            \includegraphics[width=0.4\linewidth]{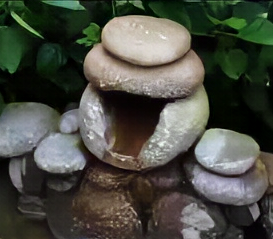} \\
            $\alpha=0.9$ & GAN \\ 
            \includegraphics[width=0.4\linewidth]{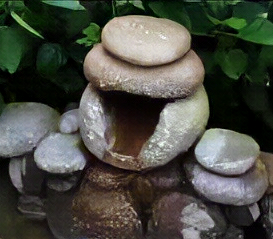} & 
            \includegraphics[width=0.4\linewidth]{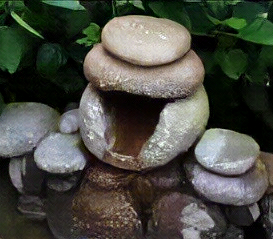} \\ \\

            Regression & $\alpha=0.7$ \\ 
            \includegraphics[width=0.5\linewidth]{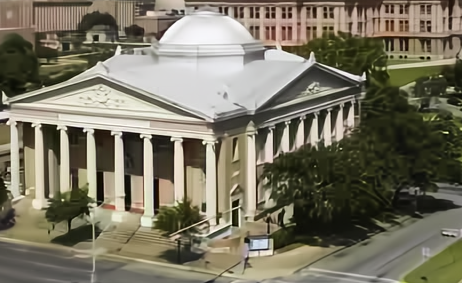} & 
            \includegraphics[width=0.5\linewidth]{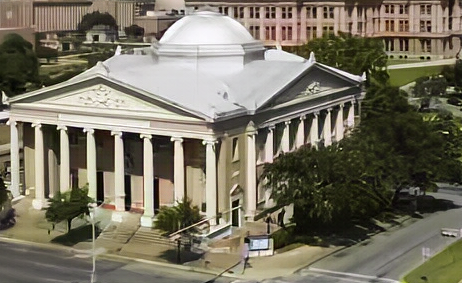} \\
            $\alpha=0.9$ & GAN \\ 
            \includegraphics[width=0.5\linewidth]{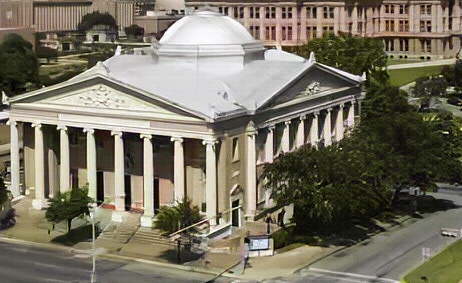} & 
            \includegraphics[width=0.5\linewidth]{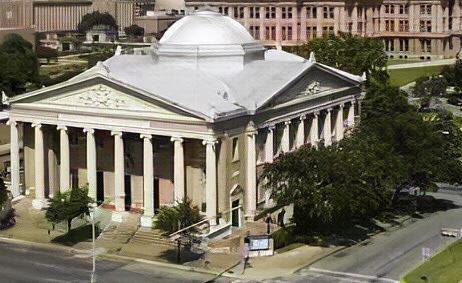}
        \end{tabular}
    }
    \caption{Model interpolation results 1/2}
\end{figure}

\begin{figure}
    \ContinuedFloat
    \centering
    \resizebox{\columnwidth}{!}{
        \begin{tabular}{cc}
            Regression & $\alpha=0.7$ \\ 
            \includegraphics[width=0.5\linewidth]{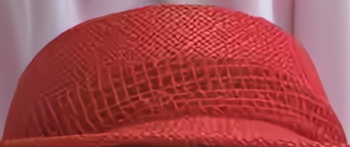} & 
            \includegraphics[width=0.5\linewidth]{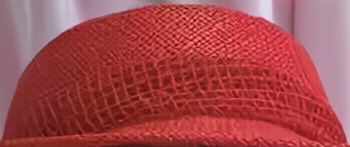} \\
            $\alpha=0.9$ & GAN \\ 
            \includegraphics[width=0.5\linewidth]{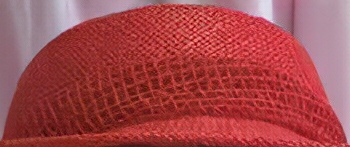} & 
            \includegraphics[width=0.5\linewidth]{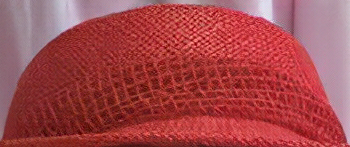} \\ \\

            Regression & $\alpha=0.7$ \\ 
            \includegraphics[width=0.5\linewidth]{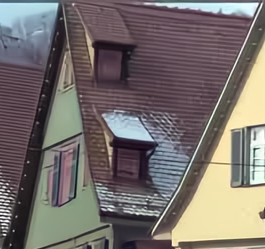} & 
            \includegraphics[width=0.5\linewidth]{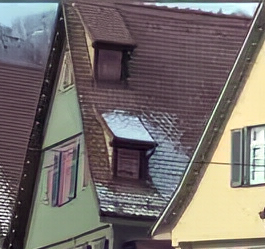} \\
            $\alpha=0.9$ & GAN \\ 
            \includegraphics[width=0.5\linewidth]{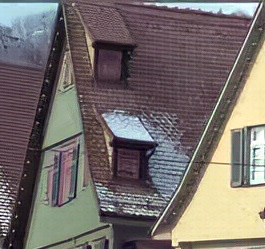} & 
            \includegraphics[width=0.5\linewidth]{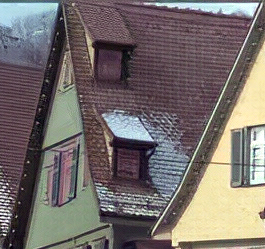} 
        \end{tabular}
    }
    \caption{Model interpolation results 2/2}
    \label{fig:model_interp}
\end{figure}

\subsection{Equivalent Quality}

One major motivation for JPEG artifact correction is that space or bandwidth can be saved by transmitting a small 
low quality JPEG and algorithmically correcting it before display. We explore how effective our model is at this
by computing the equivalent quality JPEG file for a restored image. Our argument is that a system can get the 
storage space savings of the lower quality JPEG and the visual fidelity of a higher quality JPEG by 
using our model.

\begin{figure}
    \begin{minipage}{0.5\linewidth}
        \centering
        \includegraphics[width=\linewidth]{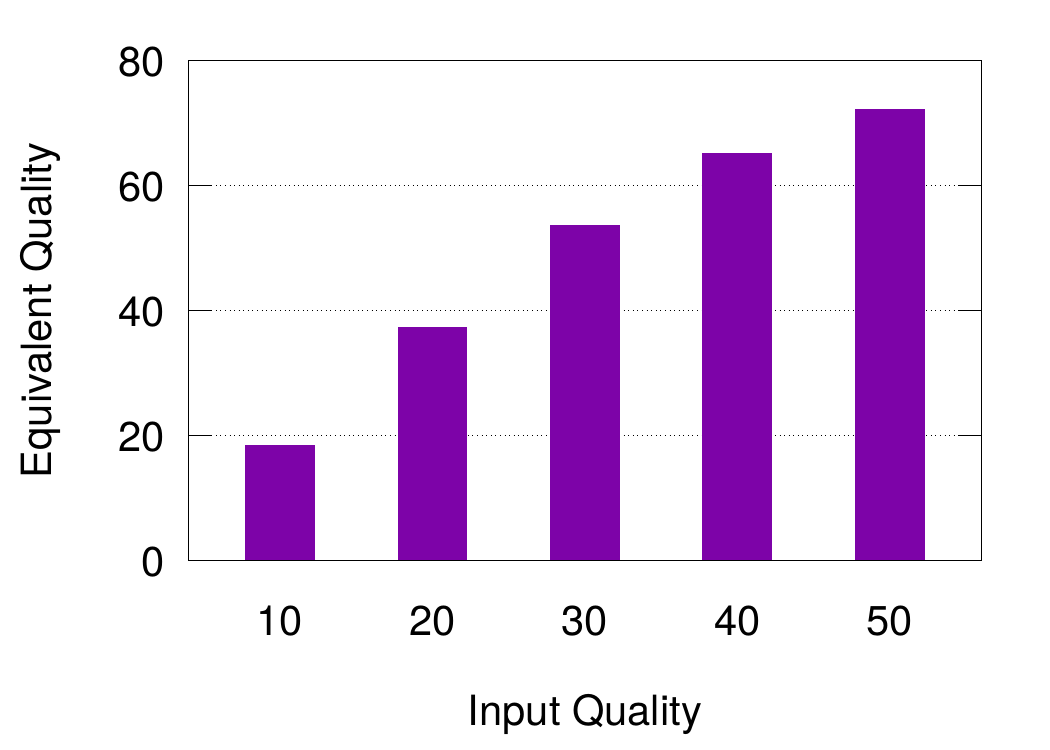}
    \end{minipage}
    \begin{minipage}{0.5\linewidth}
        \centering
        \includegraphics[width=\linewidth]{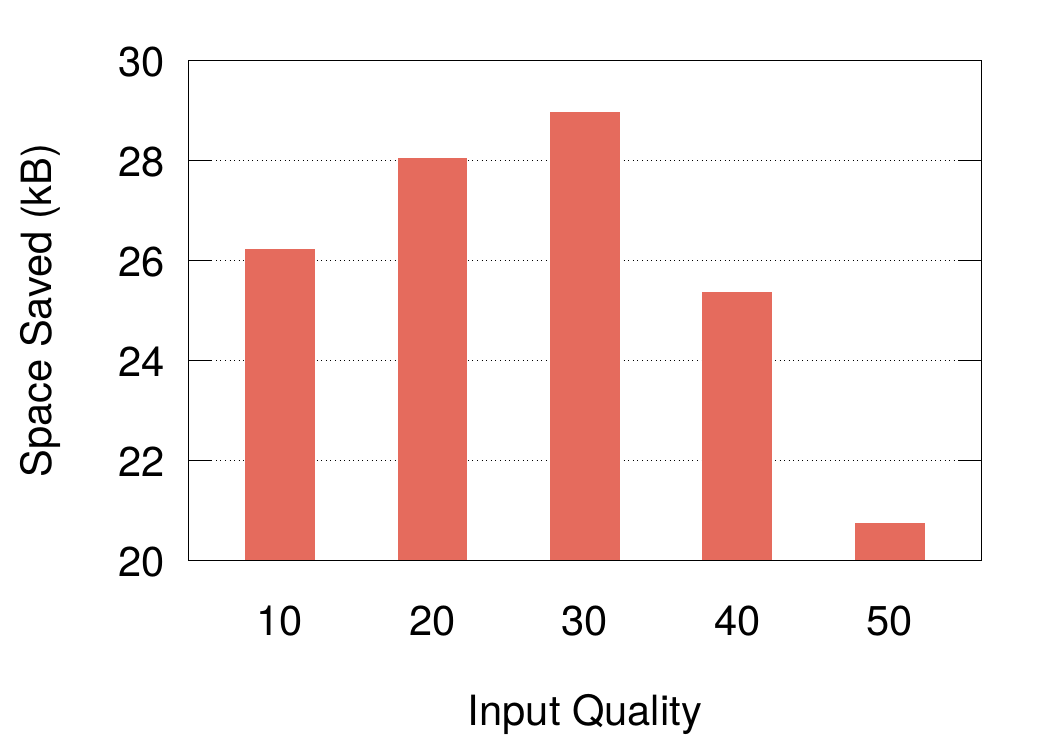}
    \end{minipage}
    \caption{Equivalent quality and space savings for Live-1 dataset.}
    \label{fig:eq_qual_live1}
\end{figure}

\begin{figure}
    \centering
    \resizebox{\columnwidth}{!}{
        \begin{tabular}{ccc}
            Input & Equivalent Quality JPEG & Ours \\ 
            \includegraphics[width=0.3\linewidth]{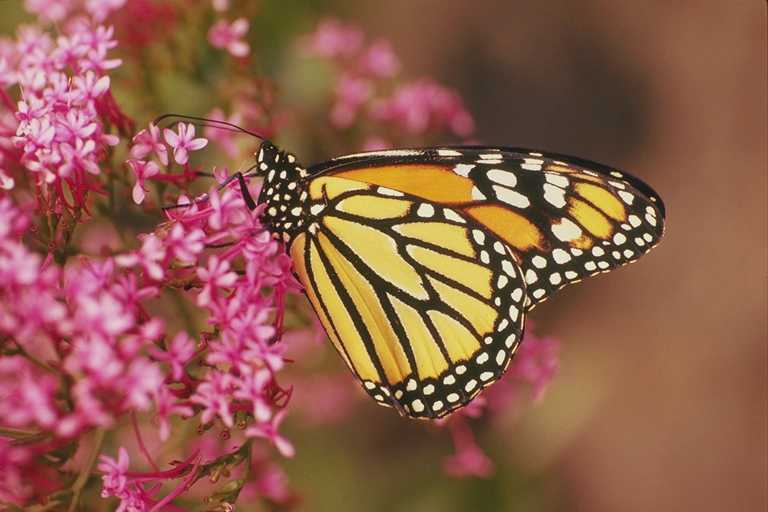} & 
            \includegraphics[width=0.3\linewidth]{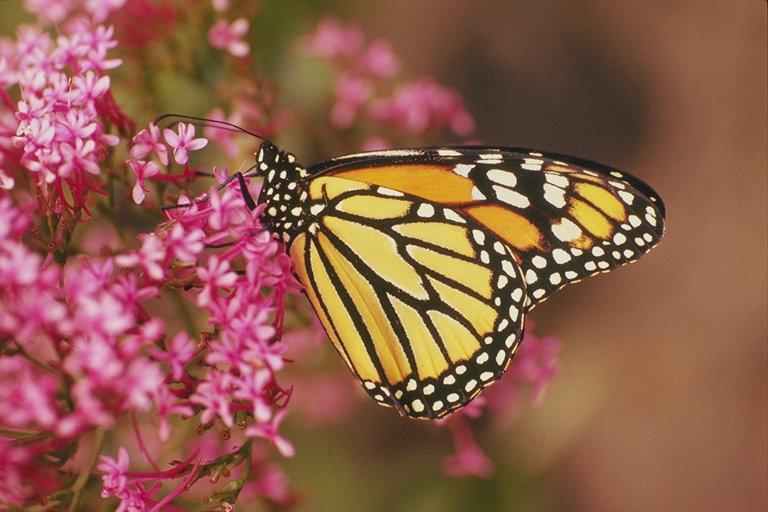} &
            \includegraphics[width=0.3\linewidth]{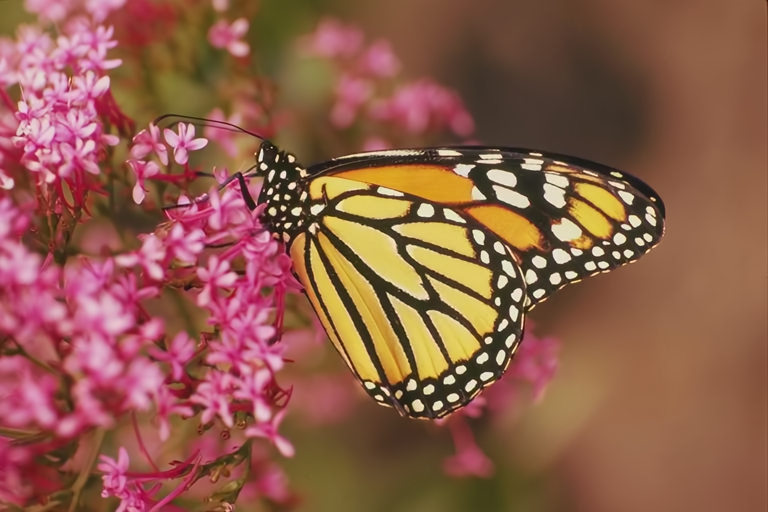} \\
            Quality: 50 & 
            Quality: 85 &
            29.5kB Saved  \\ \\
            Input & Equivalent Quality JPEG & Ours \\ 
            \includegraphics[width=0.3\linewidth]{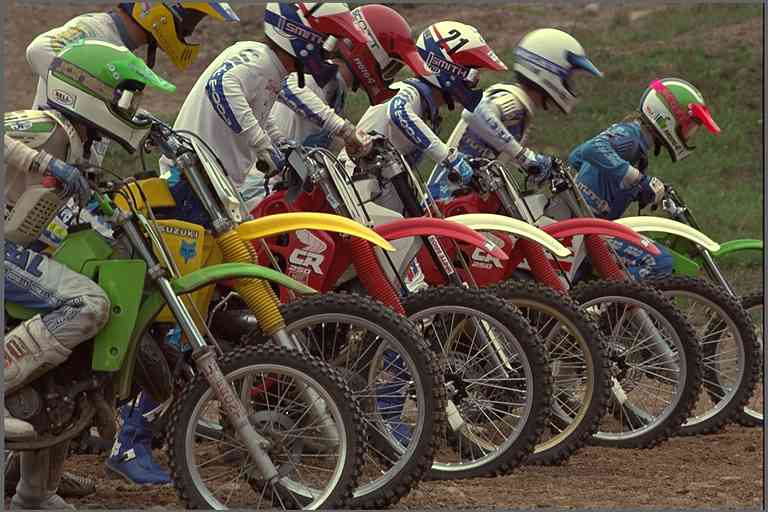} & 
            \includegraphics[width=0.3\linewidth]{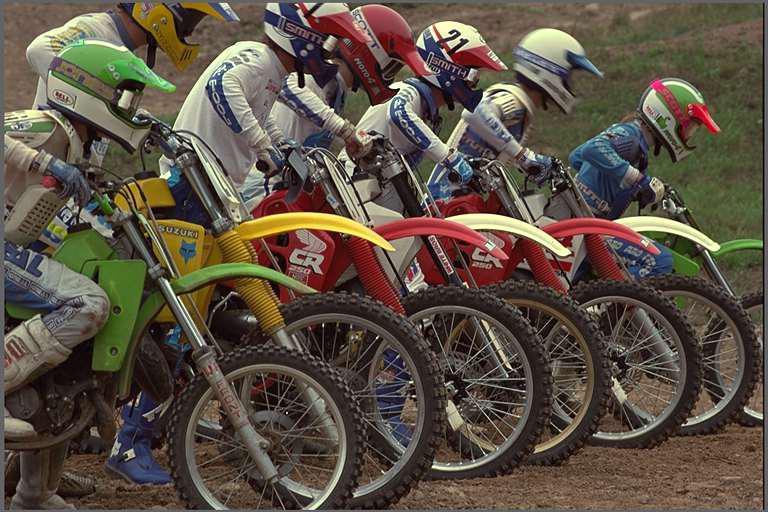} &
            \includegraphics[width=0.3\linewidth]{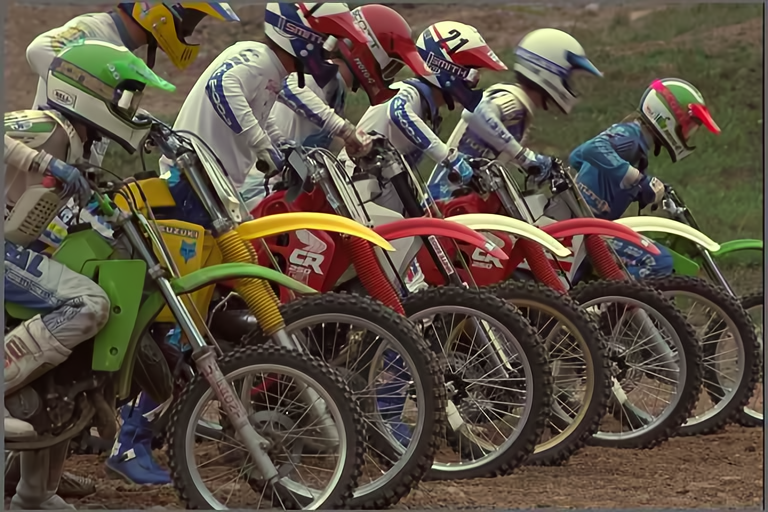} \\
            Quality: 30 & 
            Quality: 58 &
            46.8kB Saved \\ \\
            Input & Equivalent Quality JPEG & Ours \\ 
            \includegraphics[width=0.3\linewidth]{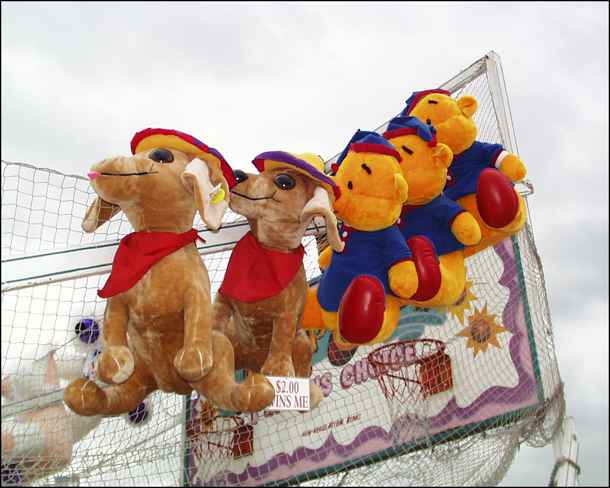} & 
            \includegraphics[width=0.3\linewidth]{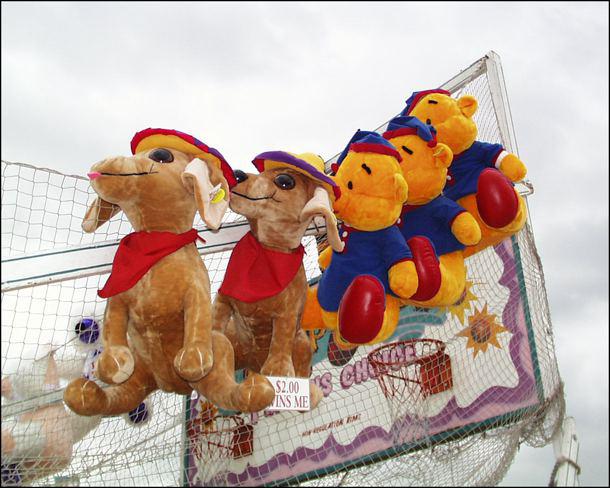} &
            \includegraphics[width=0.3\linewidth]{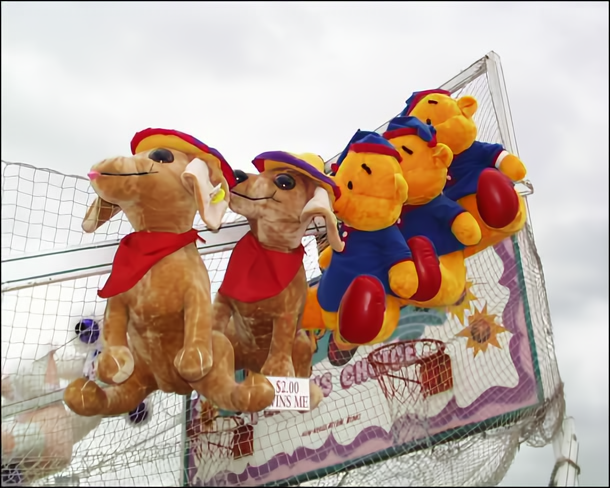} \\
            Quality: 40 & 
            Quality: 78 &
            25kB Saved
        \end{tabular}
    }
    \caption{Equivalent quality visualizations. For each image we show the input JPEG, the JPEG with equivalent SSIM
    to our model output, and our model output.}
    \label{fig:eq_qual_res}
\end{figure}

To show this we use the Live-1 dataset. For qualities in [10, 50] in steps of 10, we compute the average increase in 
JPEG quality incurred by our model. We do this by compressing the input image at higher and higher qualities until we
find the first quality with SSIM greater than or equal to our restoration's SSIM. We then save the low quality JPEG
and the equivalent quality JPEG and measure the size difference in kilobytes. We average the quality increase and 
space savings over the entire dataset, to show the amount of space saved by using our method over using the higher quality JPEG 
directly. This result is shown in Figure \ref{fig:eq_qual_live1}. We also show qualitative examples for several images in
Figure \ref{fig:eq_qual_res}. Note that because the SSIM measure is not perfect, often our model outputs images that look
better than the equivalent quality JPEG.

\subsection{Frequency Domain Analysis}

In this section we show results in the DCT frequency domain. A well known phenomenon of JPEG compression is the removal
of high frequency information. To check how well our model restores this information, we take the Y channel from several
images and show the colormapped DCT of the original image, the JPEG at quality 10, the image as restored by our regression
model, and the image restored by our GAN model. Next, for each image, we plot the probability that each of the 15 spatial
frequencies in a DCT block are set (\eg has a magnitude greater than 0). This is shown in Figure \ref{fig:freq}. While our regression model is able to fill in 
high frequencies, our GAN model nearly matches the original images in terms of frequency saturation. Additionally since our 
network operates in the DCT domain, these outputs serve as an interesting qualitative result.

\begin{figure}
    \centering
    \resizebox{\columnwidth}{!}{
        \begin{tabular}{cc}
            Original & Plot \\ 
            \includegraphics[width=0.3\linewidth]{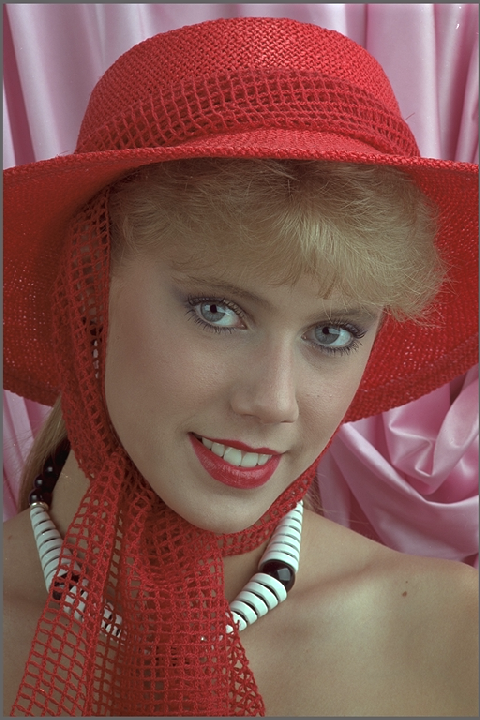} & \includegraphics[width=0.7\linewidth]{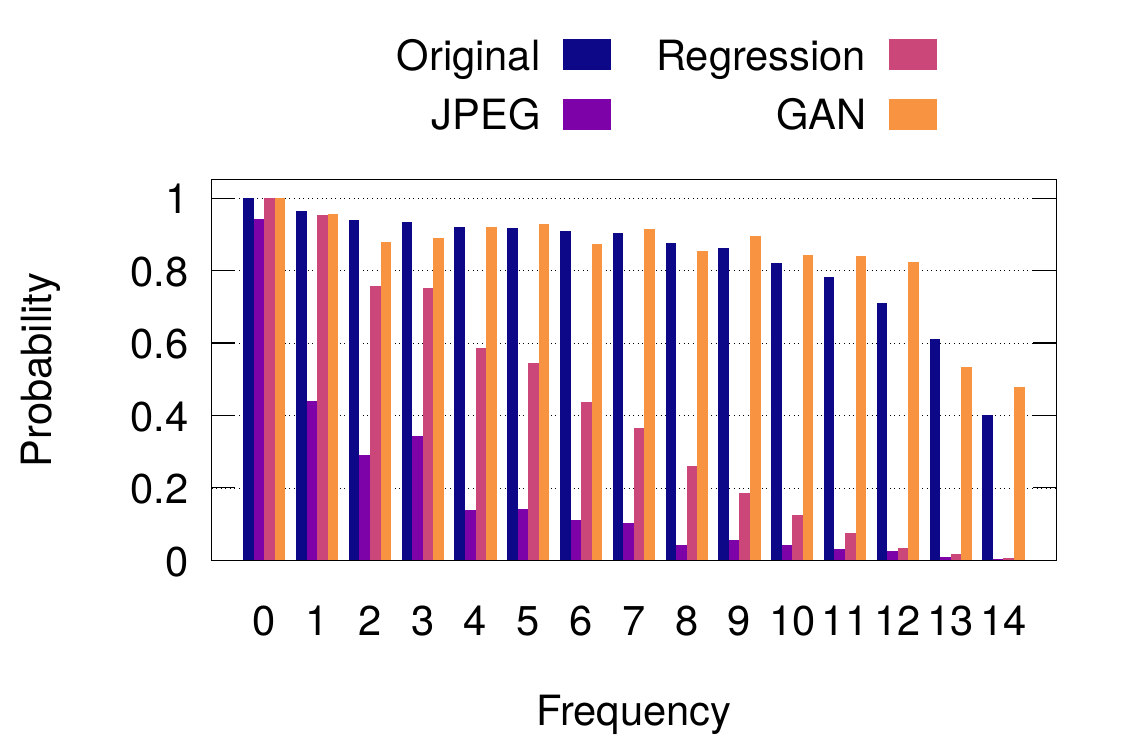}
        \end{tabular}
    }
    \resizebox{\columnwidth}{!}{
        \begin{tabular}{cccc}
            DCT & JPEG Q=10 & Regression & GAN \\
            \includegraphics[width=0.25\linewidth]{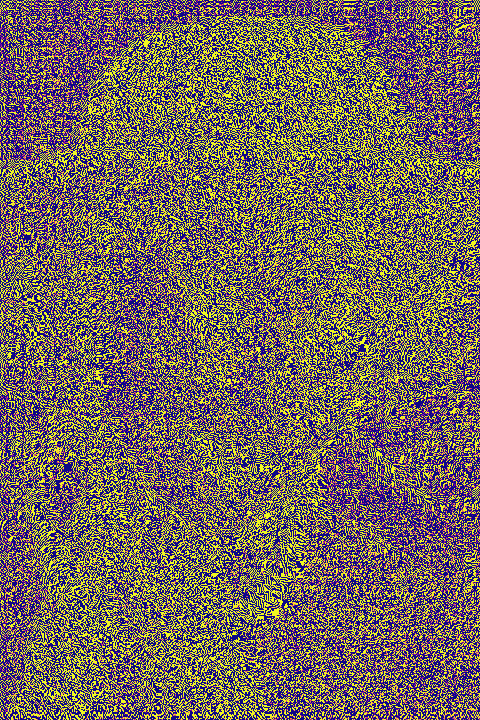} & \includegraphics[width=0.25\linewidth]{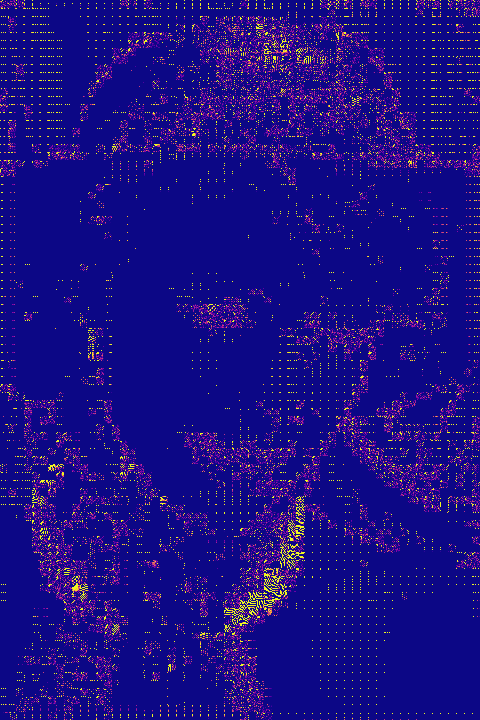} & \includegraphics[width=0.25\linewidth]{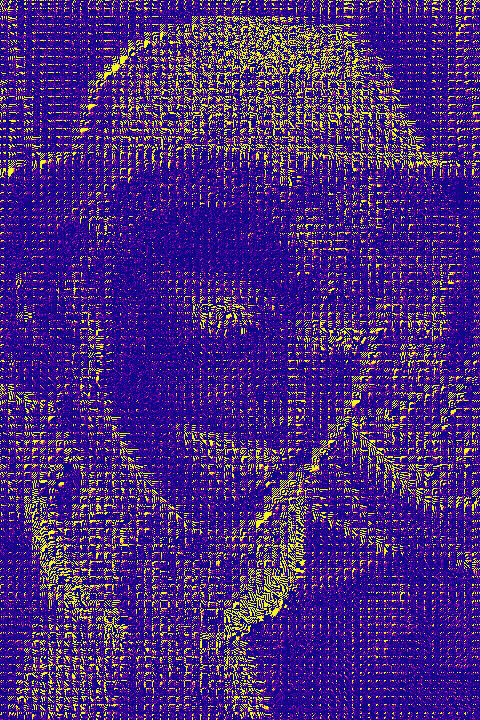} & \includegraphics[width=0.25\linewidth]{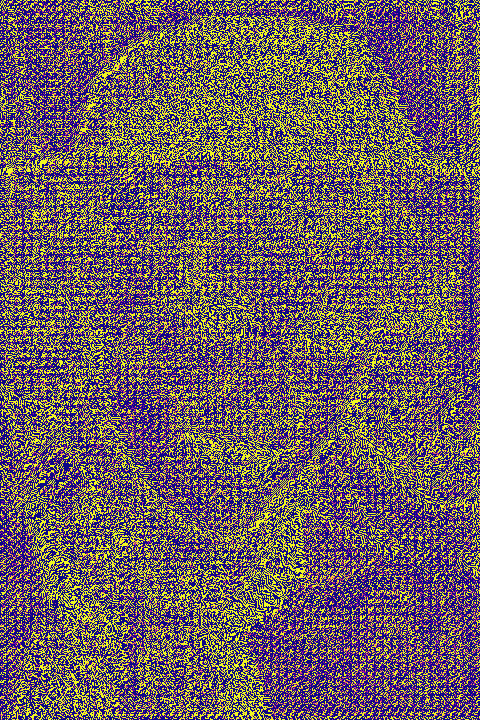}
            
        \end{tabular}
    }
    \resizebox{\columnwidth}{!}{
        \begin{tabular}{cc}
            Original & Plot \\ 
            \includegraphics[width=0.5\linewidth]{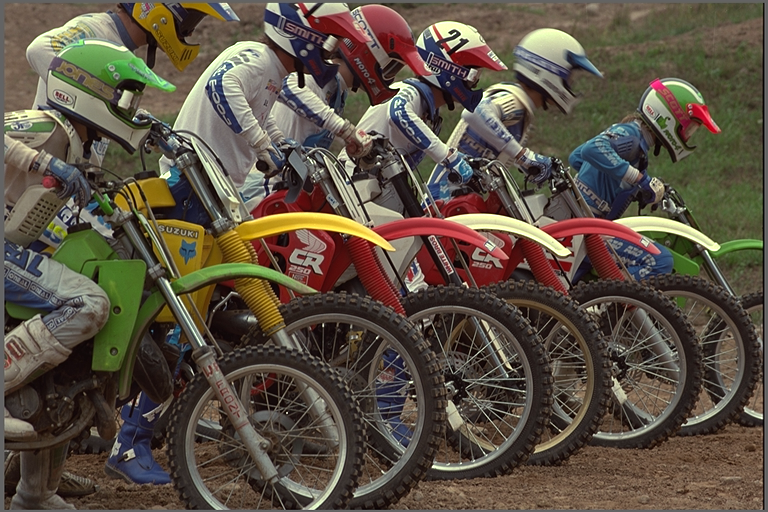} & \includegraphics[width=0.5\linewidth]{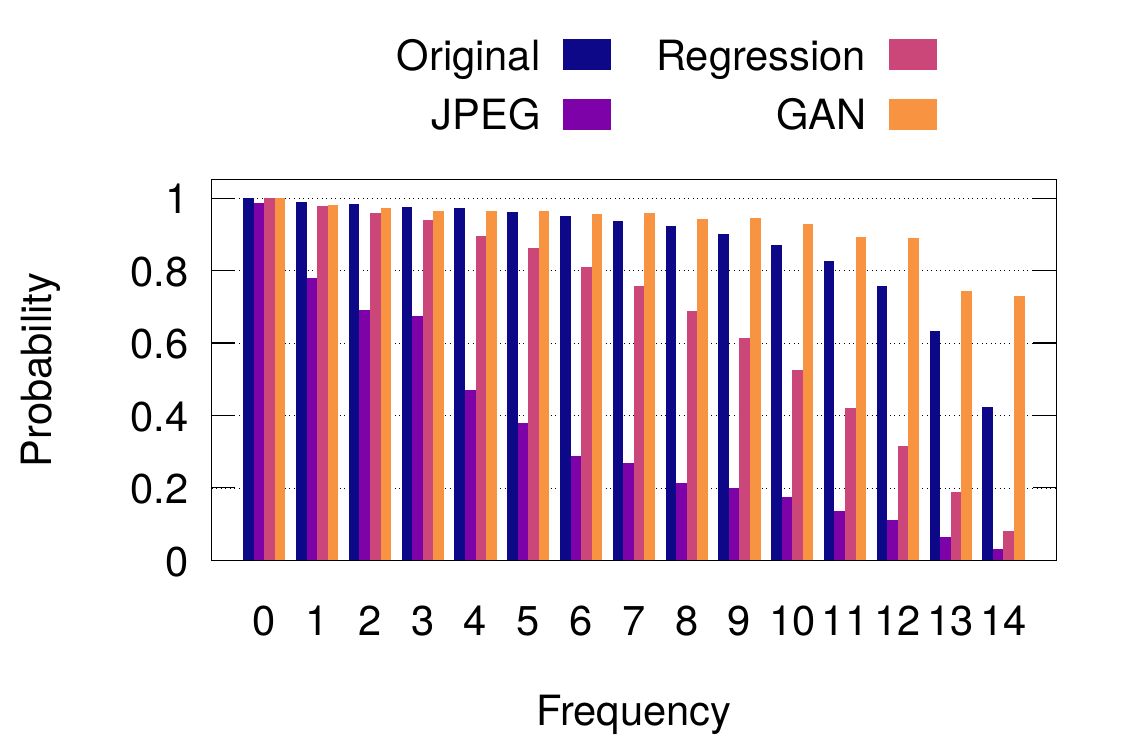}
        \end{tabular}
    }
    \resizebox{\columnwidth}{!}{
        \begin{tabular}{cccc}
            DCT & JPEG Q=10 & Regression & GAN \\
            \includegraphics[width=0.25\linewidth]{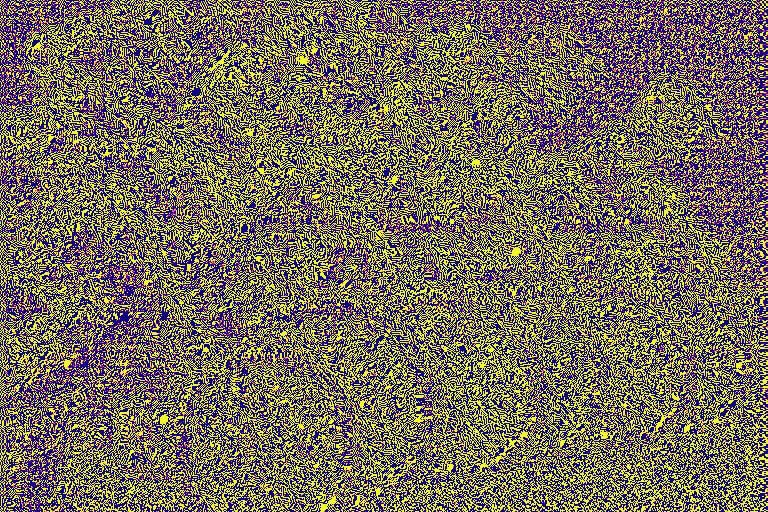} & \includegraphics[width=0.25\linewidth]{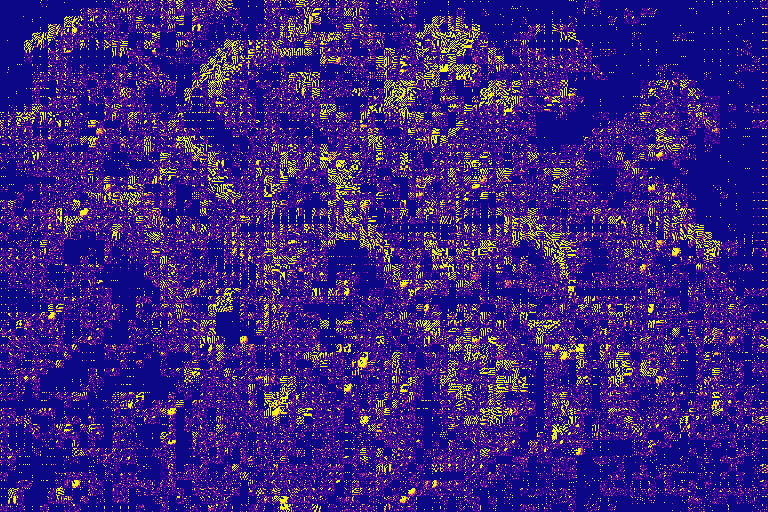} & \includegraphics[width=0.25\linewidth]{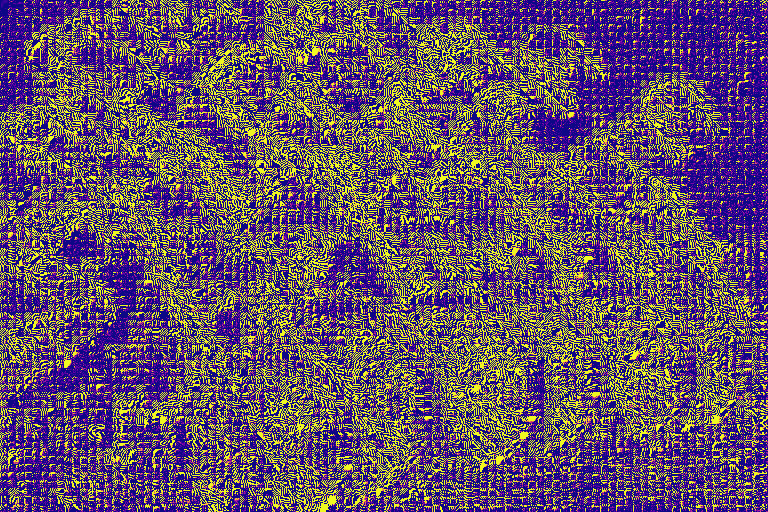} & \includegraphics[width=0.25\linewidth]{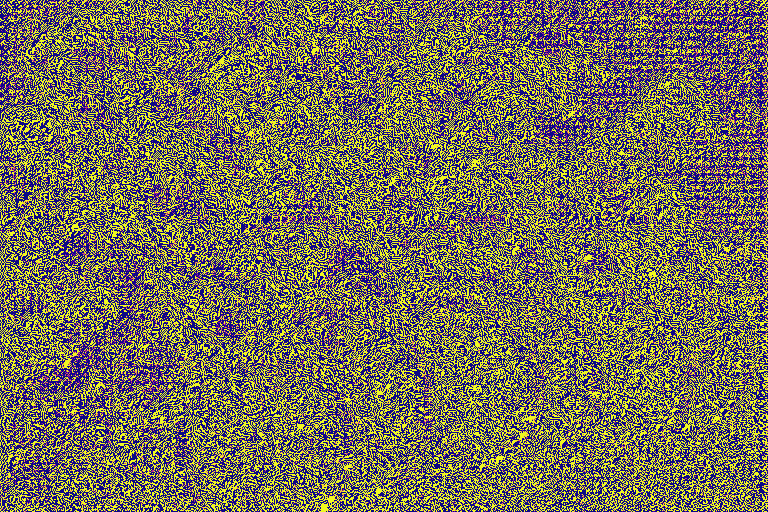}
            
        \end{tabular}
    }
    \caption{Frequency domain results 1/2.}
    \label{fig:freq}
\end{figure}
\begin{figure}
    \ContinuedFloat
    \centering
    \resizebox{\columnwidth}{!}{
        \begin{tabular}{cc}
            Original & Plot \\ 
            \includegraphics[width=0.3\linewidth]{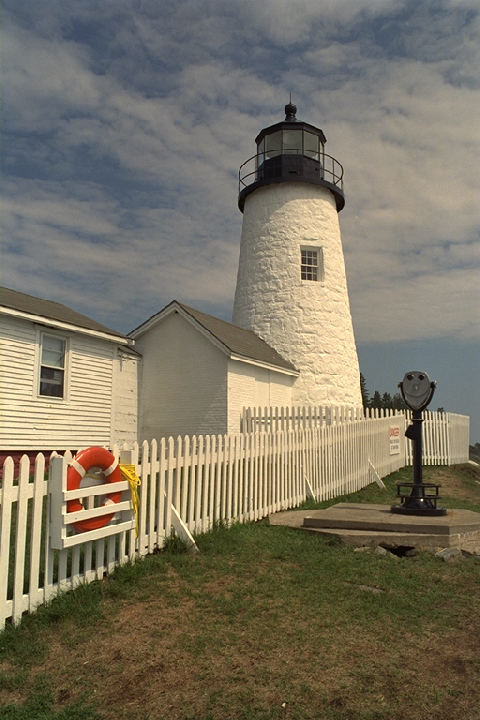} & \includegraphics[width=0.7\linewidth]{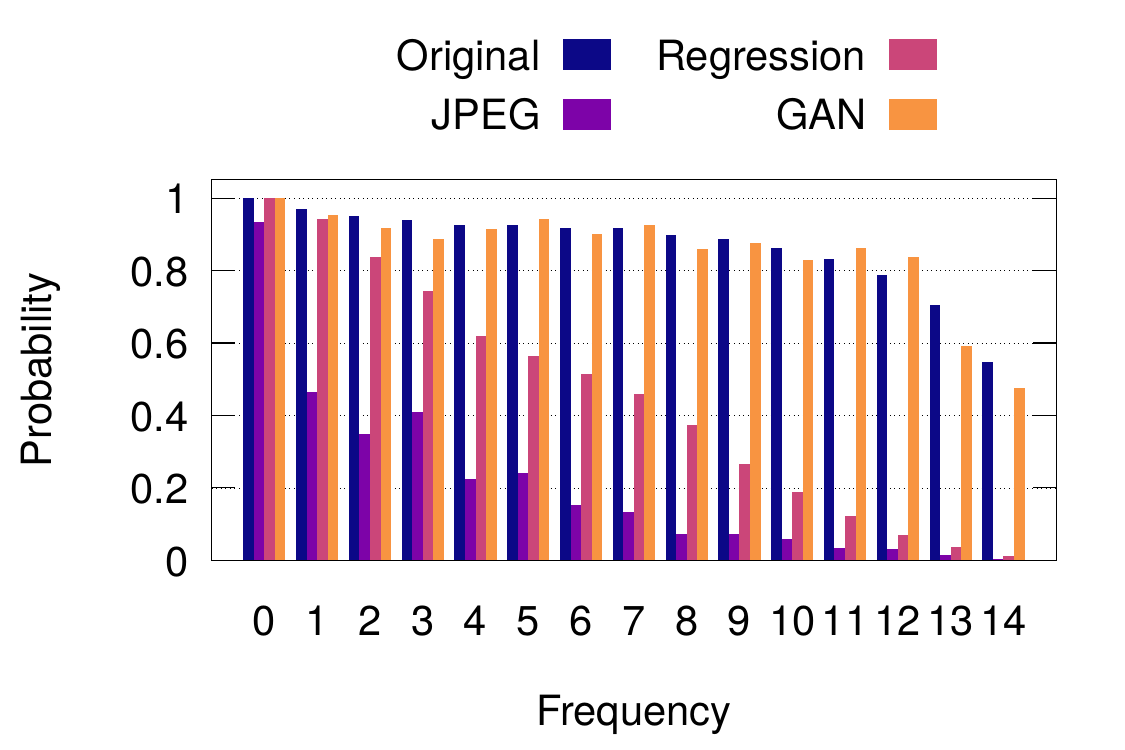}
        \end{tabular}
    }
    \resizebox{\columnwidth}{!}{
        \begin{tabular}{cccc}
            DCT & JPEG Q=10 & Regression & GAN \\
            \includegraphics[width=0.25\linewidth]{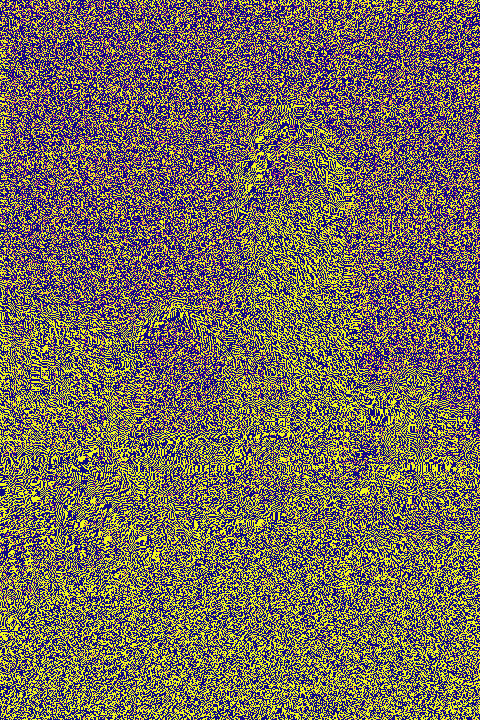} & \includegraphics[width=0.25\linewidth]{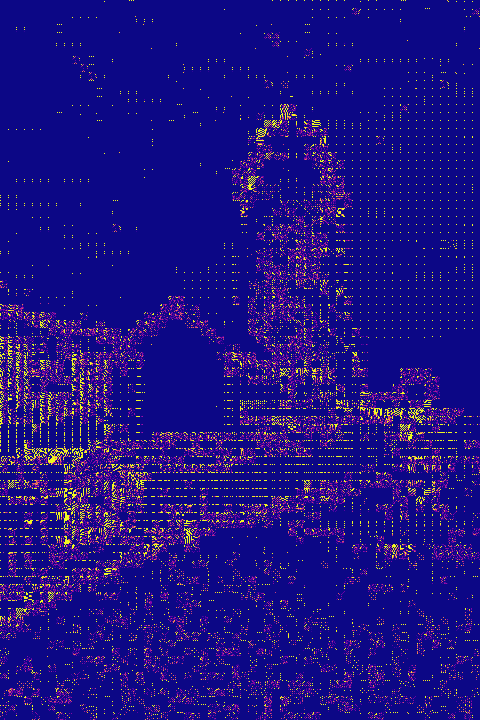} & \includegraphics[width=0.25\linewidth]{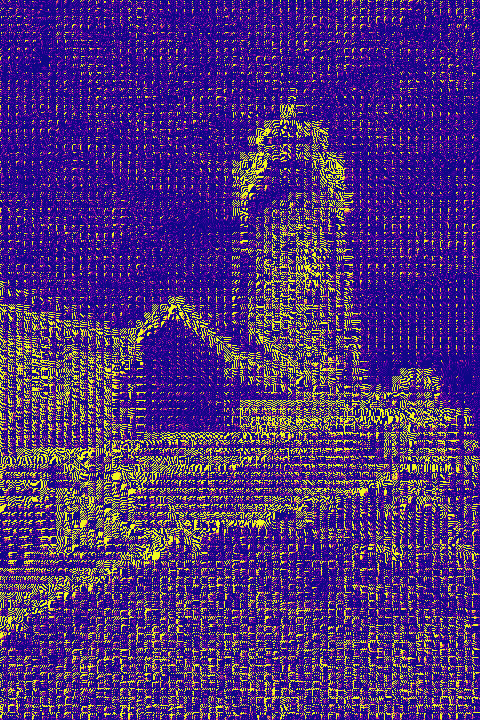} & \includegraphics[width=0.25\linewidth]{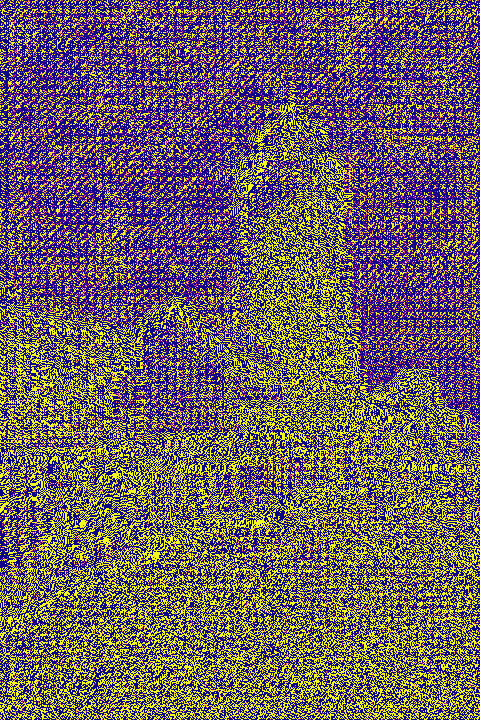}
            
        \end{tabular}
    }
    \resizebox{\columnwidth}{!}{
        \begin{tabular}{cc}
            Original & Plot \\ 
            \includegraphics[width=0.5\linewidth]{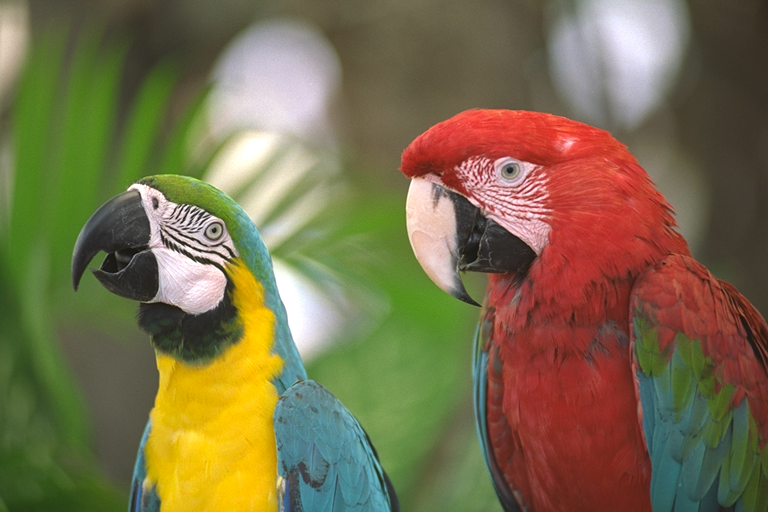} & \includegraphics[width=0.5\linewidth]{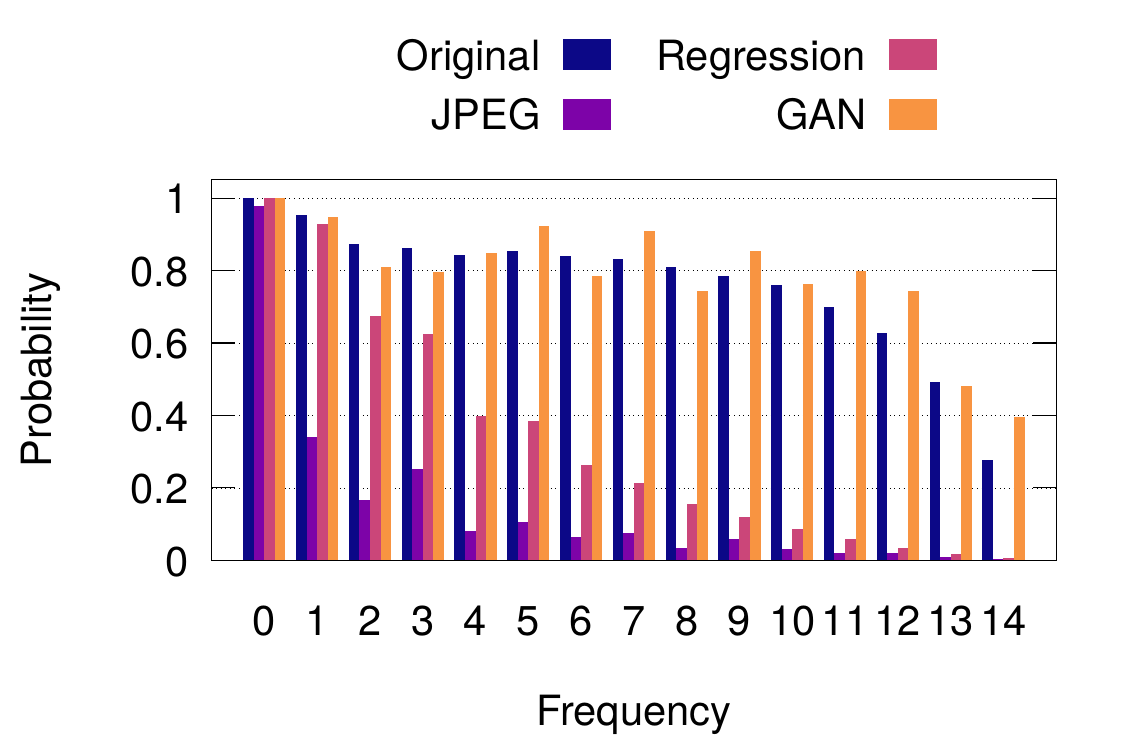}
        \end{tabular}
    }
    \resizebox{\columnwidth}{!}{
        \begin{tabular}{cccc}
            DCT & JPEG Q=10 & Regression & GAN \\
            \includegraphics[width=0.25\linewidth]{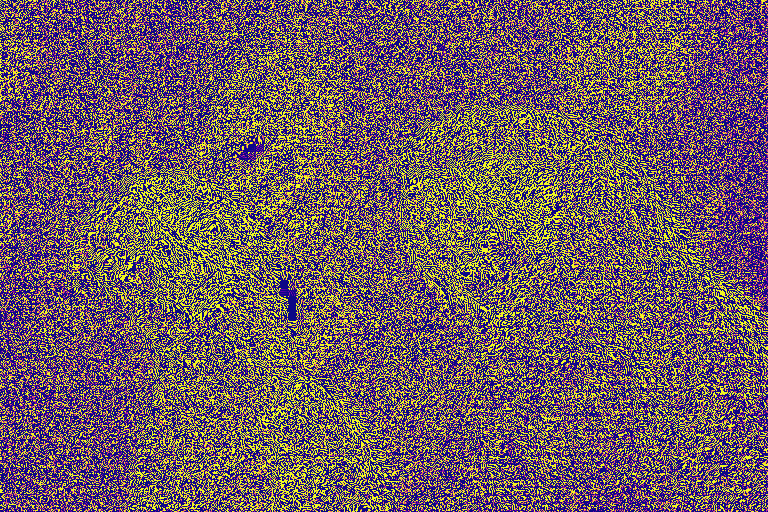} & \includegraphics[width=0.25\linewidth]{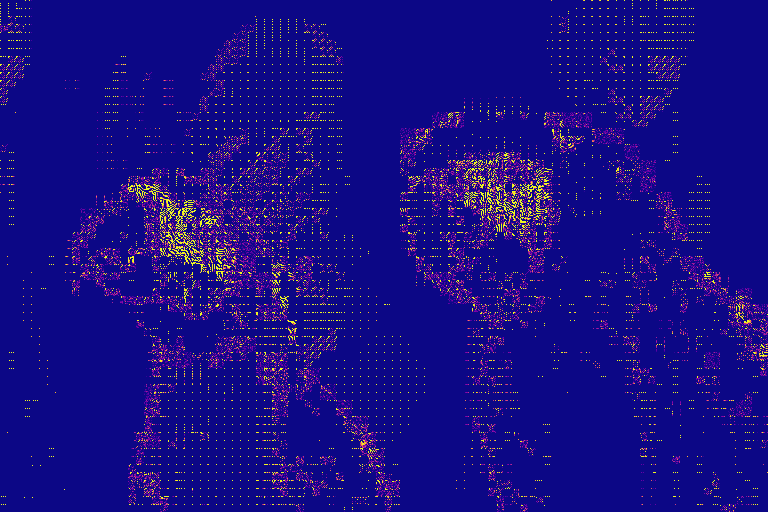} & \includegraphics[width=0.25\linewidth]{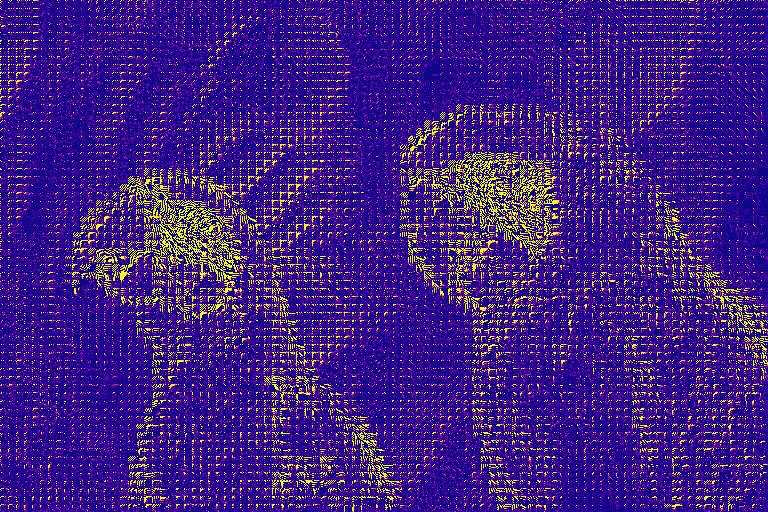} & \includegraphics[width=0.25\linewidth]{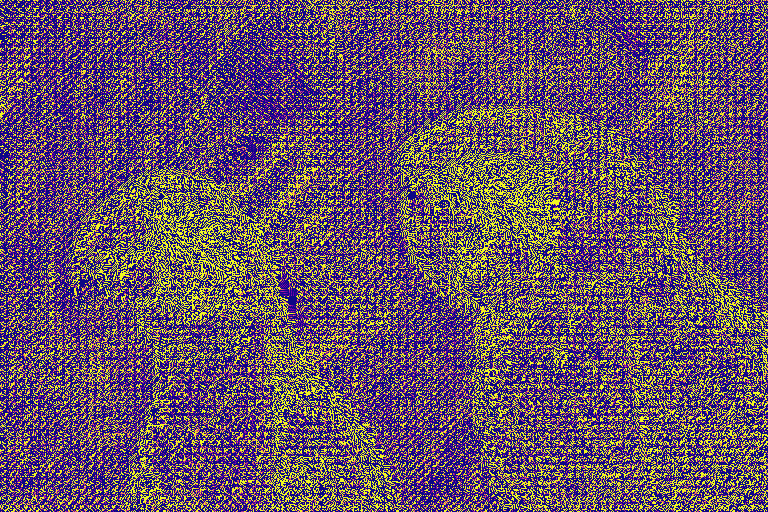}
            
        \end{tabular}
    }
    \caption{Frequency domain results 2/2.}
\end{figure}

\subsection{Runtime analysis}

\begin{figure}
    \centering
    \begin{minipage}{0.7\linewidth}
        \centering
        \includegraphics[width=\linewidth]{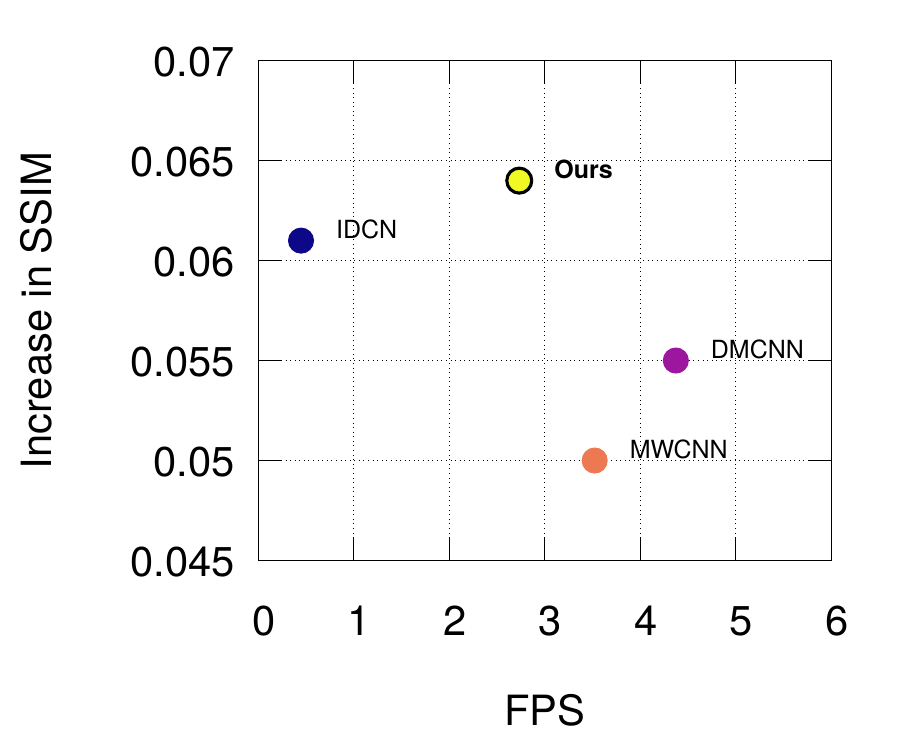}
    \end{minipage}
    \caption{Increase in SSIM vs FPS. Our result is highlighted.}
    \label{fig:fps}
\end{figure}

We show the runtime inference performance of our network compared to the other networks we ran against. We measure FPS
on our NVIDIA Pascal GPU for 100 720p ($1280 \times 720$) frames and plot frames per second vs SSIM increase for quality 10 
Live-1 images in Figure \ref{fig:fps}. We do not include ARCNN in this figure as the authors do not provide
GPU accelerated inference code. For grayscale
only models we only use single channel test images (we not not run the model three times as would be required to produce
an RGB output).

\section{Qualitative Results}

In this section we show qualitative results on Quality 10 and 20 images for our regression network. These results
are in Figure \ref{fig:qual}.

\begin{figure}
    \centering
    \resizebox{\columnwidth}{!}{
        \begin{tabular}{ccc}
            JPEG Q=10 & Ours & Original \\ 
            \includegraphics[width=0.3\linewidth]{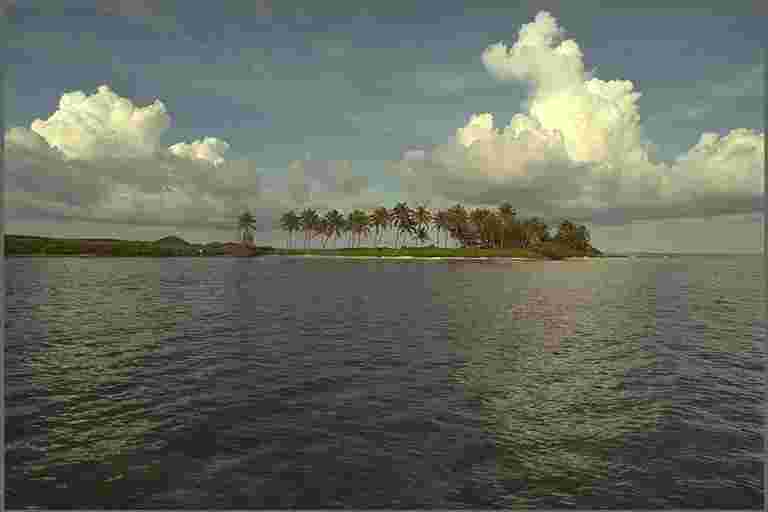} & 
            \includegraphics[width=0.3\linewidth]{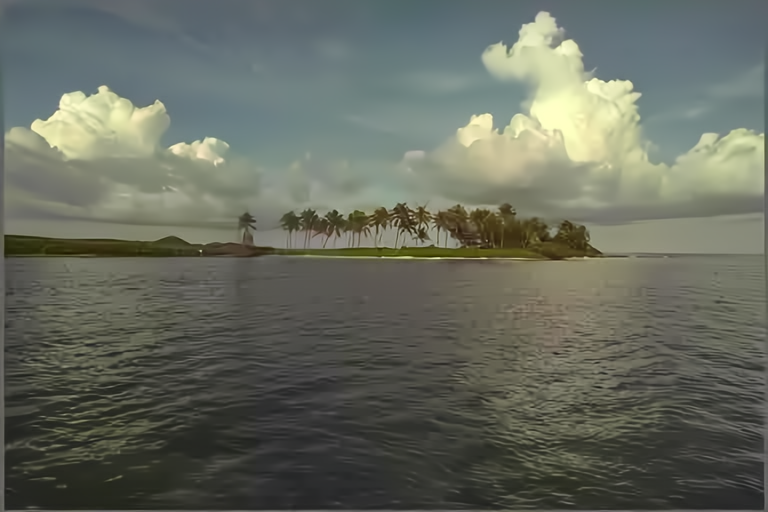} &
            \includegraphics[width=0.3\linewidth]{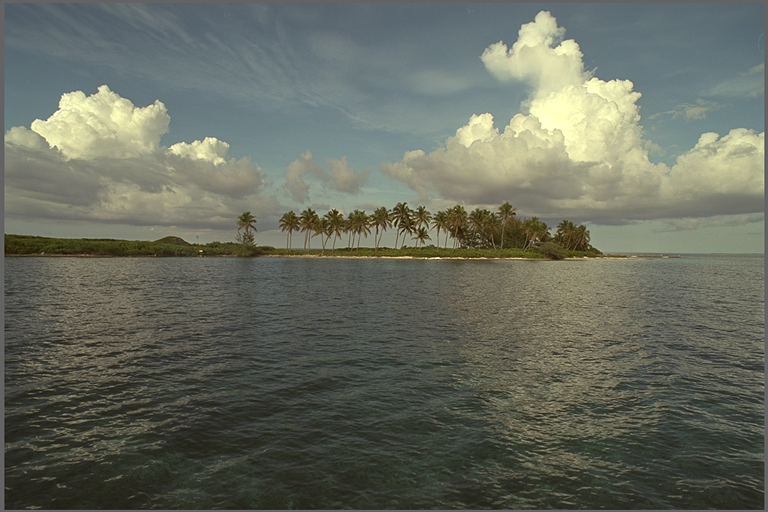} \\
            JPEG Q=20 & Ours & Original \\ 
            \includegraphics[width=0.3\linewidth]{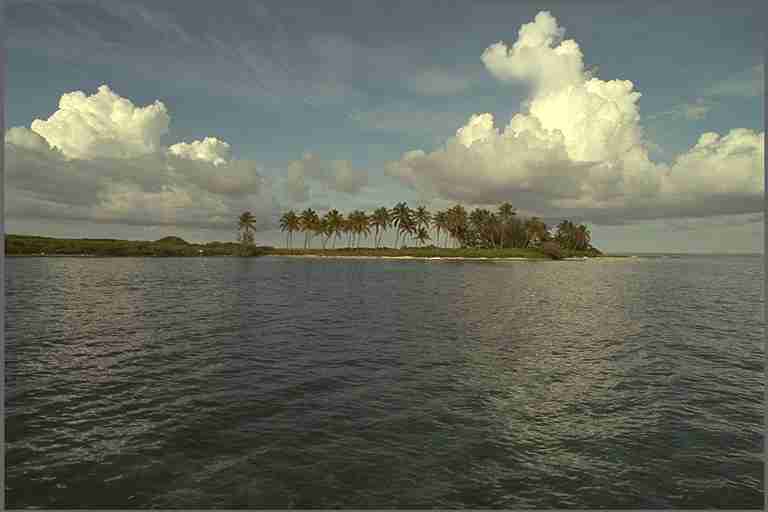} & 
            \includegraphics[width=0.3\linewidth]{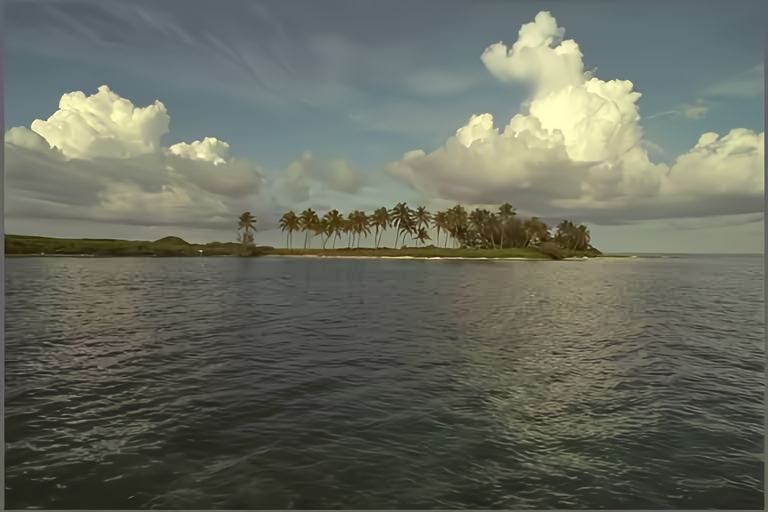} &
            \includegraphics[width=0.3\linewidth]{figures/qual/ocean/original.png} \\

            JPEG Q=10 & Ours & Original \\ 
            \includegraphics[width=0.3\linewidth]{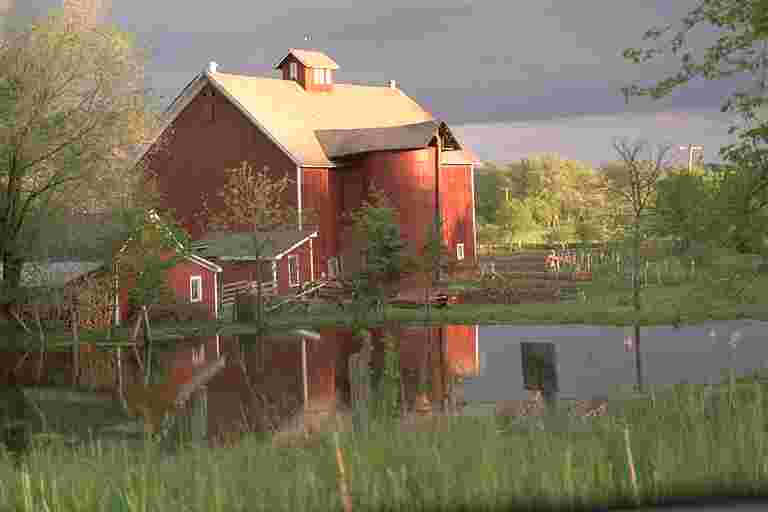} & 
            \includegraphics[width=0.3\linewidth]{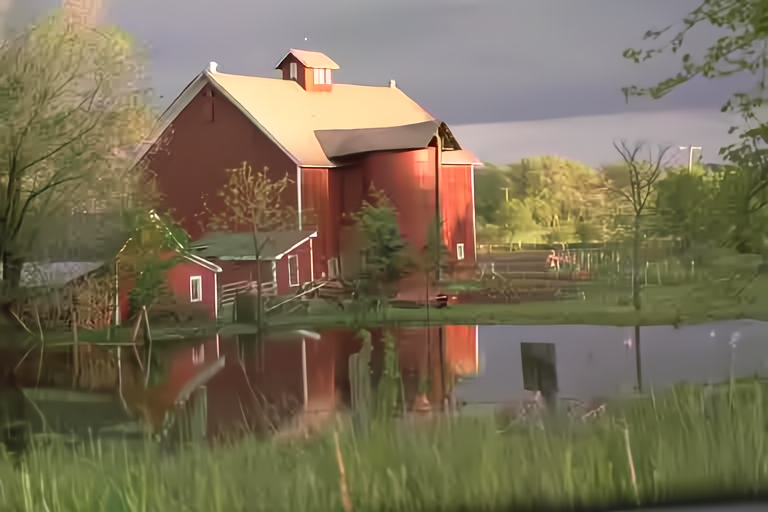} &
            \includegraphics[width=0.3\linewidth]{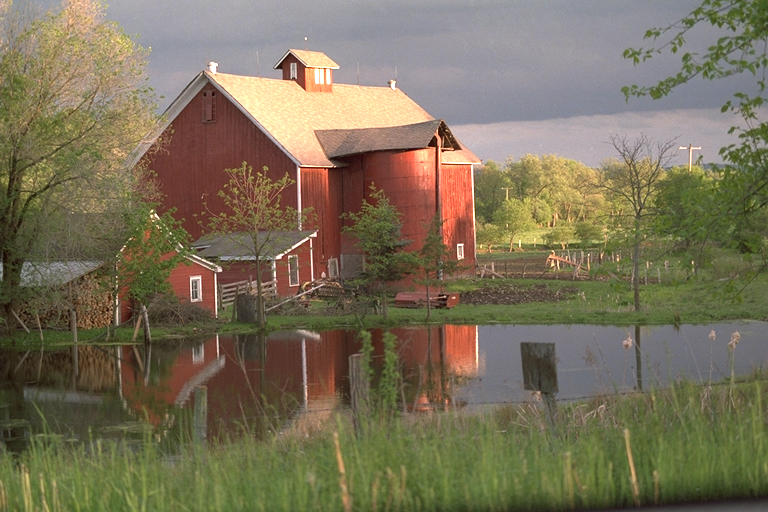} \\
            JPEG Q=20 & Ours & Original \\ 
            \includegraphics[width=0.3\linewidth]{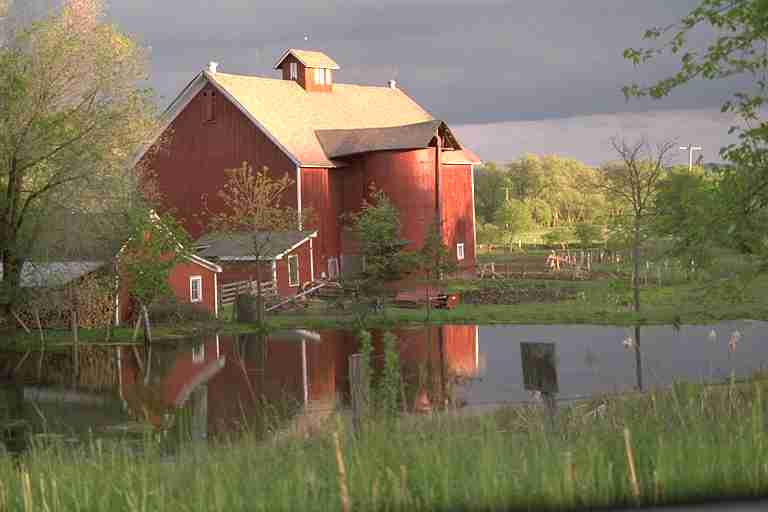} & 
            \includegraphics[width=0.3\linewidth]{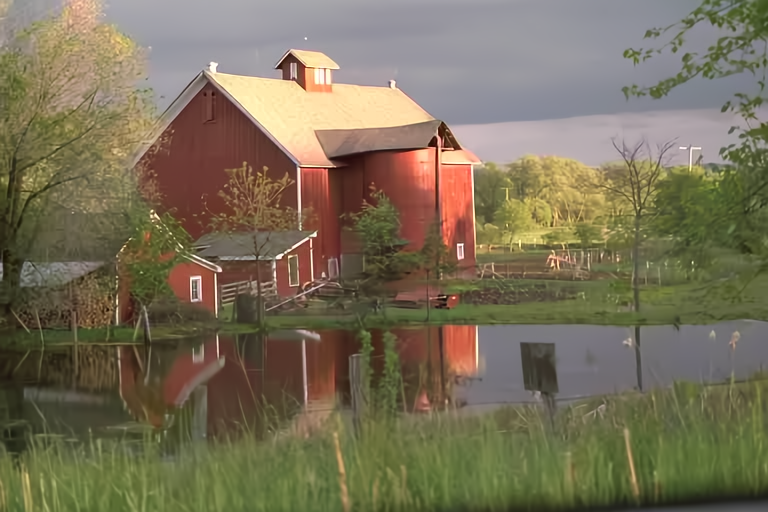} &
            \includegraphics[width=0.3\linewidth]{figures/qual/house/original.png} 
        \end{tabular}
    }
    \caption{Qualitative results 1/2. Live-1 images.}
\end{figure}

\begin{figure}
    \ContinuedFloat
    \centering
    \resizebox{\columnwidth}{!}{
        \begin{tabular}{ccc}
            JPEG Q=10 & Ours & Original \\ 
            \includegraphics[width=0.3\linewidth]{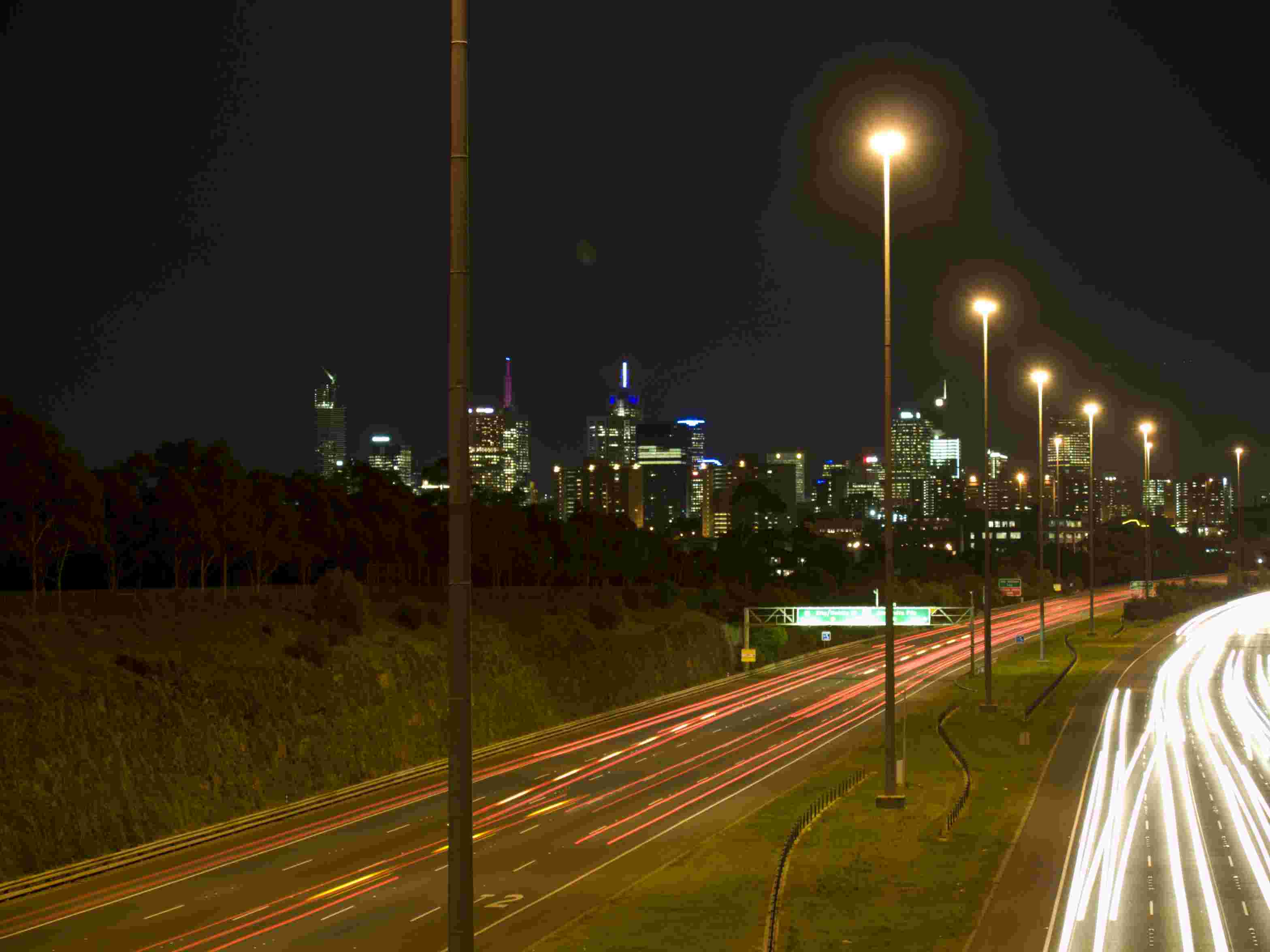} & 
            \includegraphics[width=0.3\linewidth]{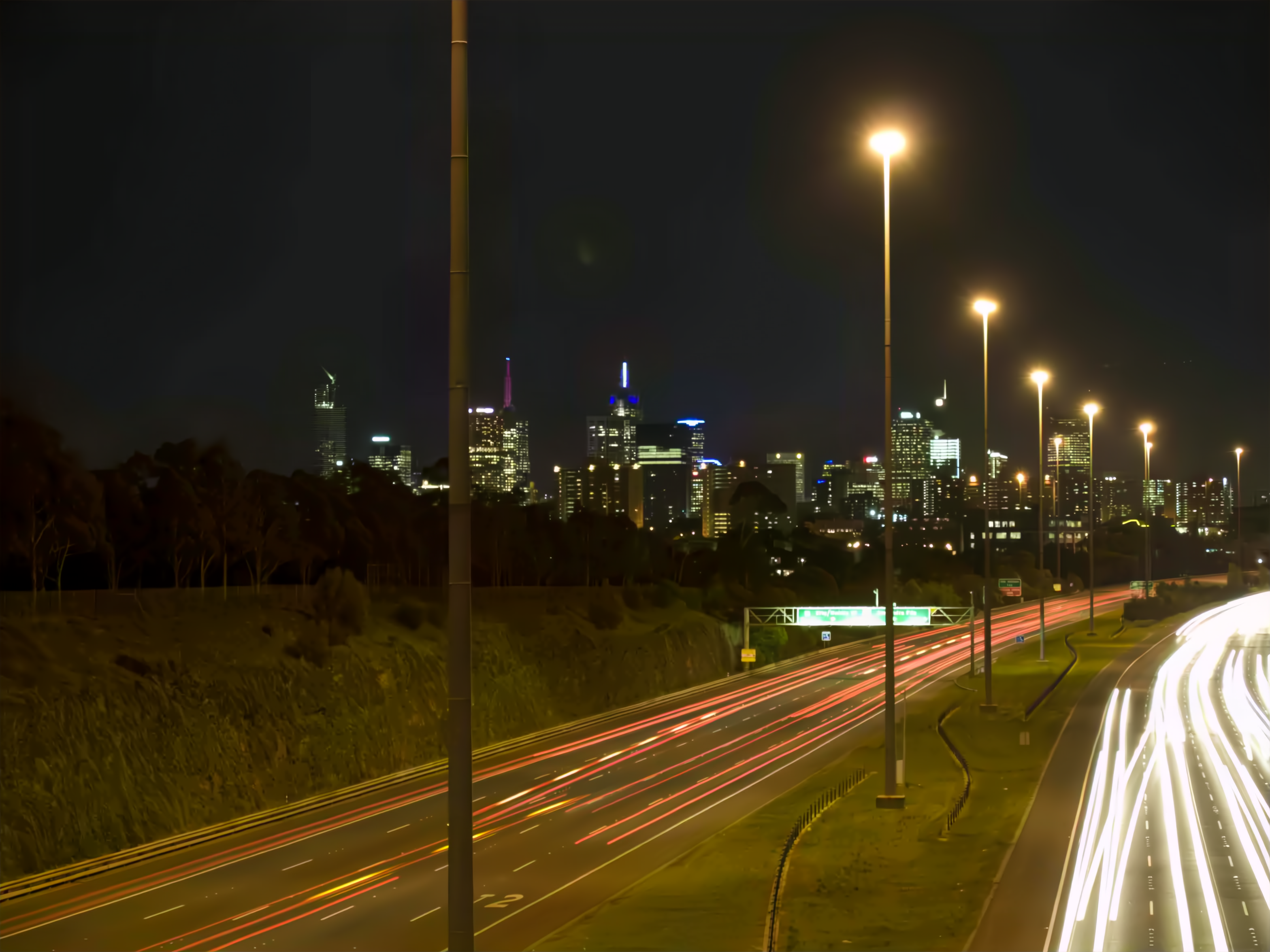} &
            \includegraphics[width=0.3\linewidth]{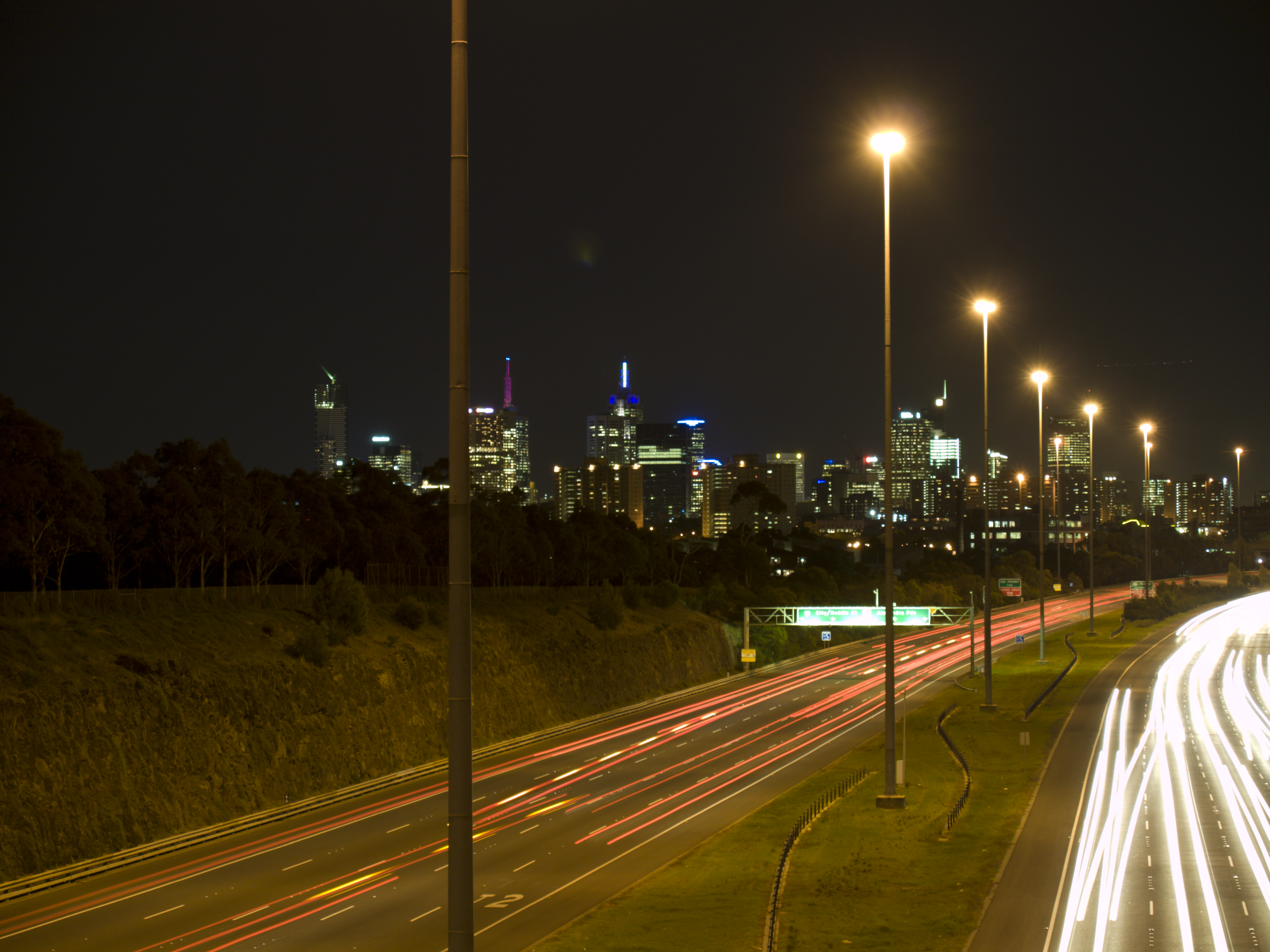} \\
            JPEG Q=20 & Ours & Original \\ 
            \includegraphics[width=0.3\linewidth]{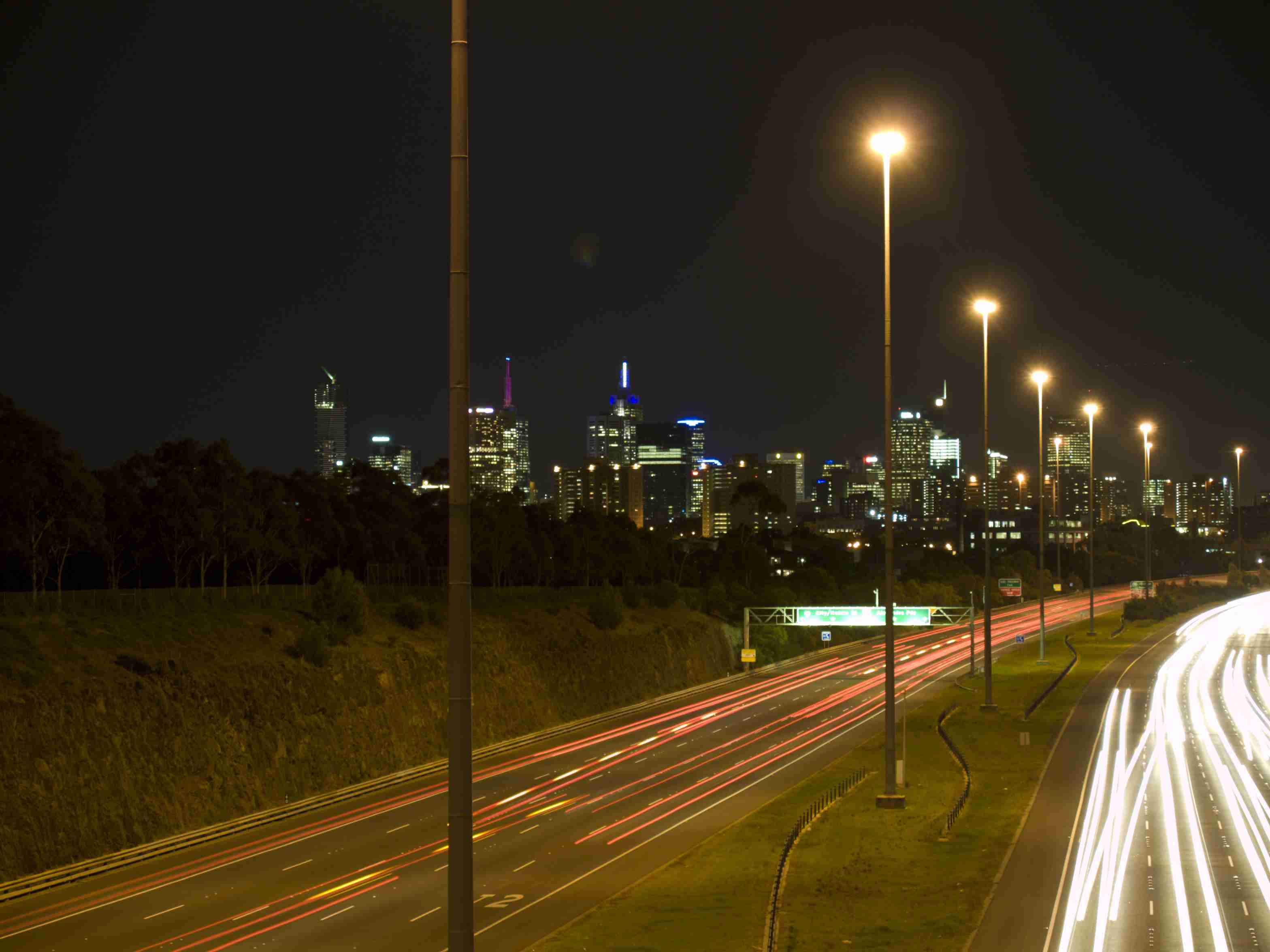} & 
            \includegraphics[width=0.3\linewidth]{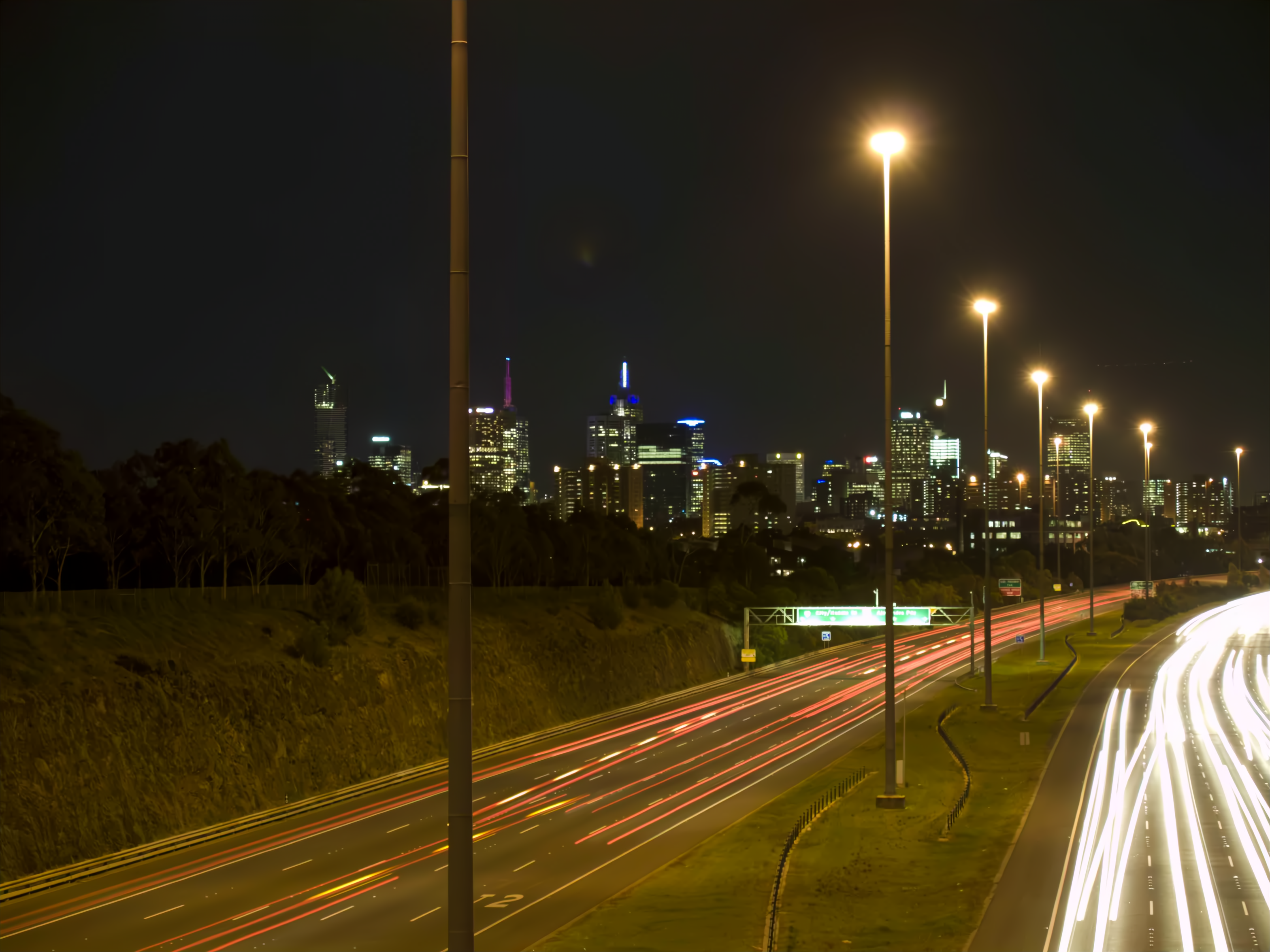} &
            \includegraphics[width=0.3\linewidth]{figures/qual/night/original.png} \\

            JPEG Q=10 & Ours & Original \\ 
            \includegraphics[width=0.3\linewidth]{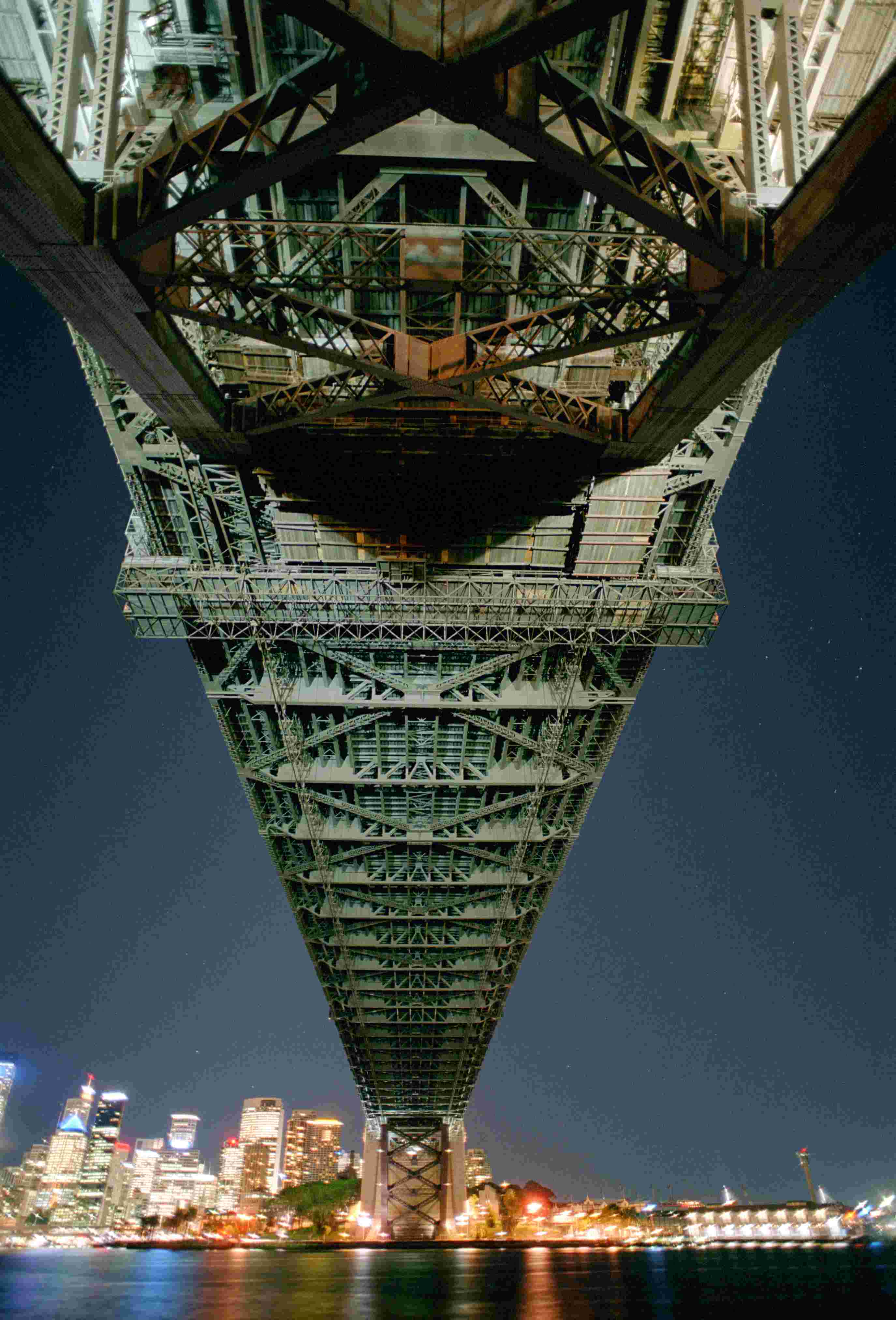} & 
            \includegraphics[width=0.3\linewidth]{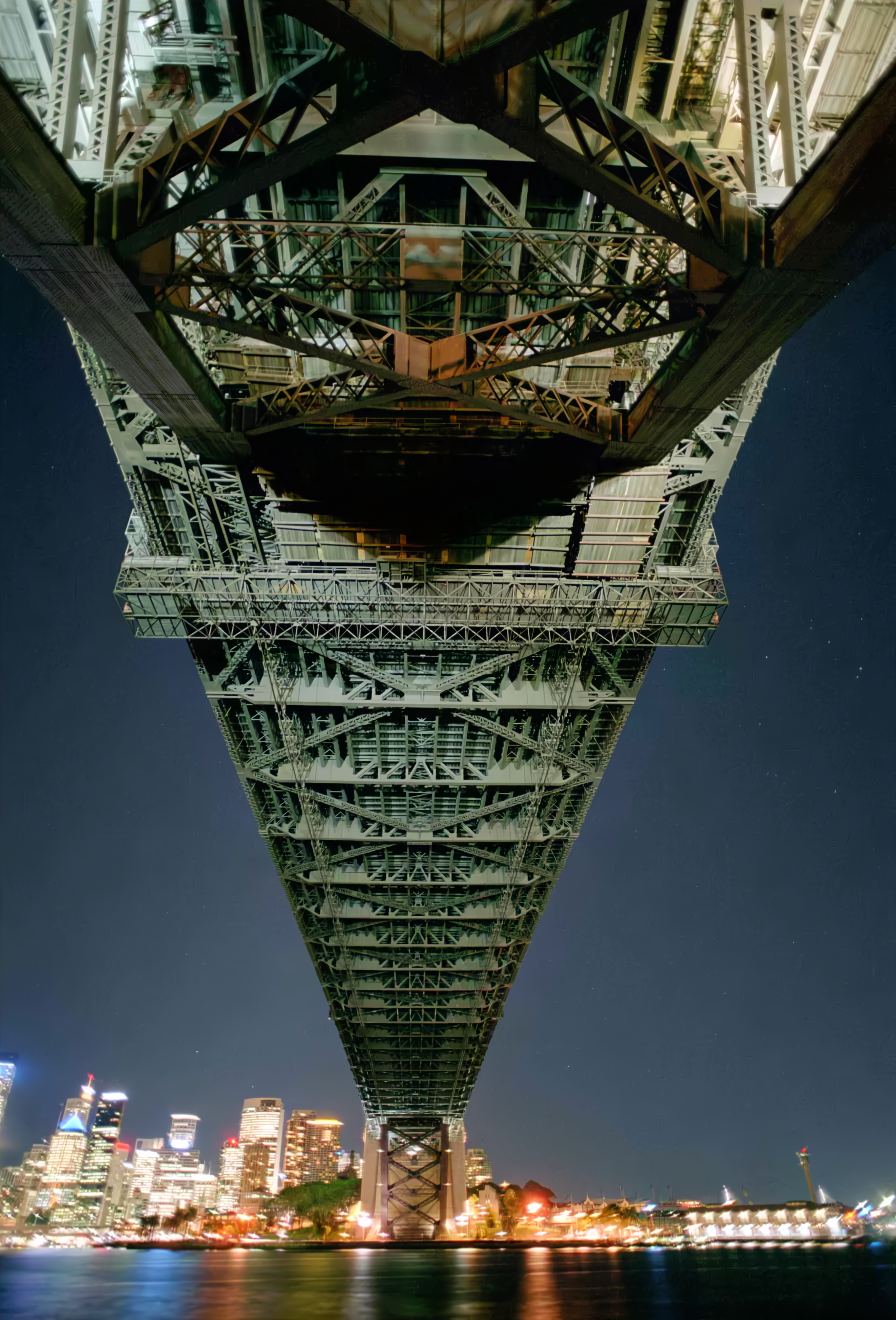} &
            \includegraphics[width=0.3\linewidth]{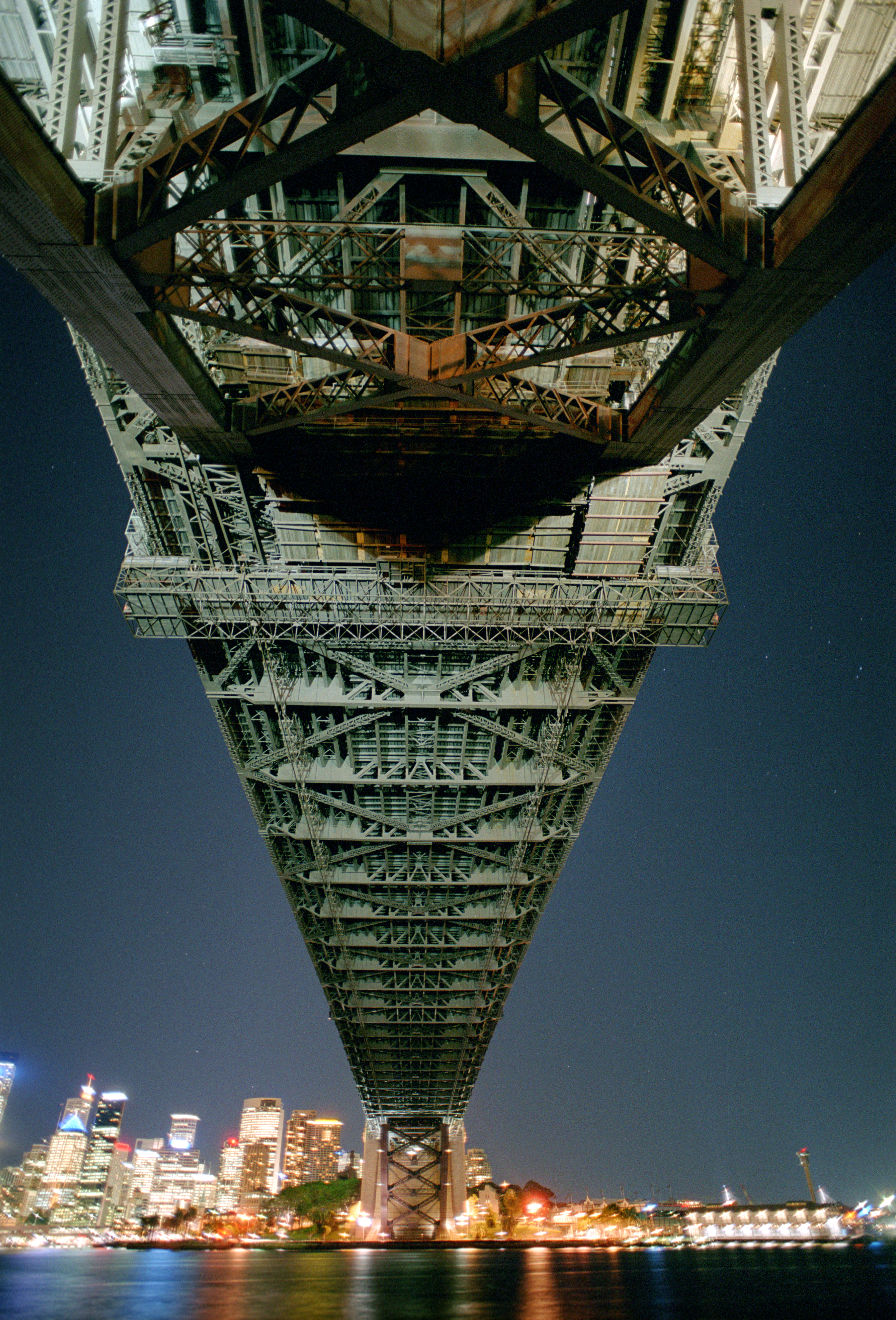} \\
            JPEG Q=20 & Ours & Original \\ 
            \includegraphics[width=0.3\linewidth]{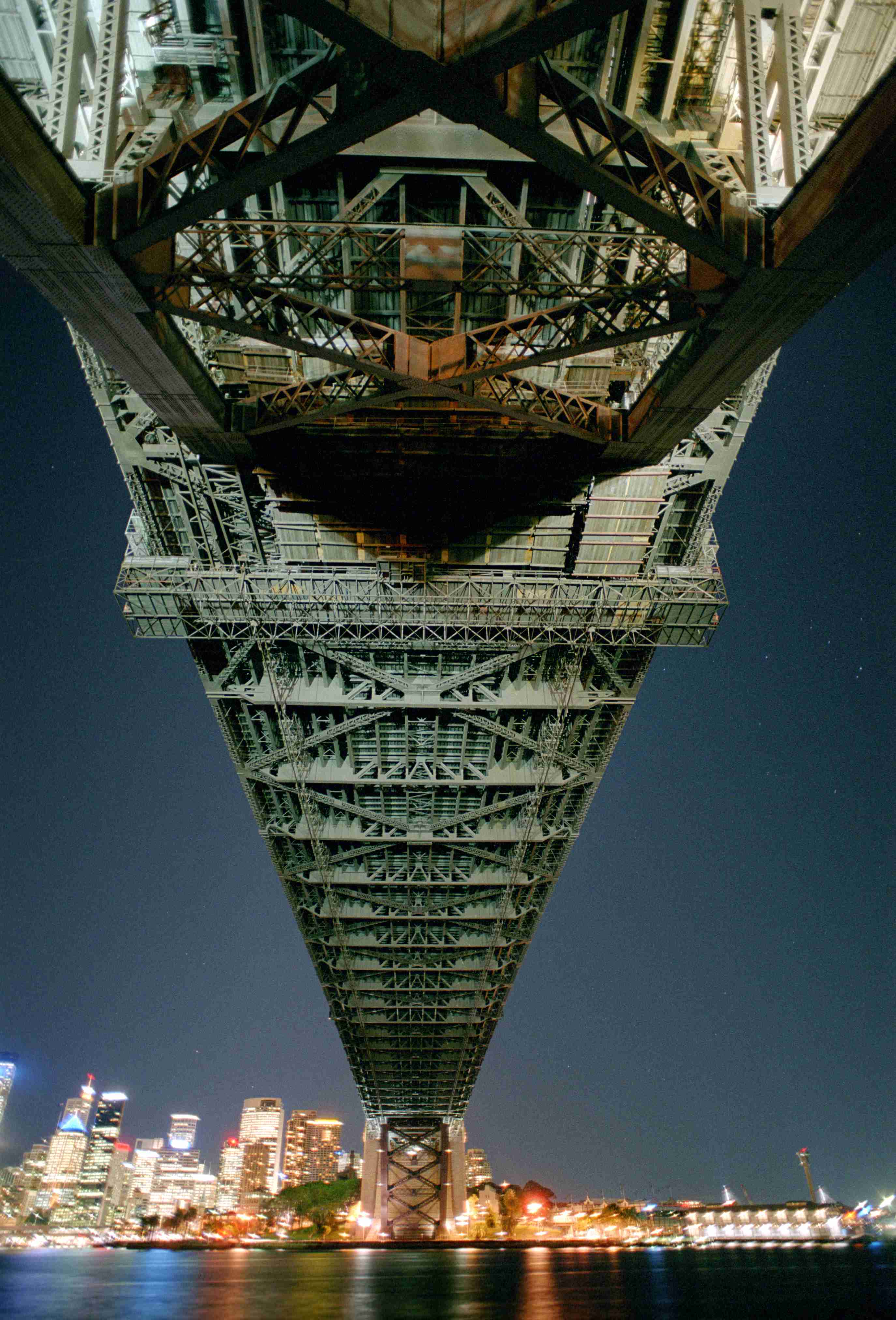} & 
            \includegraphics[width=0.3\linewidth]{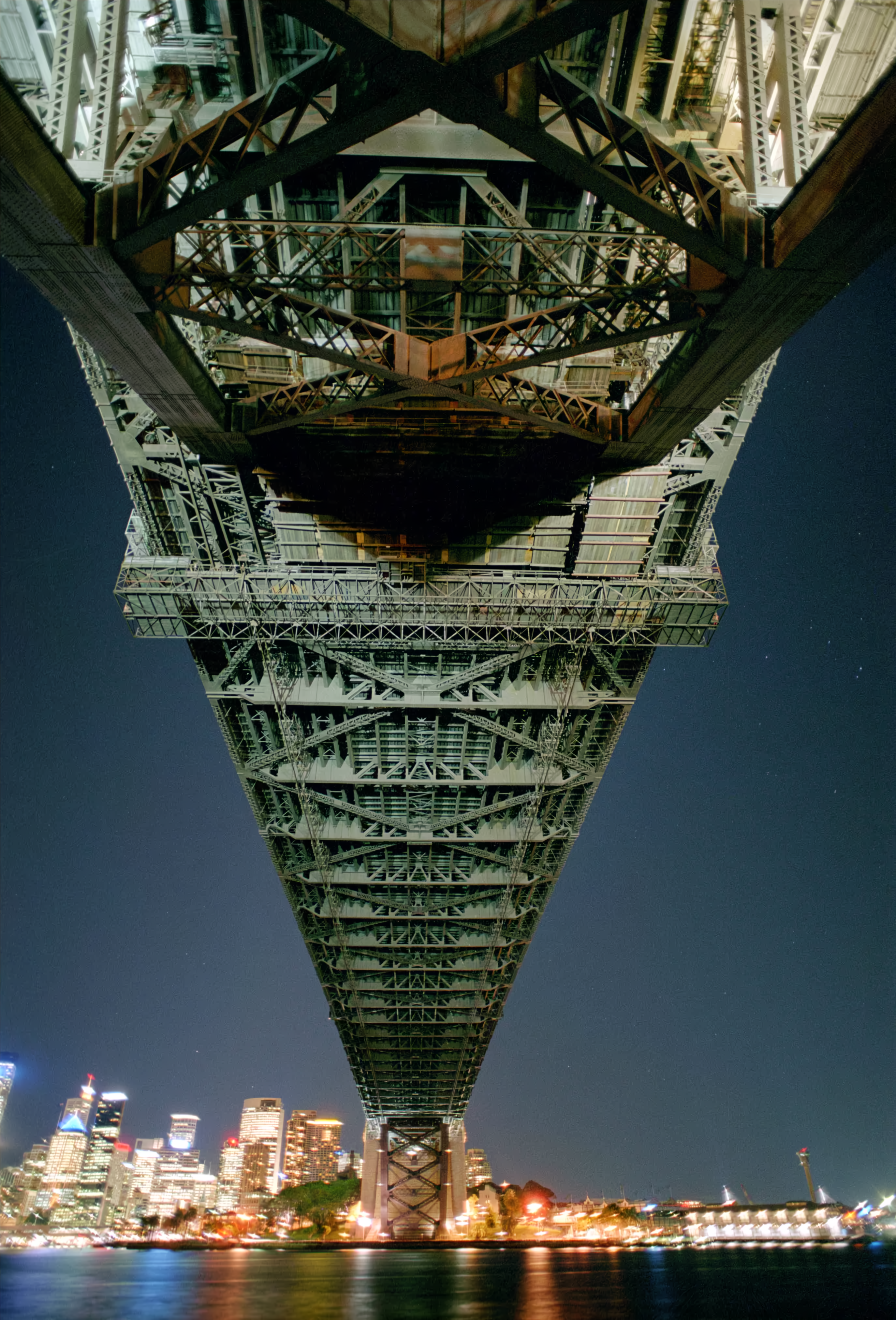} &
            \includegraphics[width=0.3\linewidth]{figures/qual/bridge/original.png} 
        \end{tabular}
    }
    \caption{Qualitative results 2/2. ICB images.}
    \label{fig:qual}
\end{figure}

\section{JPEG Compression Algorithm}

Since the JPEG algorithm is core to the operation of our method, we describe it here in detail. Where the JPEG standard is ambiguous or lacking in guidance, we defer to the Independent JPEG Group's libjpeg software.

\subsubsection{Compression}

JPEG compression starts with an input image in RGB color space (for grayscale images the procedure is the same using only the Y channel equations) where each pixel uses the 8-bit unsigned integer represenation (\eg the pixel value is an integer in [0, 255]). The image is then converted to the YCbCr color space using the full 8-bit 
represenation (pixel values again in [0, 255], this is in contrast to the more common ITU-R BT.601 standard YCbCr color conversion) using the equations:
\begin{align}
    Y = 2.99R + 0.587B + 0.114G \\
    Cb = 128 - 0.168736R - 0.331264B + 0.5G \nonumber \\
    Cr = 128 + 0.5R - 0.418688B - 0.081312G \nonumber
\end{align}

Since the DCT will be taken on non-overlapping $8 \times 8$ blocks, the image is then padded in both dimensions to a multiple of 8. Note that if the color channels will be chroma subsampled, as is usually the case, then the image must be padded to the scale factor of the smallest channel times 8 or the subsampled channel will not be an even number of blocks. In most cases, chroma subsampling will be by half, so the image must be padded to a multiple of 16, this size is referred to as the minimum coded unit (MCU), or macroblock size. The padding is always done by repeating the last pixel value on the right and bottom edges. The chroma channels can now be subsampled.

Next the channels are centered around zero by subtracing 128 from each pixel, yielding pixel values in [-128, 127]. Then the 2D Discrete type 2 DCT is take on each non-overlapping $8 \times 8$ block as follows:
\begin{align}
    D_{i, j} = \frac{1}{4}C(i)C(j)\sum_{x=0}^7\sum_{y=0}^7P_{x,y}\cos\left[\frac{(2x+1)i\pi}{16}\right]\cos\left[\frac{(2y+1)j\pi}{16}\right] \\
    C(u) = \left\{\begin{array}{lr}
            \frac{1}{\sqrt{2}} & u = 0 \\
            1 & \text{otherwise}
        \end{array}\right. \nonumber
\end{align}
Where $D_{i,j}$ gives the coefficient for frequency $i, j$, and $P_{x,y}$ gives the pixel value for image plane $P$ at position pixel position $x,y$. Note that $C(u)$ is a scale factor that ensures the basis is orthonormal. 

\begin{figure}
    \centering
    \begin{minipage}{0.4\linewidth}
    \resizebox{\columnwidth}{!}{
        \includegraphics{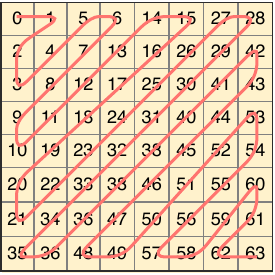}
    }
    \end{minipage}
    \caption{\textbf{Zigzag Ordering}}
    \label{fig:zig}
\end{figure}

The DCT coefficients can now be quantized. This follows the same procedure for the Y and color channels but with different quanitzation tables. We encourage readers to refer to the libjpeg software for details on how the quantization tables are computed given the scalar quality factor, an integer in [0, 100] (this is not a standardized process). Given the quantization tables $Q_Y$ and $Q_C$, the quanized coeffcients of each block are computed as:
\begin{align}
    Y'_{i, j} = \text{truncate}\left[\frac{Y_{i, j}}{Q_{Y_{i,j}}}\right] \\
    Cb'_{i, j} = \text{truncate}\left[\frac{Cb_{i, j}}{Q_{C_{i,j}}}\right] \nonumber \\ 
    Cr'_{i, j} = \text{truncate}\left[\frac{Cr_{i, j}}{Q_{C_{i,j}}}\right] \nonumber
    \label{eq:quant}
\end{align}
\
The quantized coefficients for each block are then vectorized (flattened) using a zig-zag ordering (see Figure \ref{fig:zig}) that is designed to place high frequencies further towards the end of the vectors. Given that high frequencies have lower magnitude and are more heavily quanitized, this usually creates a run of zeros at the end of each vector. The vectors are then compressed using run-length encoding on this final run of zeros (information prior to the final run is not run-length encoded.). The run-length encoded vectors are then entropy coded using either huffman coding or arithmetic coding and then written to the JPEG file along with associated metadata (EXIF tags), quantization tables, and huffman coding tables.

\subsubsection{Decompression}

The decompression algorithm largely follows the reverse procedure of the compression algorithm. After reading the raw array data, huffman tables, and quantization tables, the entropy coding, run-length coding, and zig-zag ordering is reversed. We reiterate here that the JPEG file does not store a scalar quality from which the decompressor is expected to derive a quanitzation table, the decompressor reads the quanitzation table from the JPEG file and uses it directly, allowing any software to correctly decode JPEG files that were not written by it. 

Next, the $8 \times 8$ blocks are scaled using the quantization table:
\begin{align}
    Y_{i, j} = Y'_{i, j}Q_{Y_{i,j}} \\
    Cb_{i, j} = Cb'_{i, j}Q_{C_{i,j}} \nonumber \\ 
    Cr_{i, j} = Cr'_{i, j}Q_{C_{i,j}} \nonumber
\end{align}
There are a few things to note here. First, if dividing by the quantization table entry during compression (Equation \ref{eq:quant}) resulted in a fractional part (the result was not an integer), that fractional part was lost during truncation and the scaling here will recover an integer near to the true coefficient (how close it gets depends on the magnitude quantization table entry). Next, if the division in Equation \ref{eq:quant} resulted in a number in [0, 1), then that coeffient would be truncated to zero and is lost forever (it remains zero after this scaling process). This is the \textit{only} source of loss in JPEG compression, however it allows for the result to fit into integers instead of floating point numbers, and it creates larger runs of zeros which leads to significantly larger compression ratios.

Next, the DCT process for each block is reversed using the 2D Discrete type 3 DCT:
\begin{align}
    P_{x, y} = \frac{1}{4}\sum_{i=0}^7\sum_{j=0}^7C(i)C(j)D_{i,j}\cos\left[\frac{(2x+1)i\pi}{16}\right]\cos\left[\frac{(2y+1)j\pi}{16}\right] \\
    C(u) = \left\{\begin{array}{lr}
            \frac{1}{\sqrt{2}} & u = 0 \\
            1 & \text{otherwise}
        \end{array}\right. \nonumber
\end{align}
and the blocks are arranged in their correct spatial positions. The pixel values are uncentered (adding 128 to each pixel value), and the color channels are interpolated to their original size. Finally, the image is converted from YCbCr color space to RGB color space:
\begin{align}
R = Y + 1.402(Cr - 128) \\ 
G = Y - 0.344136(Cb - 128) -0.714136(Cr - 128) \nonumber \\
B = Y + 1.772(Cb - 128) \nonumber
\end{align}
and cropped to remove any block padding that was added during compression. The image is now ready for display.

\end{document}